\documentclass{aastex6}
\usepackage{amsmath}
\usepackage{multirow}
\usepackage{hyperref}
\usepackage{textcomp}
\usepackage{cleveref}[2012/02/15]
\usepackage[toc,page]{appendix}

\usepackage{fancyhdr}

\crefformat{footnote}{#2\footnotemark[#1]#3}

\newcommand{\HI}{\ion{H}{1}}
\newcommand{\et}{et al.}
\newcommand{\kms}{km~s$^{-1}$}
\newcommand{\s}{$\sim$}
\newcommand{\n}{$-$}

\begin{document}
\slugcomment{\today}

\title{The \HI\ Chronicles of LITTLE THINGS BCDs III: Gas Clouds in and around Mrk 178, VII Zw 403, and NGC 3738}

\author{Trisha Ashley\altaffilmark{1}}
\affil{Department of Physics, Florida International University\\ 11200 SW 8th Street, CP 204, Miami, FL 33199}
\email{trisha.l.ashley@nasa.gov}
\and

\altaffiltext{1}{Current Address: NASA Ames Research Center, Moffett Field, CA, 94035}

\author{Caroline E. Simpson}
\affil{Department of Physics, Florida International University\\ 11200 SW 8th Street, CP 204, Miami, FL 33199}
\email{simpsonc@fiu.edu}
\and

\author{Bruce G. Elmegreen}
 \affil{IBM T. J. Watson Research Center,\\ 1101 Kitchawan Road, Yorktown Heights, New York 10598}
\email{bge@us.ibm.com}
\and

\author{Megan Johnson}
\affil{CSIRO Astronomy \& Space Science\\ P.O. Box 76, Epping, NSW 1710 Australia}
\email{megan.johnson@csiro.au}
\and

\author{Nau Raj Pokhrel}
\affil{Department of Physics, Florida International University\\ 11200 SW 8th Street, CP 204, Miami, FL 33199}
\email{npokh001@fiu.edu}

 \begin{abstract}
In most blue compact dwarf (BCD) galaxies, it remains unclear what triggers their bursts of star formation.  We study the \HI\ of three relatively isolated BCDs, Mrk 178, VII Zw 403, and NGC 3738, in detail to look for signatures of star formation triggers, such as gas cloud consumption, dwarf-dwarf mergers, and interactions with companions.  High angular and velocity resolution atomic hydrogen (\HI) data from the Very Large Array (VLA) dwarf galaxy \HI\ survey, Local Irregulars That Trace Luminosity Extremes, The \HI\ Nearby Galaxy Survey (LITTLE THINGS), allows us to study the detailed kinematics and morphologies of the BCDs in \HI.  We also present high sensitivity \HI\ maps from the NRAO Green Bank Telescope (GBT) of each BCD to search their surrounding regions for extended tenuous emission or companions.  The GBT data do not show any distinct galaxies obviously interacting with the BCDs.  The VLA data indicate several possible star formation triggers in these BCDs. Mrk 178 likely has a gas cloud impacting the southeast end of its disk \textit{or} it is experiencing ram pressure stripping.  VII Zw 403 has a large gas cloud in its foreground or background that shows evidence of accreting onto the disk. NGC 3738 has several possible explanations for its stellar morphology and \HI\ morphology and kinematics: an advanced merger, strong stellar feedback, or ram pressure stripping.  Although apparently isolated, the \HI\ data of all three BCDs indicate that they may be interacting with their environments, which could be triggering their bursts of star formation.
\end{abstract}
 
\keywords{galaxies: dwarf -- galaxies: individual (Mrk 178, VII Zw 403, NGC 3738) -- galaxies: star formation}
 
\section{Introduction}

Dwarf galaxies are typically inefficient star-formers \citep{leroy08}.   However, blue compact dwarf (BCD) galaxies are low-shear, high-gas-mass fraction dwarfs that are known for their dense bursts of star formation in comparison to other dwarfs \citep{thuan81, gil03}.  It is often suggested that the enhanced star formation rates in BCDs come from interactions with other galaxies or that they are the result of dwarf-dwarf mergers \citep{taylor97, noeske01, pustilnik01, bekki08, delgado12}.  Yet, there are still many BCDs that are relatively isolated with respect to other galaxies, making an interaction or merger scenario less likely \citep{taylor97, nicholls11, simpson11, ashley13}.  Other methods for triggering the burst of star formation in BCDs have been suggested, from accretion of intergalactic medium (IGM) to material sloshing in dark matter potentials \citep{wilcots98, brosch04, simpson11, helmi12, verbeke14}, but it remains unknown what has triggered the burst of star formation in a majority of BCDs.  

Understanding star formation triggers in BCDs is important for understanding how/whether BCDs evolve into/from other types of dwarf galaxies.  So far, attempts to observationally place BCDs on an evolutionary path between different types of dwarf galaxies have been largely unsuccessful \citep{papaderos96b, vanzee01b, tajiri02, gil05}.  BCDs have vastly different stellar characteristics than irregular and elliptical dwarf galaxies.  Interaction at a distance between dwarf galaxies has been suggest as a possible pathway to BCD formation \citep{pustilnik01, bekki08}, however, recent studies show observationally that dwarf galaxy pairs do not have enhanced star formation, only extended neutral gas components \citep{brosch04, pearson16}. Some authors have had success modeling the formation of BCDs through consumption of IGM \citep{verbeke14} and dwarf-dwarf mergers \citep{bekki08} and there is some evidence that these processes may contribute to the formation of individual BCDs, however, these processes have yet to be \textbf{confirmed observationally for BCDs as a group}.  

There has been evidence, in case studies of individual BCDs, that external gas clouds and mergers could be important for triggering bursts of star formation. \citet{vanzee98} found tidal features in the \HI\ of the BCD II Zw 40.  This galaxy has no known nearby companion and therefore could be an example of an advanced merger. Haro 36 is a BCD that is thought to be relatively isolated, has a kinematically distinct gas cloud in the line of sight, an \HI\ tidal tail, and may be showing some signs of an associated stellar tidal tail \citep{ashley13}.  These features led \citet{ashley13} to conclude that Haro 36 is likely the result of a merger.  IC 10 is also an interesting BCD that could be experiencing IGM accretion or is the result of a merger.  \citet{nidever13} present a new \HI\ extension that extends to the north of IC 10's  main disk.  \citet{ashley14} find that IC 10's \HI\ northern extension and the \HI\ southern plume are likely IGM filaments being accreted onto the IC 10's main disk \textit{or} extensive tidal tails that are evidence of IC 10 being an advanced merger.  For each of the above examples, the galaxies were studied as individuals rather than just one galaxy in a very large sample.  Studying the properties of BCDs as individual galaxies may therefore be the key to understanding what has triggered their burst of star formation, since each BCD is morphologically and kinematically so distinct.

In this paper we present high angular and high velocity resolution Very Large Array\footnote{\label{note1}The National Radio Astronomy Observatory is a facility of the National Science Foundation operated under cooperative agreement by Associated Universities, Inc.} (VLA) \HI\ data from the LITTLE THINGS\footnote{Local Irregulars That Trace Luminosity Extremes, The \HI\ Nearby Galaxy Survey; \url{https://science.nrao.edu/science/surveys/littlethings}} project \citep{hunter12} in order to investigate the internal \HI\ morphologies and kinematics of Mrk 178, VII Zw 403, and NGC 3738.  For basic information on these galaxies see Table~\ref{tab:galinfo}.  We also present higher sensitivity \HI\ data from the Robert C. Byrd Green Bank Telescope\cref{note1} (GBT) encompassing a total area of 200 kpc$\times$200 kpc and a velocity range of 2500 \kms. These data are used to study each BCD individually in order to look for evidence of a star formation triggers.

%\begin{landscape}
\begin{deluxetable}{lccccccc}
\tablecaption{Basic Galaxy Information\label{tab:galinfo}}
\tabletypesize{\footnotesize}
\tablewidth{0pt}
\tablehead{
\colhead{Galaxy} &  \colhead{RA (2000.0)} &  \colhead{Dec (2000.0)} &  \colhead{Distance\tablenotemark{a}} & \colhead{Systemic} & \colhead{$\rm{R}_{\rm{D}}$\tablenotemark{b}} & \colhead{$\rm{log\ SFR}_{\rm{D}}$\tablenotemark{c}} & \colhead{$\rm{M}_{\rm{V}}$\tablenotemark{a}} \\ \colhead{Name} & \colhead{(hh mm ss.s)} & \colhead{(dd mm ss)} & \colhead{(Mpc)} & \colhead{Velocity (\kms)} & \colhead{(kpc)} & \colhead{($\rm{M}_{\sun}\ \rm{yr}^{-1}\ \rm{kpc}^{-2}$)} &  \colhead{(mag)}}

\startdata
Mrk 178 & 11 33 29.0 & 49 14 24 & 3.9 & 250 & $0.33\pm0.01$ & $-1.60\pm0.01$ & $-14.1$ \\  
VII Zw 403 & 11 27 58.2 & 78 59 39 & 4.4 & -103 & $0.52\pm0.02$ & $-1.71\pm0.01$ & $-14.3$ \\  
NGC 3738 & 11 35 49.0 & 54 31 23 & 4.9 & 229 & $0.78\pm0.01$ & $-1.66\pm0.01$ & $-17.1$ \\  
\enddata

\tablenotetext{a}{\citet{hunter12}}
\tablenotetext{b}{$\rm{R}_{\rm{D}}$ is the V-band disk exponential scale length \citep{hunter06}.}
\tablenotetext{c}{$\rm{SFR}_{\rm{D}}$ is the star formation rate, measured from H$\alpha$ data, normalized to an area of $\pi \rm{R}_{\rm{D}}^{2}$ \citep{hunter12}}
\end{deluxetable}
%\end{landscape}

\section{Sample}
\subsection{Mrk 178}

Mrk 178 (=UGC 6541) is a galaxy that has a confusing classification record in the literature; it has been classified as a merger \citep{pustilnik01} and two separate articles have suggested that Mrk 178 has a nearby companion (a different companion in each paper) at a velocity close to its own velocity \citep{peterson79, thuan81}.  Upon further investigation, none of these claims can be verified \citep[as also pointed out by][]{schulte00}.  \citet{pustilnik01} cite \citet{mazzarella91} to support the idea that Mrk 178 is a merger, however, \citet{mazzarella91} do not claim that it is a merger, merely a dwarf irregular. \citet{thuan81} suggest that UGC 6538 is a companion to Mrk 178, however, the velocity difference of these two galaxies is almost 2800 \kms\ (velocities taken from NED\footnote{NASA/IPAC Extragalactic Database (NED) http://ned.ipac.caltech.edu/}), making UGC 6538 more likely a background galaxy.  \citet{peterson79} suggest that Mrk 178 is a close pair with UGC 6549, however, the velocity difference between these two galaxies is more than 9000 \kms\ (velocities taken from NED), making it a very unlikely candidate for a companion to Mrk 178 and probably a spatially coincident background galaxy.  The classification of Mrk 178 in \citet{peterson79} has led NED to label Mrk 178 as a galaxy in a pair.  \citet{hunter04} find, from a NED search, that the closest galaxy to Mrk 178 within $\pm$150 \kms\  is NGC 3741 at a distance of 410 kpc and a velocity difference of 20 \kms.  Assuming that the line of sight velocity difference between Mrk 178 and NGC 3741 is comparable to their transverse velocity, this distance is too large for Mrk 178 to have recently interacted with NGC 3741; using their relative velocity and their approximate distance, they would require 20 Gyr to meet, making an interaction between the two highly unlikely.  Mrk 178 is located roughly in the Canes Venetici I group of galaxies \citep{karachentsev03}.

The stellar and gas components of Mrk 178 have been previously studied \citep{stil02a, roychowdhury09, lelli13, kehrig13}.  \citet{stil02a} studied Mrk 178's \HI\ distribution using the Westerbork Synthesis Radio Telescope (WSRT) .  Their integrated \HI\ intensity map, at a resolution of 13\arcsec, shows a broken ring-like structure.  Mrk 178 is also part of a survey that uses the Giant Metrewave Radio Telescope (GMRT) called
the Faint Irregular Galaxies GMRT Survey (FIGGS).  \citet{roychowdhury09} present Mrk 178's FIGGS data, but do not discuss the morphology or kinematics of Mrk 178 as an individual galaxy.  Their integrated \HI\ intensity maps, at resolutions of 22\farcs68 and 11\farcs71, show that Mrk 178 has a highly irregular shape to its gaseous disk.

Mrk 178's stellar population is well known for its young Wolf-Rayet (WR) features \citep{schulte00, guseva00, brinchmann08, stevens98, kehrig13}.  A detailed study of Mrk 178's WR population by \citet{kehrig13} revealed a large number of WR stars in its brightest stellar component.  Mrk 178's stellar population was studied in detail by \citet{schulte00}.  Their results indicate that Mrk 178 had a higher star formation rate \s0.5 Gyr ago when compared to its current star formation rate.  Their results also indicate that Mrk 178 has an old underlying stellar population.

\subsection{VII Zw 403}

VII Zw 403 (=UGC 6456) is a well known isolated BCD without obvious signatures of tidal interaction \citep{schulte98, noeske01, pustilnik01, simpson11}.  It sits just beyond the M81 group and is falling in towards the M81 group \citep{karachentsev02}.  \citet{hunter04} found that the closest galaxy to VII Zw 403 within $\pm$150 \kms\ is KDG 073 at a distance of 900 kpc and a velocity difference of 32 \kms.  Companions that are currently interacting are not likely to be more than \s100 kpc away from each other and are likely to be close in velocity \citep{taylor95, pustilnik01, thilker04, westmeier08, chynoweth11}.  Using the relative velocity and distance of VII Zw 403 and KDG 073 to estimate the time it would take for them to have passed each other last and assuming that their line of sight velocity difference is comparable to their transverse velocity, we note that it would be 28 Gyr since their last interaction or twice the age of the universe, making it highly unlikely that they have interacted.  

VII Zw 403's \HI\ has been studied using the GBT, Nan\c{c}ay, the GMRT, and the VLA \citep{tully81, paturel03, begum08, roychowdhury09, thuan04, simpson11}.  The \HI\ data reveal a galaxy that has a disturbed velocity field and irregular morphology.  The VLA data presented in \citet{simpson11} reveal a break in the major axis of the isovelocity contours and a possible \HI\ hole.  \citet{simpson11} conclude that VII Zw 403 may have experienced an \HI\ accretion event in its past, which is now difficult to detect.  

\citet{schulte01} modeled the star formation history of VII Zw 403 using far infrared Hubble Space Telescope data.  They concluded that VII Zw 403's star formation has been continuous and not episodic, with an increased star formation rate over the past Gyr.  \citet{lynds98} also modeled VII Zw 403's star formation history and showed a starburst occurring in the 600-800 Myr interval of VII Zw 403's past, followed by a lower star formation rate.  X-ray emission was detected in VII Zw 403 using the PSPC instrument on the ROSAT satellite \citep{papaderos94}.  There was apparent extended X-ray emission in the form of three diffuse arms, which all emanated from a central source.  A point source located at the central X-ray source was later confirmed by two studies \citep{lira00, ott05}, but the diffuse emission seen in \citet{papaderos94} was not detected by Chandra or ROSAT's HRI instrument.  

\subsection{NGC 3738}

NGC 3738 (=UGC 6565) is a galaxy that has not received a significant amount of individual attention.  NGC 3738 is not always classified as a BCD; however, \citet{vaduvescu06} suggest that the light profile properties of NGC 3738 are those of a BCD.  \citet{hunter06} also support this idea, with their light profile of NGC 3738 being most similar to the other eight BCDs in their sample. NGC 3738 has been included in several other large \HI\ surveys \citep{stil02a, paturel03}, however, it is not discussed in detail.  The map from \citet{stil02a} reveals a very clumpy and irregular \HI\ morphology at a resolution of 13\farcs5.  \citet{hunter04} find that the closest companion to NGC 3738 within $\pm$ 150 \kms\ is NGC 4068, at a distance of 490 kpc and a velocity difference of 19 \kms.  With these parameters as rough estimates (assuming their line of sight velocity difference is comparable to their transverse velocity), it is unlikely that these two galaxies would have met, since they would require \s25 Gyr to travel to each other.  NGC 3738 is located roughly in the Canes Venetici I group of galaxies \citep{karachentsev03}.

 \section{Observations and Data Reduction}\label{obs}
 
 \subsection{The Very Large Array Telescope}
 
 The VLA data of Mrk 178, VII Zw 403, and NGC 3738 were collected as part of the LITTLE THINGS project.  LITTLE THINGS is a survey of 41 dwarf galaxies; each galaxy in the survey has high angular (\s6\arcsec) and high velocity ($\le$2.6 \kms) resolution \HI\ data from the B, C, and D configurations of the VLA. For more information about LITTLE THINGS see \citet{hunter12}.  Basic observational parameters for Mrk 178, VII Zw 403, and NGC 3738 are given in Table~\ref{tab:vobsinfo}.   
 
A detailed description of data calibration and mapping techniques used for LITTLE THINGS can be found in \citet{hunter12}.  The VLA maps were made using the Multi-Scale (M-S) \textsc{clean} algorithm as opposed to the classical \textsc{clean} algorithm.  M-S \textsc{clean} convolves the data to several beam sizes (0\arcsec, 15\arcsec, 45\arcsec, 135\arcsec\ for LITTLE THINGS maps) and then searches the convolved data for the region of highest flux amongst all of the convolutions.  That region is then used for clean components.  The larger angular scales will map the tenuous structure, while the smaller angular scales will map the high resolution details of the \HI. Therefore, M-S \textsc{clean} allows us to recover tenuous emission while maintaining the high angular resolution details in the images.  Basic information on the VLA \HI\ maps of Mrk 178, VII Zw 403's, and NGC 3738 can be found in Table~\ref{vlamapinfo}.  For more information on the advantages of M-S Clean see \citet{cornwell08}.  

 \floattable 
\begin{deluxetable}{lcccc}
\tablecaption{VLA Observing Information \label{tab:vobsinfo}}
\tabletypesize{\footnotesize}
\tablewidth{\textwidth}
\tablehead{
\colhead{Galaxy Name} & \colhead{Configuration} &  \colhead{Date Observed} &  \colhead{Project ID} & \colhead{Time on Source (hours)}}
\startdata
\multirow{3}{*}{Mrk 178} & B & 08 Jan 15, 08 Jan 21, 08 Jan 27 & AH927 & 10.4\\
& C & 08 Mar 23, 08 Apr 15 & AH927 & 5.75\\
& D & 08 Jul 8, 08 Jul 24, 08 Jul 25 & AH927 &  1.75\\ \hline
\multirow{3}{*}{VII Zw 403} & B & 06 Sep 10 & AH907 & 8.55\\
& C & 92 Apr 11 & AH453 & 3.7\\
& D & 97 Nov 10 & AH623 &  4\\ \hline
\multirow{3}{*}{NGC 3738} & B & 08 Jan 15, 08 Jan 21, 08 Jan 27 & AH927 & 10.87\\
& C & 08 Mar 23, 08 Apr 15 & AH927 & 5.22\\
& D & 08 Jul 8, 08 Jul 24, 08 Jul 25 & AH927 &  1.69\\ 
\enddata
\end{deluxetable}

 \floattable 
%\begin{landscape}
\thispagestyle{empty}
\begin{deluxetable}{llcccc}
\tablecaption{VLA Map Information\label{vlamapinfo}}
\tabletypesize{\scriptsize}
\tablewidth{0pt}
\tablehead{
\colhead{Galaxy} & \colhead{Weighting} &  \colhead{Synthesized} &  \colhead{Linear} & \colhead{Velocity Resolution} & \colhead{RMS over \s10 \kms}\\  \colhead{Name} & \colhead{Scheme} & \colhead{Beam Size (\arcsec)} & \colhead{Resolution (pc)} & \colhead{(\kms)} & \colhead{($10^{19}$ atoms cm$^{-2}$)}}

\startdata
\multirow{2}{*}{Mrk 178} & Robust (r=0.5) & 6.19 $\times$ 5.48 & 120 & 1.29 & 4.6\\  
& Natural & 12.04 $\times$ 7.53 & 230 & 1.29 & 1.6 \\ \hline
\multirow{2}{*}{VII Zw 403} & Robust (r=0.5) & 9.44 $\times$ 7.68 & 200 & 2.58 & 3.7\\  
& Natural & 17.80 $\times$ 17.57 & 380 & 2.58 & 0.74\\  \hline
%& Convolved & 25 $\times$ 25 & 530 & 2.58 & 0.50\\ \hline
\multirow{2}{*}{NGC 3738} & Robust (r=0.5) & 6.26 $\times$ 5.51 & 150 & 2.58 & 4.7\\  
& Natural & 13.05 $\times$ 7.79 & 310 & 2.58 & 1.5 \\
%& Convolved & 25 $\times$ 25 & 590 & 2.58 & 0.61 \\
\enddata

\end{deluxetable}
%\end{landscape}

\subsection{The Green Bank Telescope}\label{gbtsection}

Mrk 178, VII Zw 403, and NGC 3738 were also observed in \HI\ with the GBT by two projects.  The GBT data are higher sensitivity and lower resolution than the VLA data, therefore, the GBT maps were used to search the surrounding regions for companion galaxies and extended, tenuous \HI\ emission, while the VLA maps were used to see the detailed morphology and kinematics of the \HI.  For basic GBT observing information, see Table~\ref{tab:gobsinfo}. The first project (Proposal ID GBT/12B-312; P.I. Johnson) covered a 2\degr$\times$2\degr\ field around each galaxy (140 kpc$\times$140 kpc for Mrk 178, 150 kpc$\times$150 kpc for VII Zw 403, and 170 kpc$\times$170 kpc for NGC 3738).  These data were combined with the second project (P.I. Ashley; Proposal ID GBT13A-430) which covered a 200 kpc$\times$200 kpc region around each BCD and a total velocity range of 2500 \kms\ to search for any extended emission and nearby companions \citep[a reasonable distance for interacting companions and extended \HI\ emission:][]{taylor95, pustilnik01, thilker04, westmeier08, chynoweth11}.   In order to make the maps' sensitivity uniform throughout a 200 kpc$\times$200 kpc region, the second project also observed the regions around the  2\degr$\times$2\degr\ maps from the first project to fill in the 200 kpc$\times$200 kpc region.

On-the-fly mapping was used for both projects, scanning in a raster pattern in Galactic latitude and longitude, and sampling at the Nyquist rate.  With a 12.5 MHz bandwidth, in-band frequency switching with a central frequency switch of 3.5 MHz was implemented to calibrate the data.  Using a code written by NRAO staff, the data were first corrected for stray radiation (with the exception of Mrk 178's data, see below) and then the data were Hanning smoothed to increase the signal-to-noise ratio.  Next, standard calibration was done using the GBTIDL\footnote{Developed by NRAO; for documentation see \url{http://gbtidl.sourceforge.net}.} task \textsc{getfs}.  Radio frequency interference (RFI) spikes were then manually removed by using the values of neighboring channels to linearly interpolate over the spike in frequency space\footnote{Mrk 178's data suffered significant RFI throughout most of the integrations; some integrations were flagged entirely.  We removed as much RFI as possible from other integrations, however, with about 50 hours of data and a 3 second integration time, there are about 60,000 integrations each with 2 polarizations, therefore we were unable to remove all RFI.  These RFI also often remained at low levels in individual integrations until scans were averaged together, making them difficult to find with available RFI finding visual tools and requiring manual removal.}.  After further smoothing, the spectra baselines were fit using third or fourth ordered polynomials to remove residual instrumental effects\footnote{NGC 3738's data contained two sources that are not near NGC 3738 in frequency space, but due to the frequency switching, appeared close to NGC 3738 in frequency space.   For more information on how this source was calibrated see Appendix~\ref{appendix:n3738}.}.  

Mrk 178's baselines contained low level sinusoidal waves throughout each session possibly due to a resonance in the receiver.  In order to remove the sinusoidal features, prior to any other calibration steps, we used a code written by Pisano, Wolfe, and Pingel (Pingel, private communications) that uses the ends of each row in the GBT map as faux off-positions.  These faux off-positions in each row are subtracted from the rest of the row, resulting in a stable baseline.  This procedure should not result in any loss of extended flux around Mrk 178 since it is using only blank pixels on the edges of each row in the map to subtract baselines from the corresponding row of pixels.  The code also removed the need for stray radiation corrections as much of the Milky Way effects were removed through subtraction of the faux off-positions.

After calibration, the data were imaged in AIPS.  \textsc{dbcon} was used to combine all of the sessions from both projects and \textsc{sdgrd} was then used to spatially grid the data.  The final GBT \HI\ maps were made in \textsc{xmom} and had their coordinates transformed from Galactic to Equatorial using the task \textsc{flatn} in AIPS.  Rotating the data in \textsc{flatn} to align the coordinates so that north faces up and east to the left (like the VLA maps) requires the data to be re-gridded in the process.  The re-gridding results in a slight change of flux values for each pixel, therefore, any measurements (mass, noise, etc.) obtained for the GBT data were taken prior to the rotation of the maps.  Also, each map was compared before and after the rotation occurred to look for any features in the map that may have been morphologically distorted.  The effects of re-gridding were inconsequential at the resolution of the GBT maps.  The rotation was done as a visual aid for the reader to easily compare the orientation of the VLA \HI\ maps to the GBT \HI\ maps.  For basic information on the individual GBT maps, see Table~\ref{gbtmapinfo}.

\begin{deluxetable}{lcc}
\tablecaption{GBT Observing Information \label{tab:gobsinfo}}
\tabletypesize{\small}
\tablewidth{0pt}
\tablehead{
\colhead{Galaxy Name} & \colhead{Total Time Observing\tablenotemark{a} (hours)} & \colhead{Angular Size Observed\tablenotemark{b} (degr)}}

\startdata
Mrk 178 & 58.5 & 2.9$\times$2.9\\
VII Zw 403 & 43.5 & 2.6$\times$2.6\\
NGC 3738 & 34.25 & 2.3$\times$2.3\\
\enddata

\tablenotetext{a} {Total time spent on observations including overhead time (e.g., moving the telescope, getting set up for observations, and time on calibrators.)}
\tablenotetext{b} {These angular sizes represent the 200$\times$200 kpc$^{2}$ fields of the galaxies.  However, for observing purposes these fields had 0.1 degrees added horizontally and vertically to them as a buffer.  This was done to account for the time that the telescope would start/stop moving vs. the time that the telescope began recording data.}

\end{deluxetable}

%\begin{landscape}
\thispagestyle{empty}
\begin{deluxetable}{lcccc}
\tablecaption{GBT Map Information\label{gbtmapinfo}}
\tabletypesize{\footnotesize}
\tablewidth{0pt}
\tablehead{
\colhead{Galaxy} &  \colhead{Beam Size} &  \colhead{Linear} & \colhead{Velocity Resolution} & \colhead{RMS over \s10 \kms} \\  \colhead{Name} & \colhead{(\arcsec)} & \colhead{Resolution (kpc)} & \colhead{(\kms)} & \colhead{($10^{16}$ atoms $\rm{cm}^{-2}$)}} 

\startdata
Mrk 178 & 522.85 $\times$  522.85 & 10 & 0.9661 & 6.7\\  
VII Zw 403 & 522.23 $\times$  522.23 & 11 & 0.9661 & 3.9\\  
NGC 3738 & 522.81 $\times$ 522.81 & 12 & 0.9661 & 3.6\\  
\enddata
\end{deluxetable}
%\end{landscape}
\clearpage

\section{Results: Mrk 178}
\subsection{Mrk 178: Stellar Component}

\begin{figure}[!ht]
\centering
\epsscale{1.11}
\plottwo{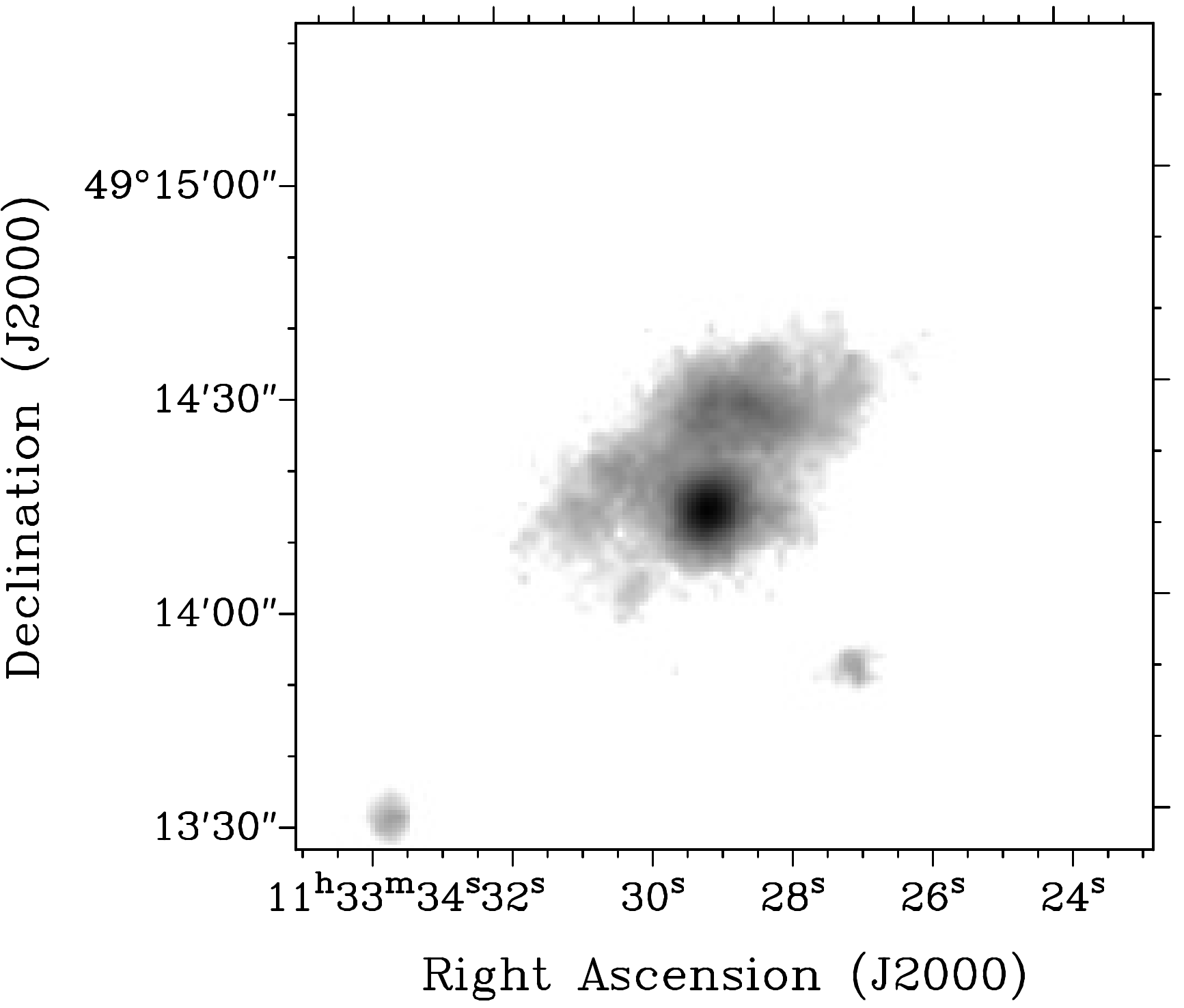}{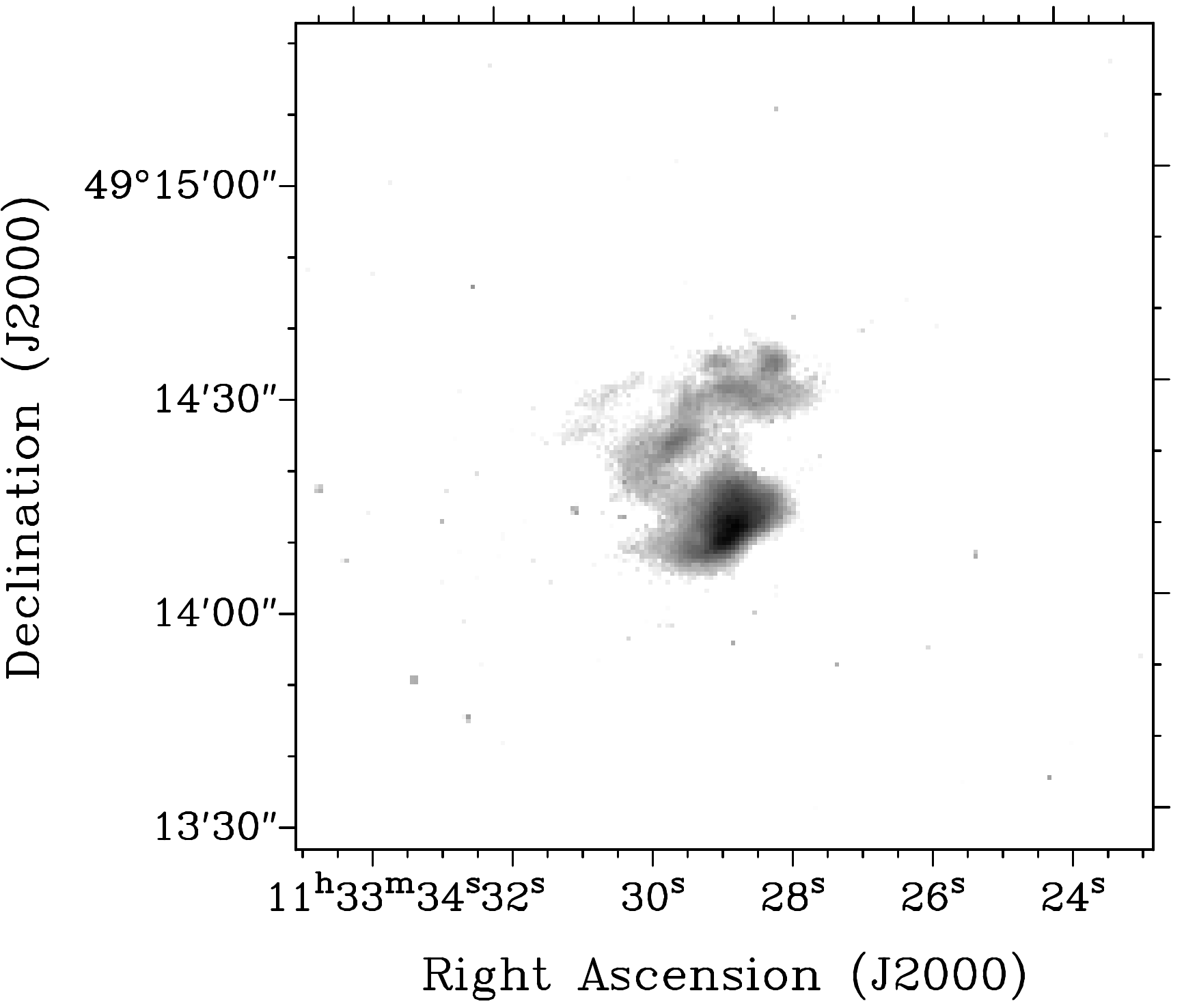}
\epsscale{0.5}
\plotone{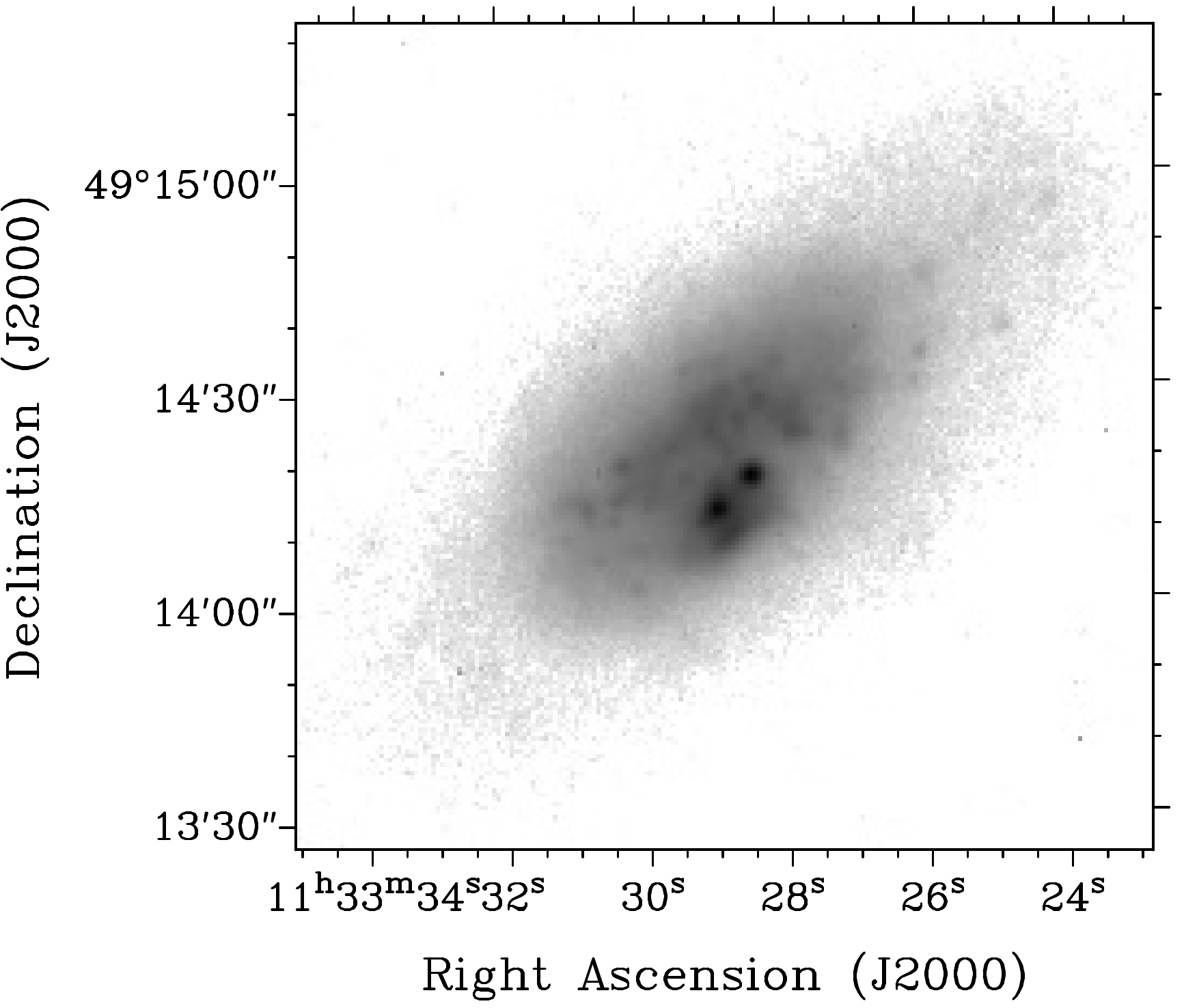}
\caption{Mrk 178's optical maps. \ \ \textit{Left:} FUV;\ \ \textit{Right:} H$\alpha$;\ \ \textit{Bottom:} V-band. \label{m178_star}}
\end{figure}

The FUV, H$\alpha$, and V-band maps of Mrk 178 are shown in Figure~\ref{m178_star}.   The FUV data were taken with GALEX \citep{hunter10}, and the H$\alpha$ and V-band data were taken with the Lowell Observatory 1.8m Perkins Telescope \citep{hunter04, hunter06}.  The FUV, H$\alpha$, and V-band surface brightness limits are \s28.5, \s28, and \s27 mag arcsec$^{-2}$, respectively \citep{herrmann13}.  All  three of these maps share a common feature in their morphology (most easily seen in the H$\alpha$ map): there appears to be a region of high stellar density to the south, then there is a curved structure of stars, which lies north of the high stellar density region and curves to the west (right).

\subsection{Mrk 178: VLA \HI\ Morphology}\label{m178vla_morph}
\begin{figure}[!ht]
\epsscale{0.49}
\plotone{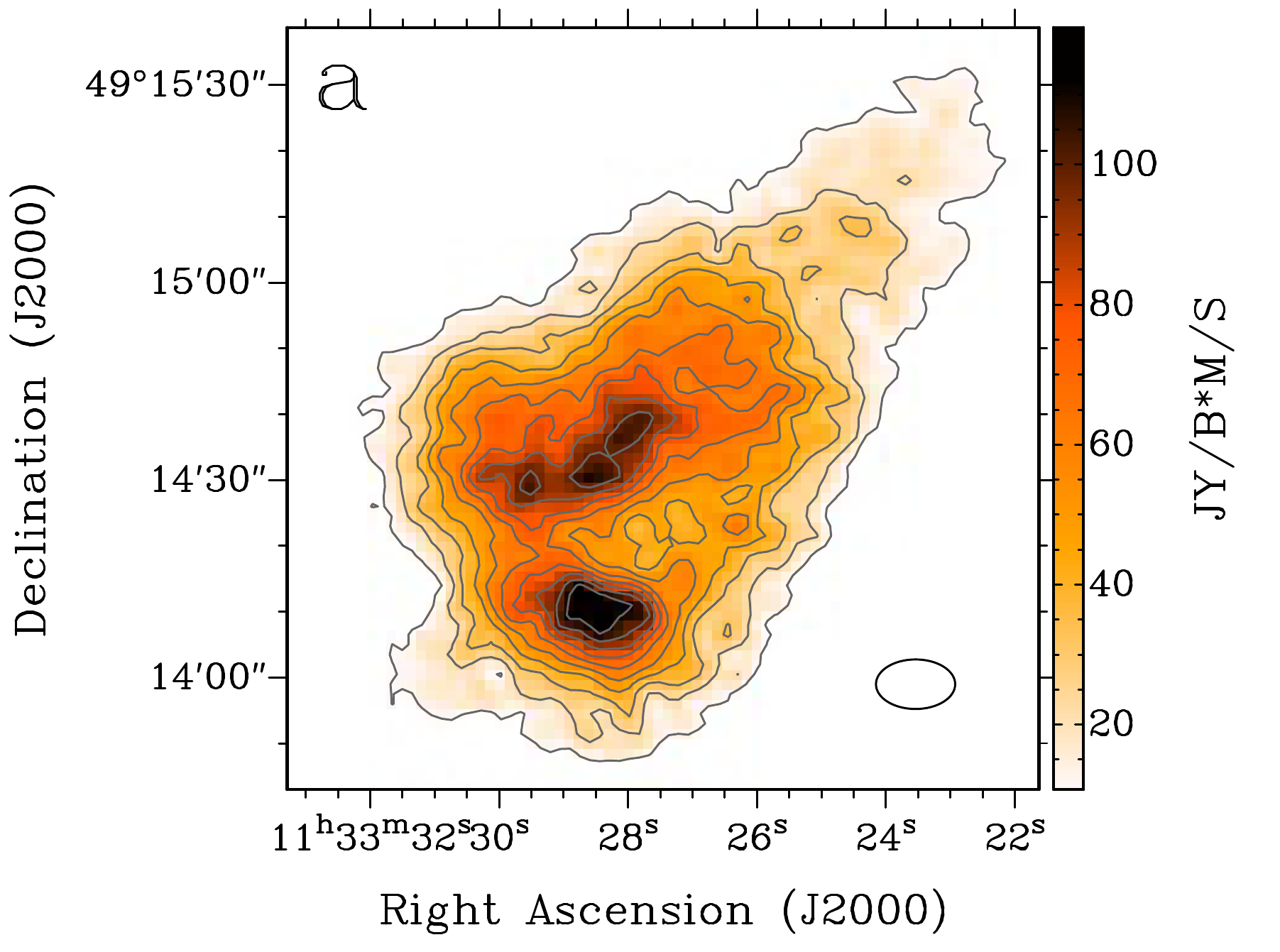}
\epsscale{0.50}
\plotone{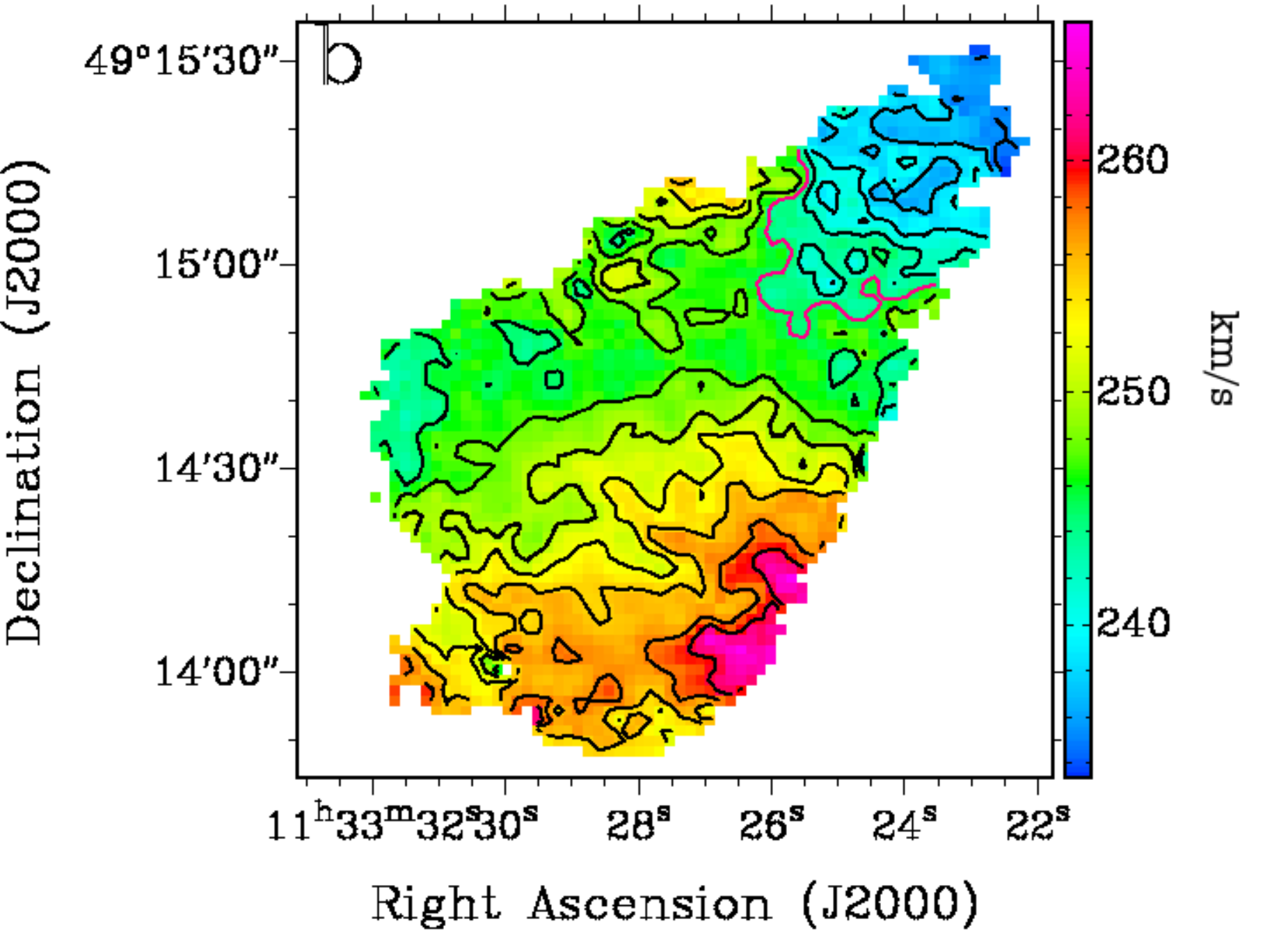}
\epsscale{0.49}
\plotone{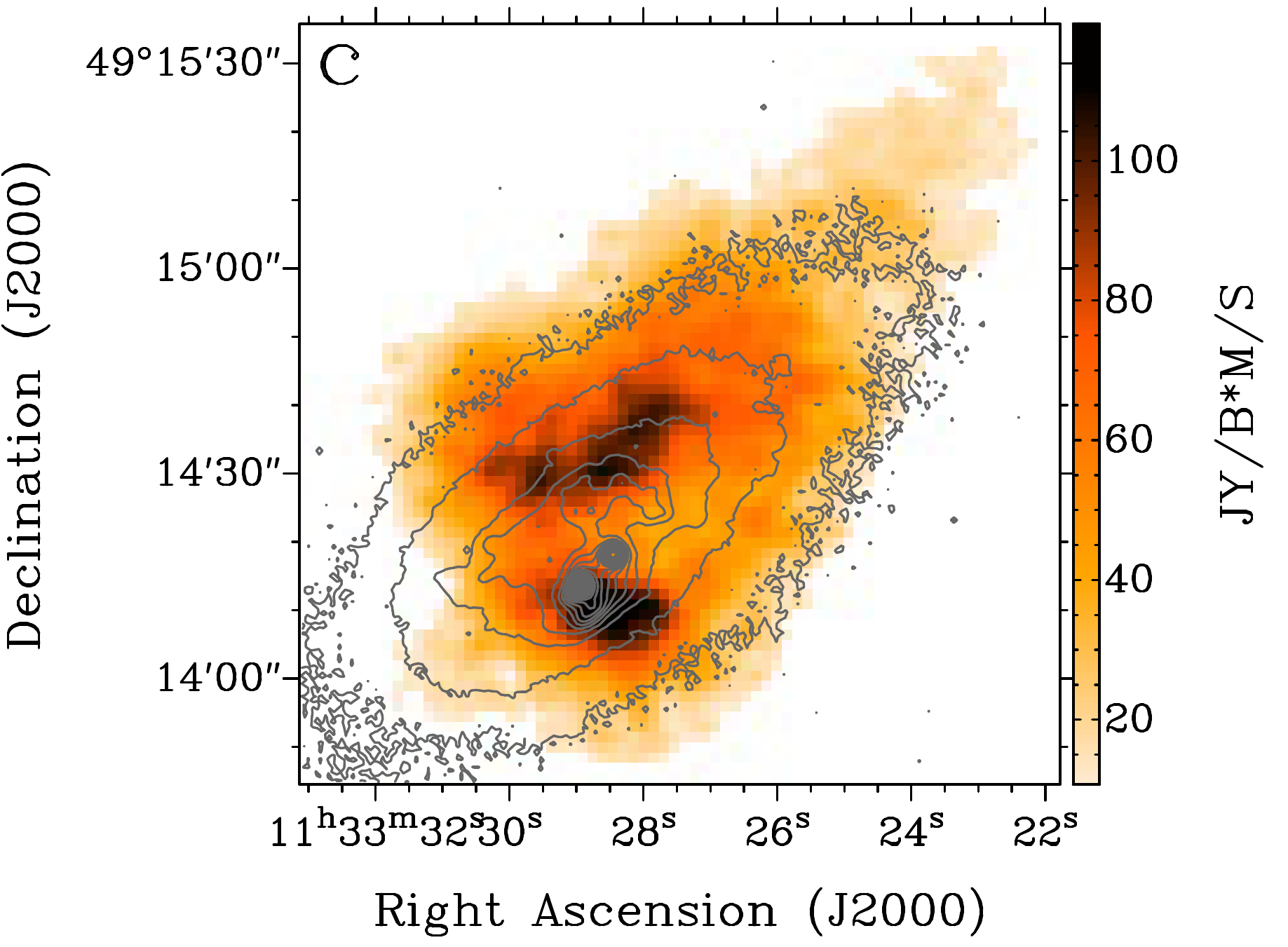}
\epsscale{0.51}
\plotone{{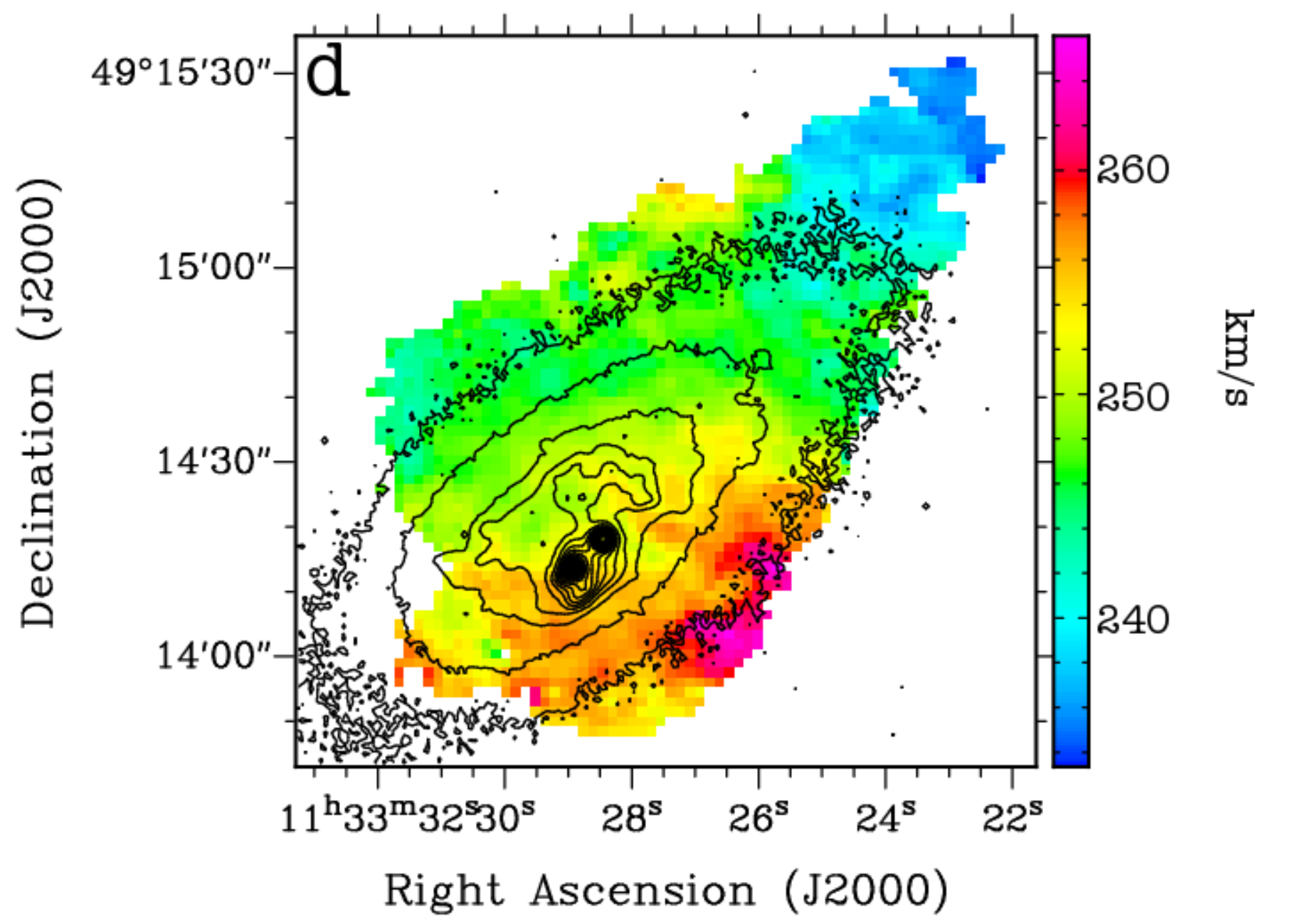}}\\
\epsscale{0.50}
\plotone{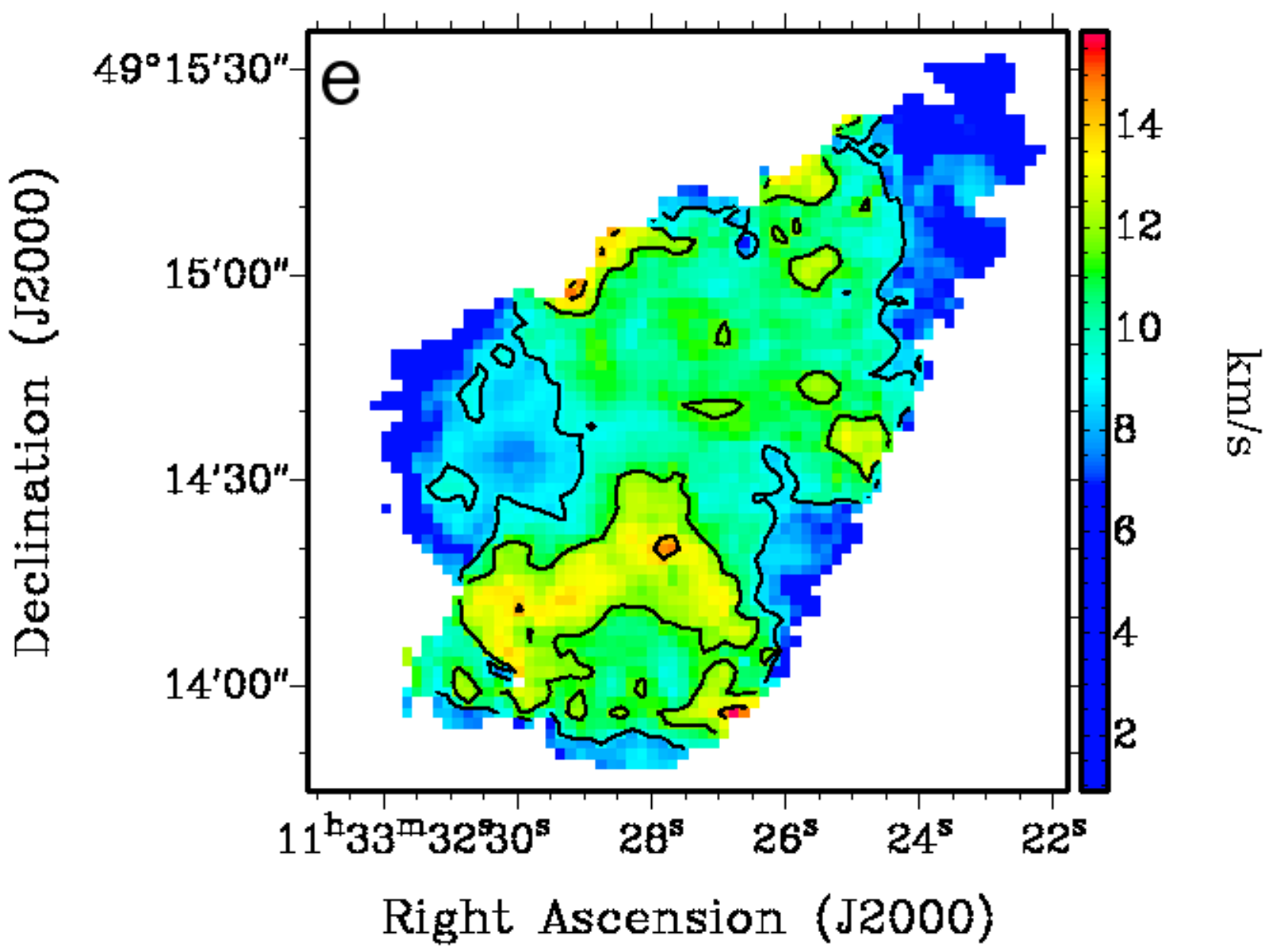}
\caption{Mrk 178's VLA natural-weighted moment maps. (a): Integrated \HI\ intensity map; contour levels are 1$\sigma\times$(2, 4, 6, 8, 10, 12, 14, 16, 18, 20) where 1$\sigma=6.68\times10^{19}\ \rm{atoms}\ \rm{cm}^{-2}$. The black ellipse in the bottom-right represents the synthesized beam. (b): Intensity-weighted velocity field; contour levels are 235 \kms\ to 260 \kms\ separated by 2.5 \kms. The red contour represents 245 \kms.  (c): The colorscale of Figure~\ref{m178vla_na}a with the V-band contours.  (d) The colorscale of Figure~\ref{m178vla_na}b with the V-band contours.  (e): Velocity dispersion field; contour levels are 9, 11.5, 14 \kms. \label{m178vla_na}}
\end{figure}

The VLA natural-weighted integrated \HI\ intensity map is shown in Figure~\ref{m178vla_na}a.  There are two distinct regions of high density; one to the north and one to the south.   The dense region to the north has three \HI\ peaks, while the region to the south has one \HI\ peak.  These two high density regions appear to be part of a ring-like structure that was also seen in \citet{stil02a}.  There is also tenuous \HI\ to the northwest that creates an overall cometary morphology in the map.

\begin{figure}[!ht]
\epsscale{0.505}
\plotone{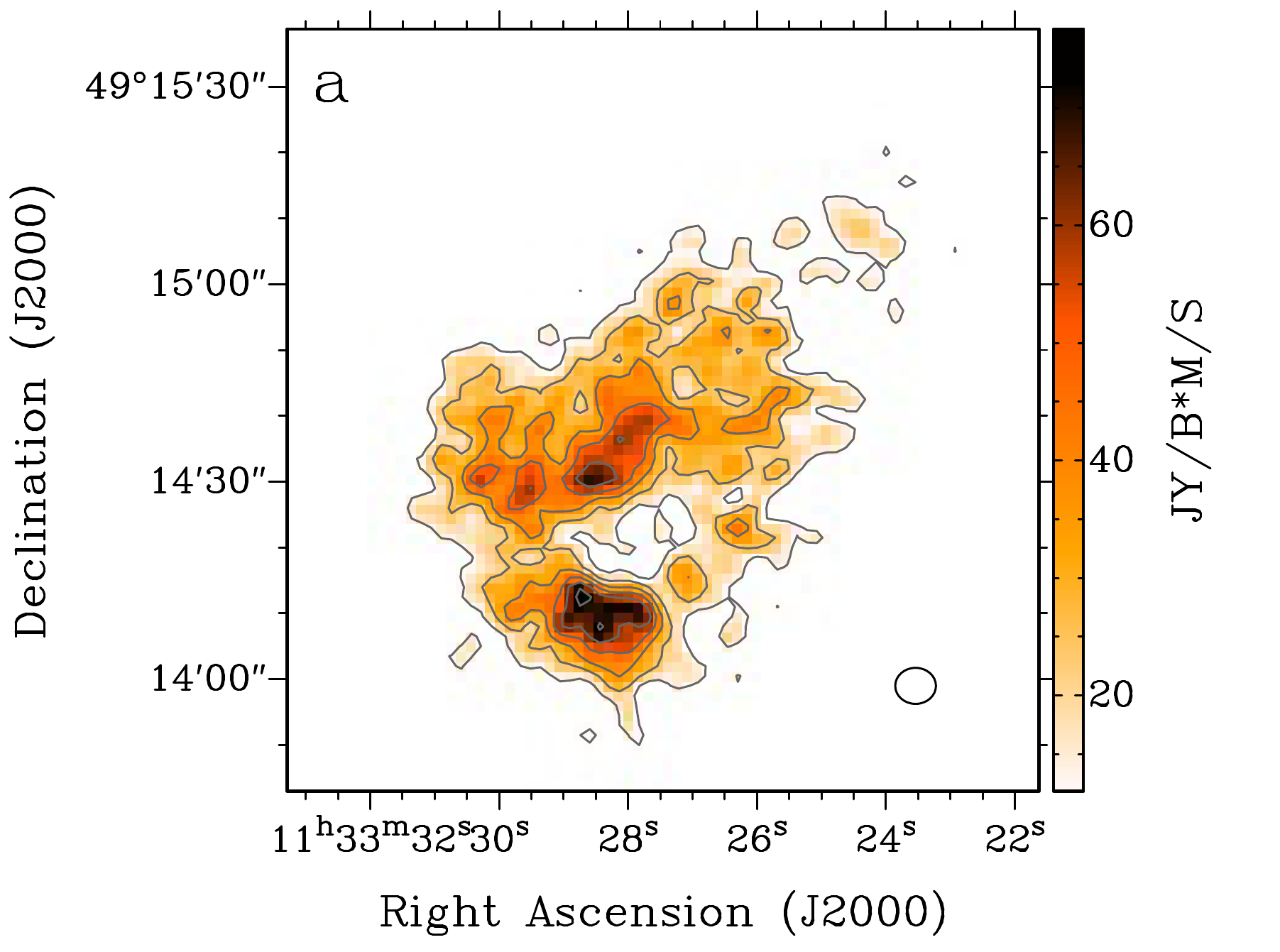}
\epsscale{0.48}
\plotone{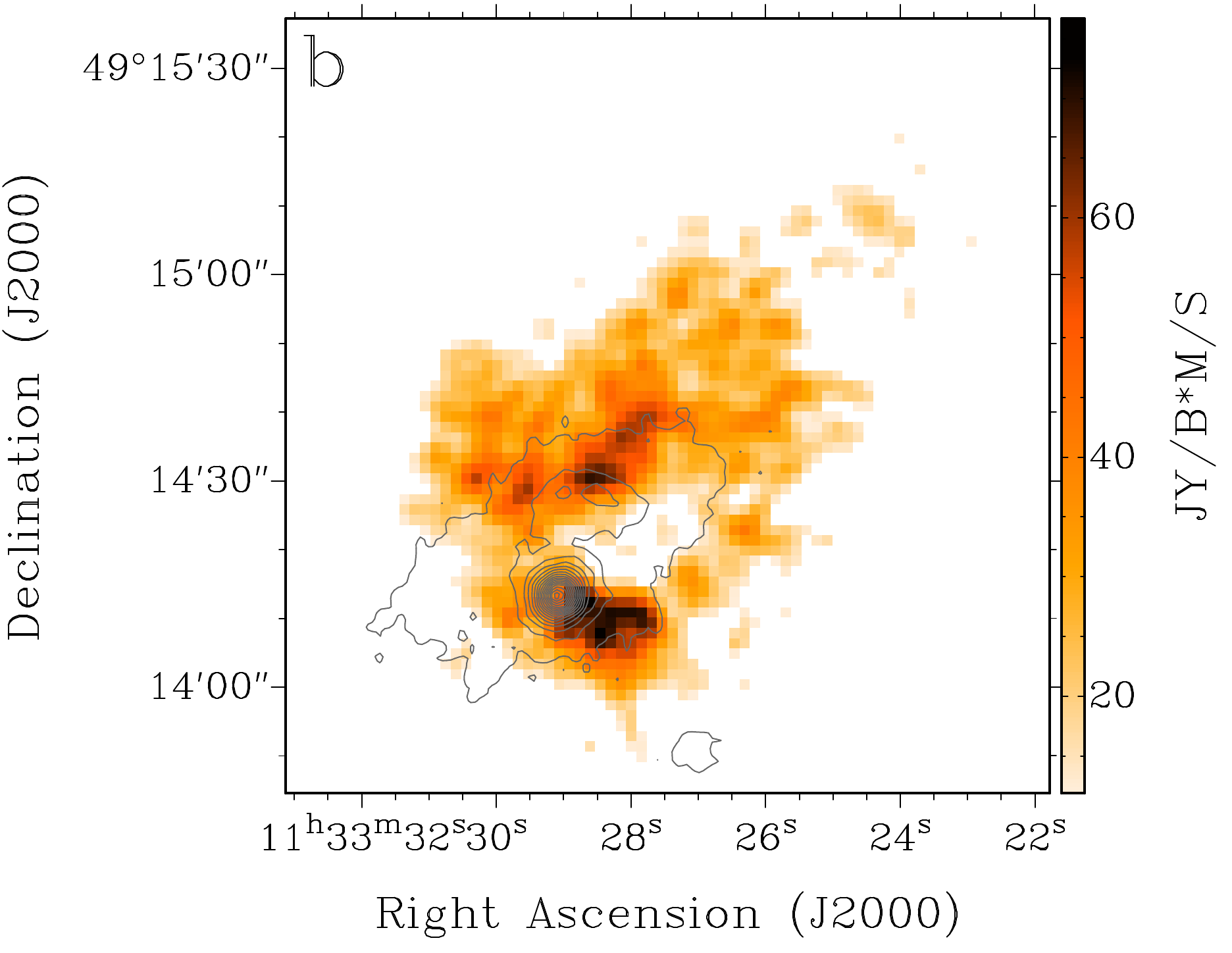}\\
\epsscale{0.96}
\plottwo{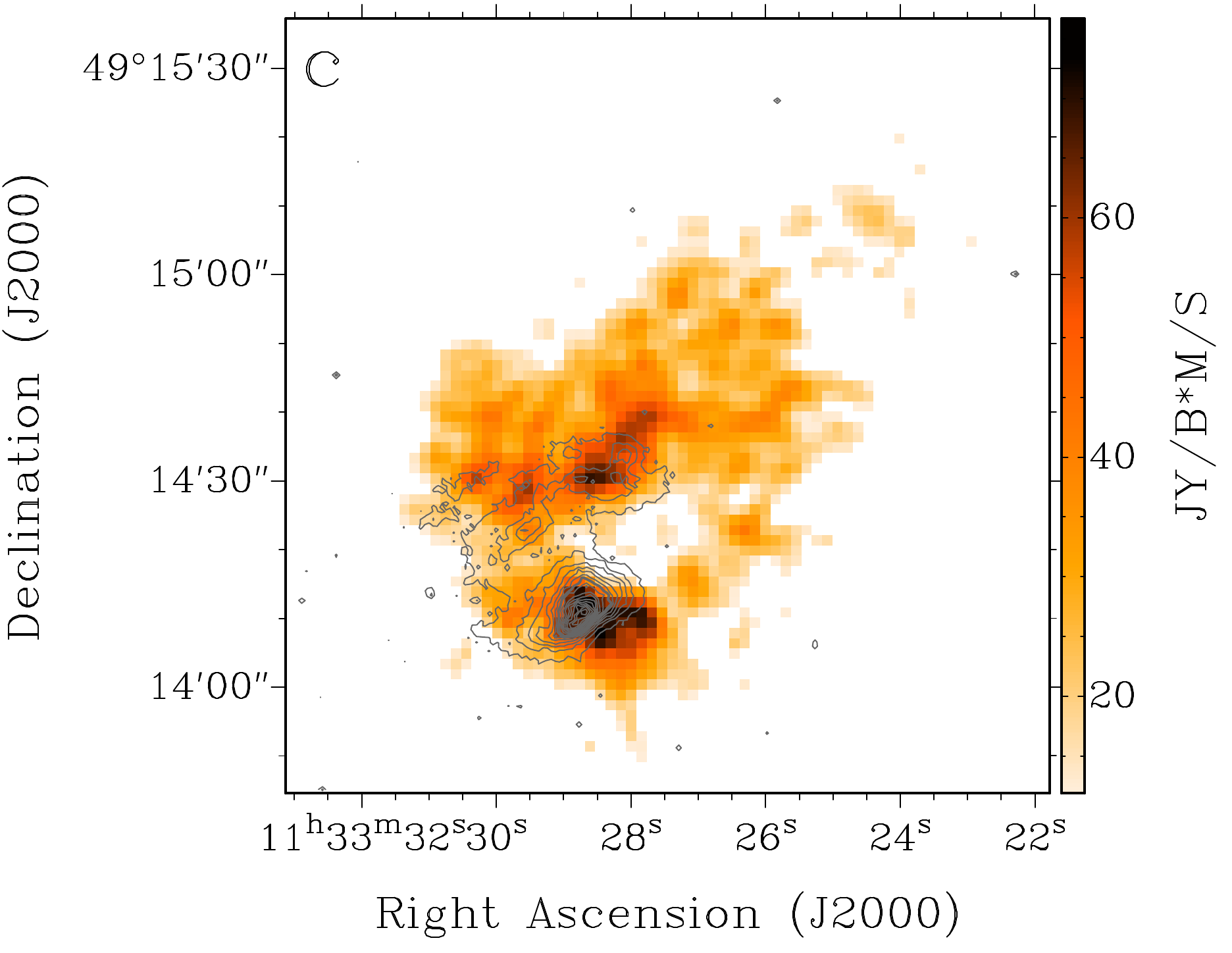}{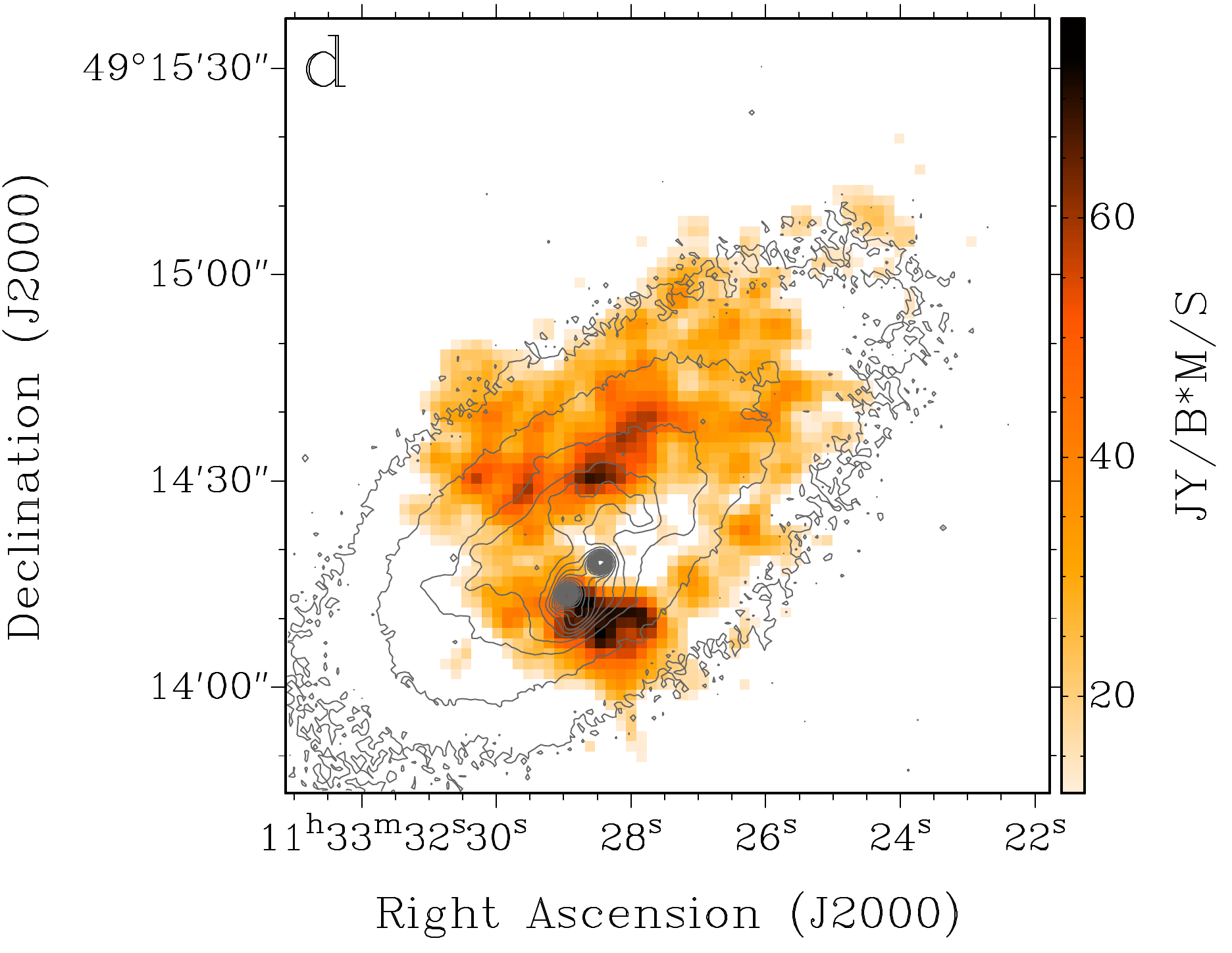}
\caption{Mrk 178's VLA robust-weighted moment maps. (a): Integrated \HI\ intensity map; contour levels are 1$\sigma\times$(2, 4, 6, 8, 10, 12) where 1$\sigma=1.95\times10^{20}\ \rm{atoms}\ \rm{cm}^{-2}$.  The black ellipse in the bottom-right represents the synthesized beam. (b): Integrated \HI\ intensity map colorscale and FUV contours. (c):  Integrated \HI\ intensity map colorscale and H$\alpha$ contours. (d): Integrated \HI\ intensity map colorscale and V-band contours.\label{m178vla_r}}
\end{figure}

The VLA robust-weighted integrated \HI\ intensity map is shown in Figure~\ref{m178vla_r}a.  This map shows the broken ring-like structure at a higher resolution.  Plots of the FUV, H$\alpha$, and V-band contours over the colorscale of the robust-weighted integrated \HI\ intensity map are also shown in Figures~\ref{m178vla_r}b-\ref{m178vla_r}d, respectively.  In all three figures the curved feature in the stellar components follows the morphology of the broken ring-like \HI\ structure in the northeast.  In both the robust-weighted data (Figure \ref{m178vla_r}d) and in the natural-weighted data (Figure \ref{m178vla_na}c) the V-band disk extends further southeast than the \HI.

\subsection{Mrk 178: VLA \HI\ Velocity and Velocity Dispersion Field}\label{m178vlavel}
\begin{figure}[!ht]
\epsscale{1.}
\plottwo{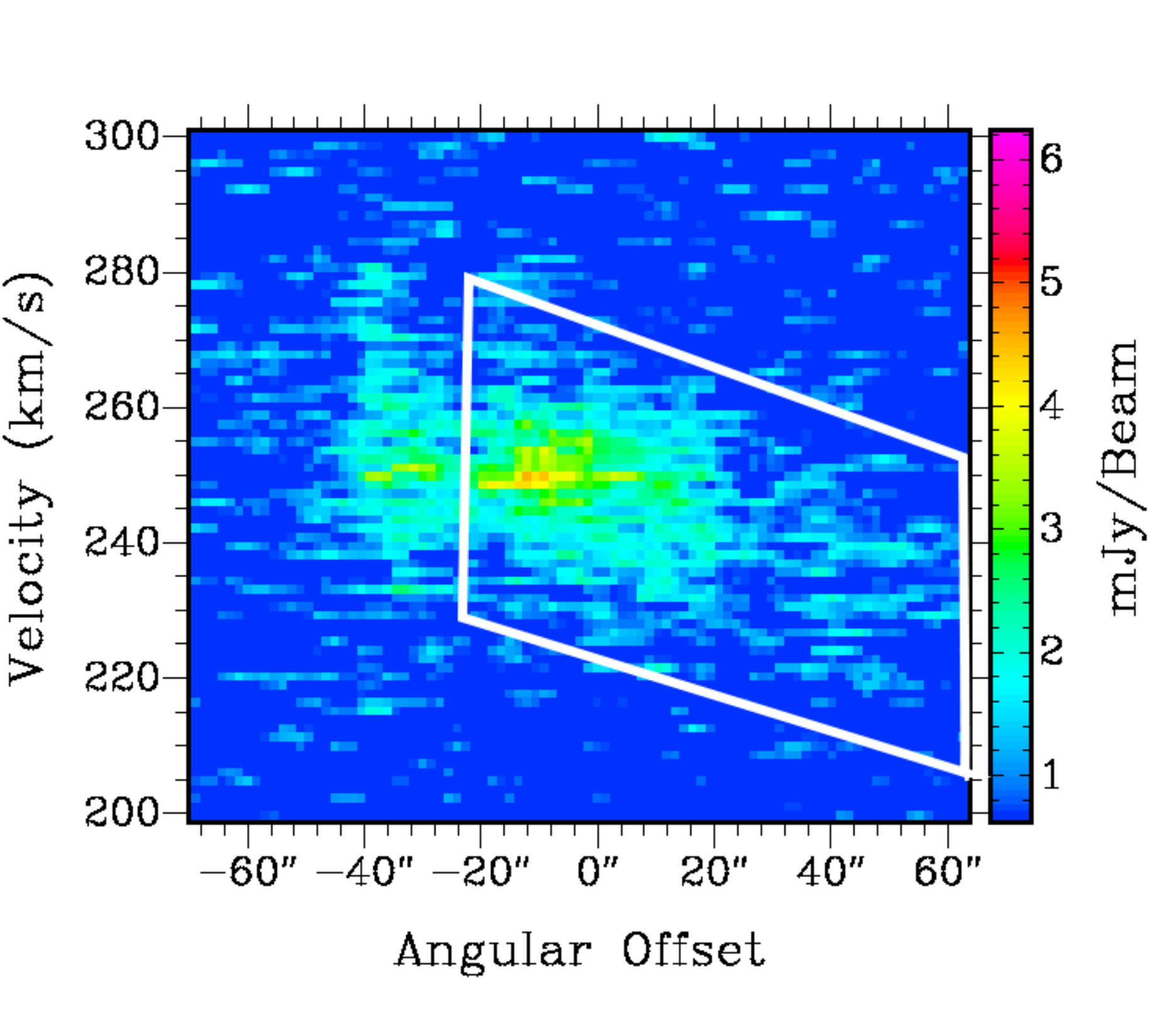}{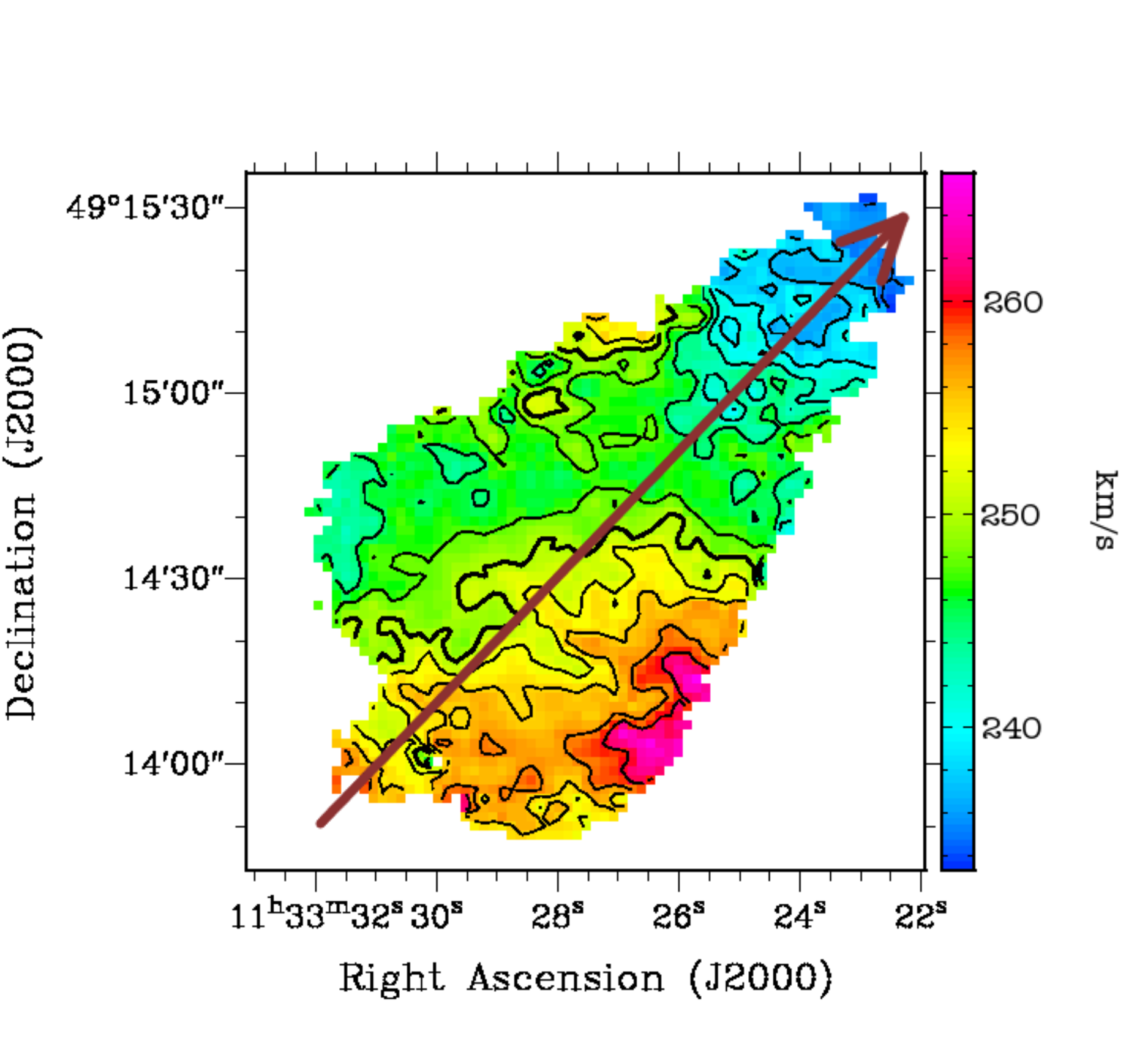}
\plottwo{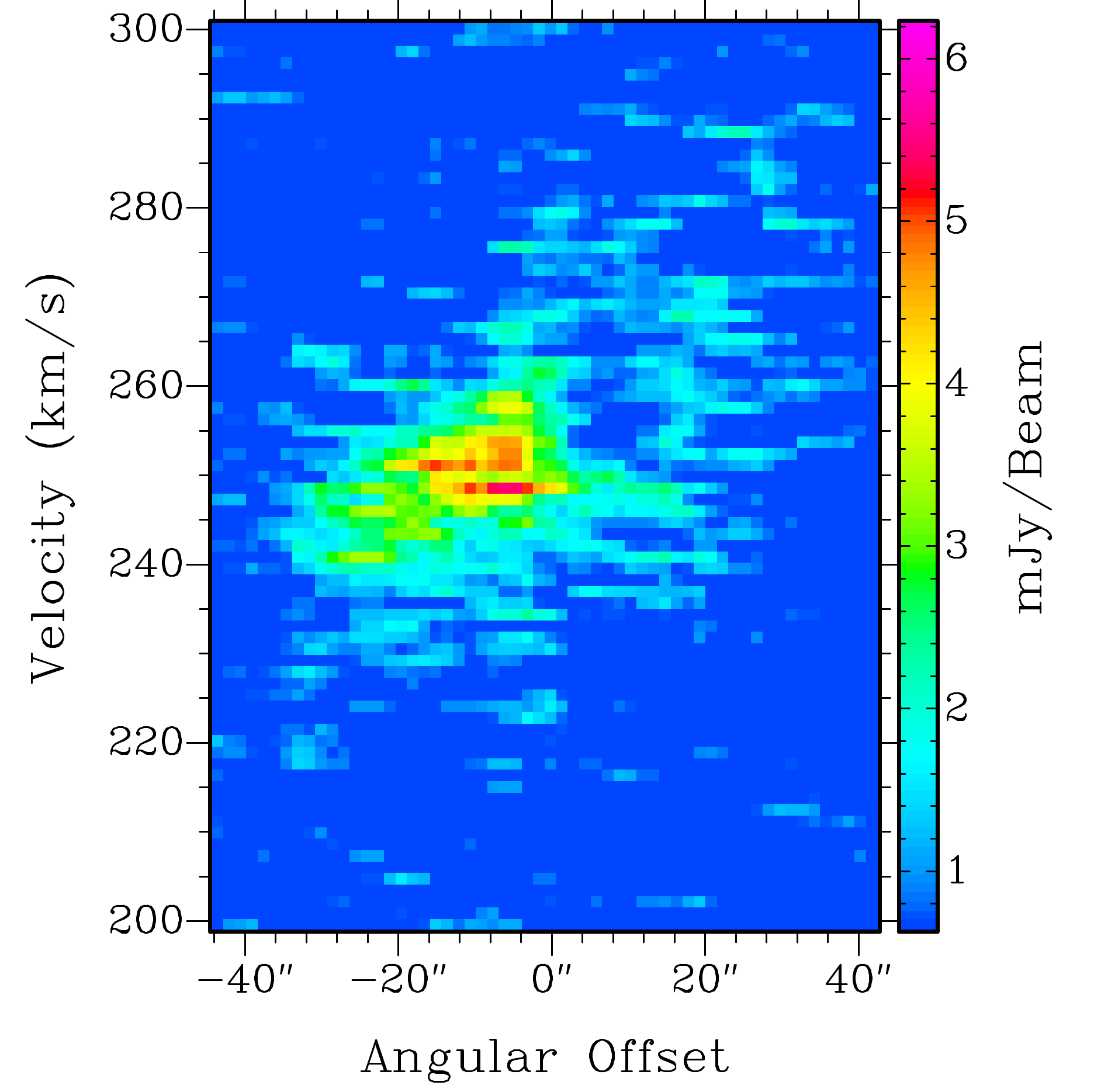}{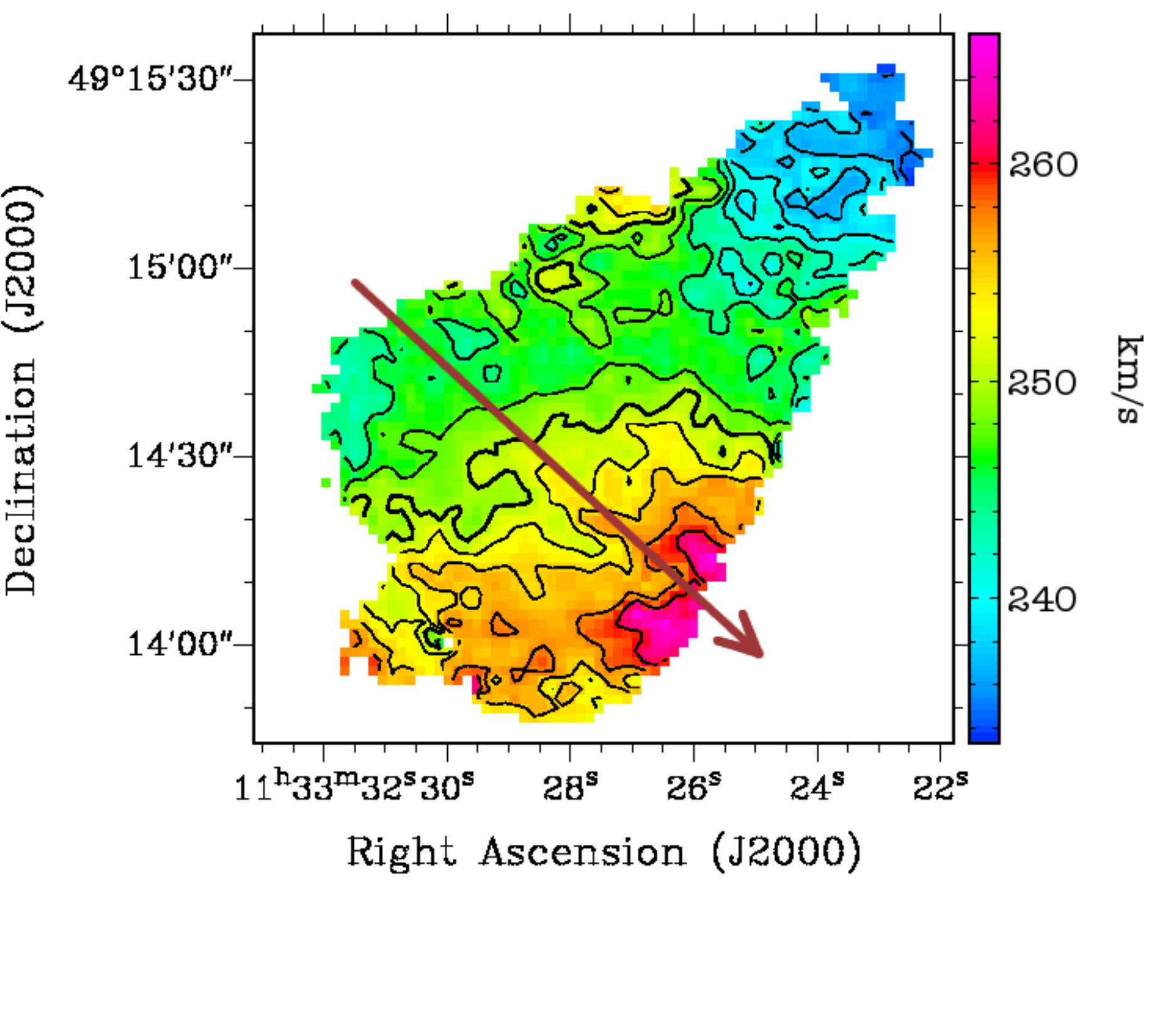}
\caption{\textit{Left:} P-V diagrams from Mrk 178's VLA natural-weighted cube, starting at a 1$\sigma$ level (0.64 mJy/beam), indicating the rotation of Mrk 178's \HI\ along two separate axes. The white box in the first P-V diagram outlines the tenuous emission in the northwest end of Mrk 178 that is moving with one of the kinematic axes. The emission to the left of the white box is from the morphological peak of \HI\ emission in the south. \textit{Right:}  The natural-weighted \HI\ velocity field with a red arrow indicating the location of the corresponding slice through the galaxy and pointing in the direction of positive angular offset. \label{2axes}}
\end{figure}

The VLA \HI\ velocity field for Mrk 178 can be seen in Figure~\ref{m178vla_na}b.  If the tenuous gas creating the \HI\ extension to the northwest is not included, then the velocity field of Mrk 178 is reminiscent of solid body rotation with a kinematic major axis at a position angle (PA) of roughly 230\degr\ (estimated by eye).  This first kinematic major axis is nearly perpendicular to the stellar morphological major axis \citep[PA of 128.7\degr;][]{hunter04}.  The tenuous \HI\ extension to the northwest has isovelocity contours that are nearly perpendicular to the isovelocity contours of the rest of the \HI\ with a kinematic major axis with a PA of roughly 135\degr\ (estimated by eye). This second kinematic major axis also nearly aligns with the stellar morphological major axis.   A  position-velocity (P-V) diagram for each of the kinematic axes in Mrk 178's velocity field can be seen in Figure~\ref{2axes}.  Figure~\ref{2axes} was created in \textsc{kpvslice}, which is part of the Karma\footnote{Documentation is located at  \url{http://www.atnf.csiro.au/computing/software/karma/}.} software package \citep{gooch96}.  In Karma the user draws a line over a map of the galaxy and Karma plots the velocity of the gas at every position along the line.  The P-V diagrams of the natural-weighted cubes have color bars that begin at 1$\sigma$ level (0.64 mJy/beam).  The white box in the top left P-V diagram encompasses the tenuous gas in the northwest end of Mrk 178 that is rotating with one of the two kinematic major axes (indicated by the red slice in the velocity map to the top right of Figure~\ref{2axes}).  The emission at higher negative angular offsets to this box is associated with the morphological peak of \HI\ emission to the south identified in Section~\ref{m178vla_morph}.  The P-V diagram in the bottom-left of Figure~\ref{2axes} shows the velocity of the gas in the head of the cometary shape increasing from the northeast to the southwest.

The northwest \HI\ extension has a length of \s920 pc.  The length of the extension was taken from the natural-weighted map from the tip of the northwest edge of the $2\sigma$ contour to the southeast tip of the 245 \kms\ contour.  The 245 \kms\ contour (the red contour in Figure~\ref{m178vla_na}b) was chosen as a cutoff for the length because it is the most southeastern isovelocity contour with a kinematic major axis that is nearly aligned with the stellar morphological major axis.  

The tenuous \HI\ component to the northwest has a maximum velocity difference  of \s17 \kms\ from the systemic velocity and the southeast kinematic component has a maximum velocity difference of \s16 \kms\ from the systemic velocity.  Mrk 178 is therefore rotating slowly and probably has a shallow potential well.  Most of the velocity dispersion map contains velocity dispersions of \s9-13 \kms\ (Figure~\ref{m178vla_na}e).

Neither of the two kinematic major axes are likely due to dispersions in the gas; the velocity dispersion map in Figure~\ref{m178vla_na}e has a gradient of 2-4 \kms\ and the velocity gradient is 10-16 \kms\ over the radius of the galaxy.  Instead, one of the major kinematic axes of Mrk 178 is likely from rotation in the disk and the other could be from an extragalactic impact or some other disturbance, as discussed in Section~\ref{m178disc}.  The V-band image of Mrk 178 (Figure~\ref{m178_star}) indicates that Mrk 178's outer stellar isophotes are elliptical, as would be expected of a disk. Disk galaxies typically have rotation associated with their disk.   As previously mentioned, the northwestern edge of the gaseous disk has a kinematic major axis that follows the stellar disk's morphological major axis (see Figure~\ref{m178vla_na}d), as would be expected from gas rotating with the disk of the galaxy. Therefore, there are likely two real kinematic major axes in the gas of Mrk 178 with the kinematic major axis to the northeast rotating like a typical disk.

\subsection{Mrk 178: GBT \HI\ Morphology And Velocity Field}

\begin{figure}
\epsscale{0.59}
\plotone{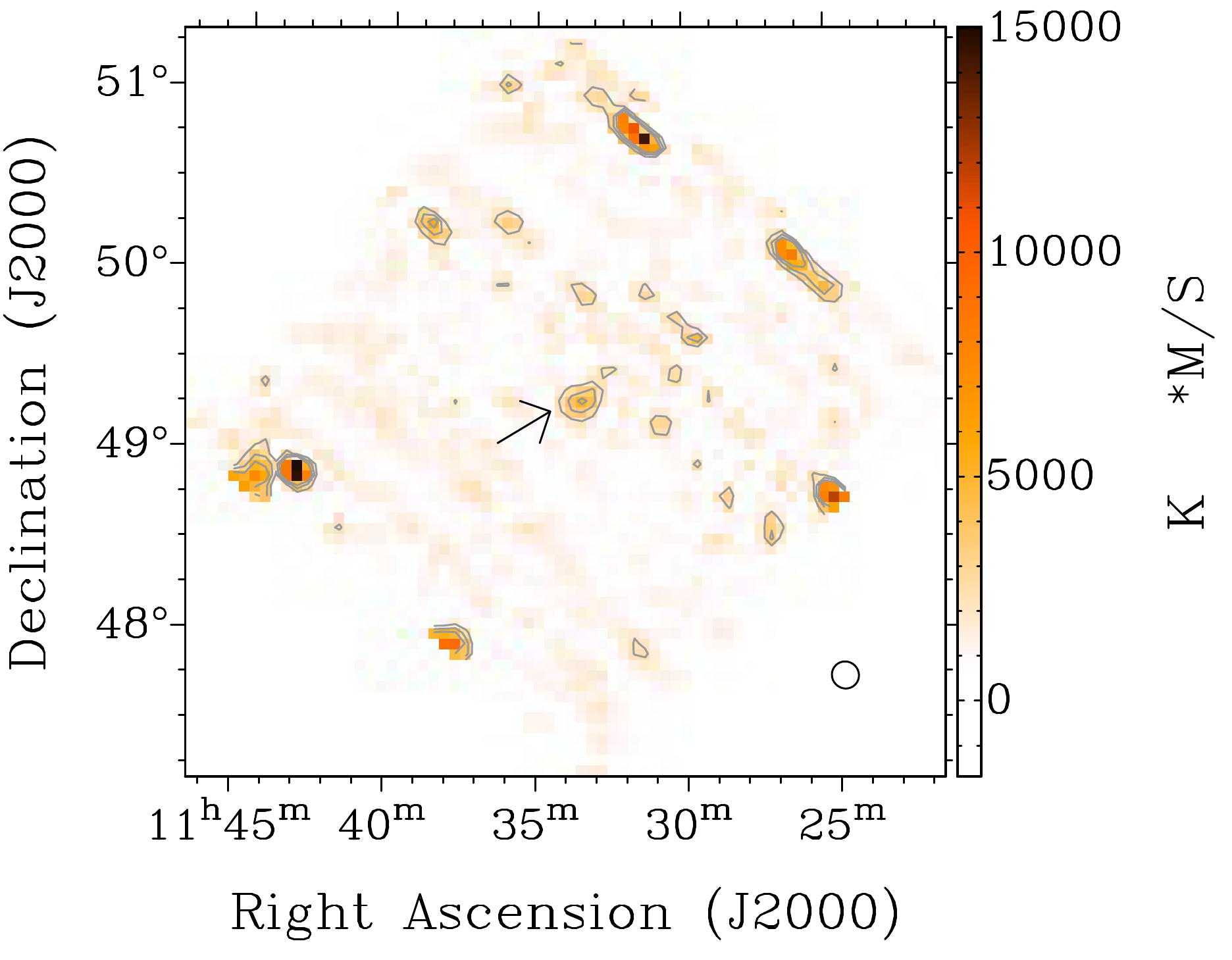}
\epsscale{0.56}
\plotone{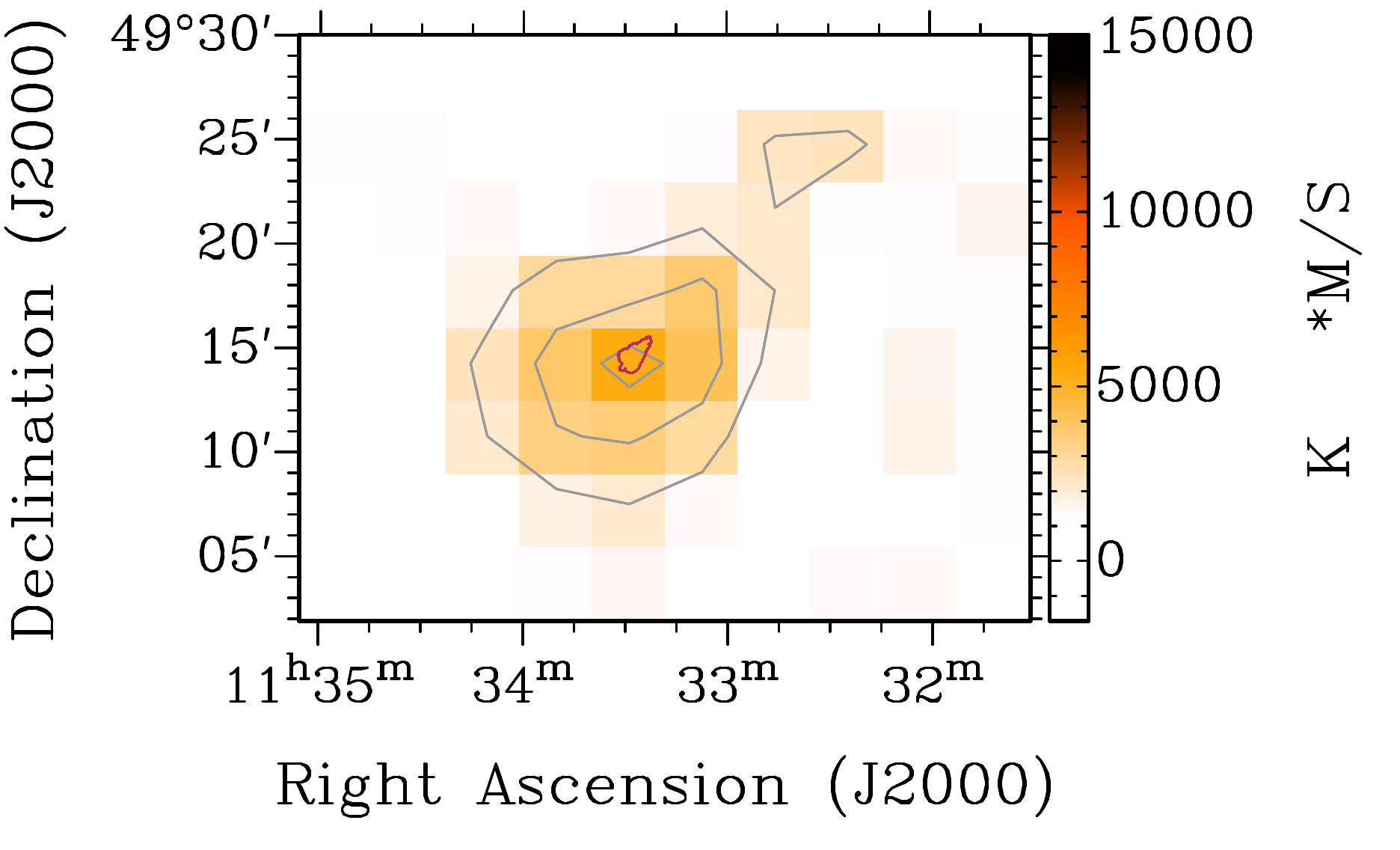}
\caption{Mrk 178's GBT integrated \HI\ intensity map. \textit{Left:} The full area mapped; contour levels are 1$\sigma\times$(25, 40, 55) where 1$\sigma=6.7\times10^{16}\ \rm{atoms}\ \rm{cm}^{-2}$.  Mrk 178 is in the center of the map as indicated by the black arrow.  The other bright regions in the map are thought to be largely due to RFI.  The black ellipse represents the GBT beam.    \textit{Right:} A close up of Mrk 178's GBT integrated \HI\ intensity map with a maroon overlay of the outer VLA \HI\ intensity contour. \label{m178gbt}}
\end{figure}

The integrated \HI\ intensity map as measured with the GBT is shown in the top of Figure~\ref{m178gbt}.  An arrow is used to identify Mrk 178 which is very difficult to distinguish from the noise in this map.  Mrk 178's GBT data had several problems including the large amounts of difficult-to-remove RFI as noted in Section~\ref{gbtsection}.  The data cube was inspected with the histogram tool in CASA's \citep{mcmullin07} viewer window.  From this inspection it was concluded that the other bright spots in Figure~\ref{m178gbt}a are likely due to RFI.  The northwestern region of the map is significantly noisier than the rest of the map (about 1.5 times noisier).  Due to the noisiness of the GBT maps, we will not discuss the results of Mrk 178's GBT maps much throughout the rest of the paper.  However, Mrk 178's \HI\ emission detected with the GBT is likely real and not noise; in the bottom left of Figure~\ref{m178gbt}, Mrk 178's VLA \HI\ outer contour is plotted on top of a close up of Mrk 178's GBT \HI\ map and these two maps have the same general morphology with an extension of gas to the northwest.  Mrk 178's GBT \HI\ velocity field was also inspected, however, due to the resolution of the map, no new information can be gleaned from the velocity map.  

\subsection{Mrk 178: \HI\ Mass}\label{m178himass}
The mass of each galaxy was calculated using the AIPS task \textsc{ispec} and masses of individual galaxy features were measured using the AIPS task \textsc{blsum}.  \textsc{ispec} will sum the flux within a user-specified box from a user-specified range of velocity channels.  The sum can then be used to calculate the mass of \HI\ using the following equation:
\begin{equation}\label{1}
M(M_{\sun})=235.6D^{2}\sum_{i}S_{i}\Delta V
\end{equation}
where $D$ is the distance of the galaxy in units of Mpc, $S_{i}$ is the flux in mJy in channel i, and $\Delta V$ is the channel width in \kms.  \textsc{blsum} also results in the sum of a flux within a given region, however, the region can be any shape drawn by the user on the \HI\ map.  Mrk 178's total \HI\ mass from the VLA data is $8.7\times10^{6}$ M$_{\sun}$.  The mass of the northwest extension in the VLA natural-weighted map was also calculated using \textsc{blsum}.  The border which separates the northwest extension from the rest of the \HI\ in Mrk 178 was again defined by the isovelocity contour of 245 \kms\ that is furthest northwest in the map.  The mass of the northwest extension is $7.5\times10^{5}$ M$_{\sun}$ or 8.6\% of the total VLA \HI\ mass.  Mrk 178's total mass from the GBT data is $1.3\times10^{7}$ M$_{\sun}$.  Mrk 178's VLA maps recovered 67\% of the GBT mass.   Using the same velocity width used to measure Mrk 178's GBT \HI\ mass, the uncertainty in Mrk 178's GBT mass is $5\times10^{5}$ M$_{\sun}$ making Mrk 178 a significant detection in the GBT maps with an \HI\ mass at 26$\sigma$.

\section{Discussion: Mrk 178}\label{m178disc}

Mrk 178 has odd VLA \HI\ and stellar morphologies, including: an overall cometary shape in its \HI\ disk, two kinematic major axes in the \HI\ velocity map, and a stellar disk that extends beyond the \HI\ natural-weighted map due to a lack of tenuous gas to the southeast of the disk.  We discuss four possible explanations for these kinematics and morphologies.  

\subsection{Mrk 178 Has A Large Hole In The \HI}\label{m178_hihole}
In Figures~\ref{m178vla_na}a and \ref{m178vla_r}a a large hole-shell structure is visible in the southern region of Mrk 178's disk. The high density \HI\ regions in the north and south of Mrk 178's \HI\ disk could be part of a shell that has been created by the hole between them.  This hole was identified as part of a LITTLE THINGS project cataloguing and characterizing all of the \HI\ holes in the 41 dwarf galaxy sample using the hole quality checks outlined in \citet{bagetakos11} (Pokhrel \et, in prep.).  To initially be included in the catalog, the hole structure must be visible in 3 consecutive velocity channels; Mrk 178's hole is visible in 6 consecutive velocity channels in the natural-weighted cube (251 \kms\ to 258 \kms).  \citet{bagetakos11} also assign each \HI\ hole with a quality value of 1-9 (low to high quality) based on the number of velocity channels that contain the hole, whether the location of the hole's center changes across the velocity channels, the difference in \HI\ surface brightness between the hole and its surroundings (at least 50\%), and how elliptical the appearance of the hole is in a P-V diagram.  Based on these criteria, Mrk 178's hole has a quality value of 6, which is average quality.   

\begin{figure}[!ht]
\epsscale{0.45}
\begin{center}
\plotone{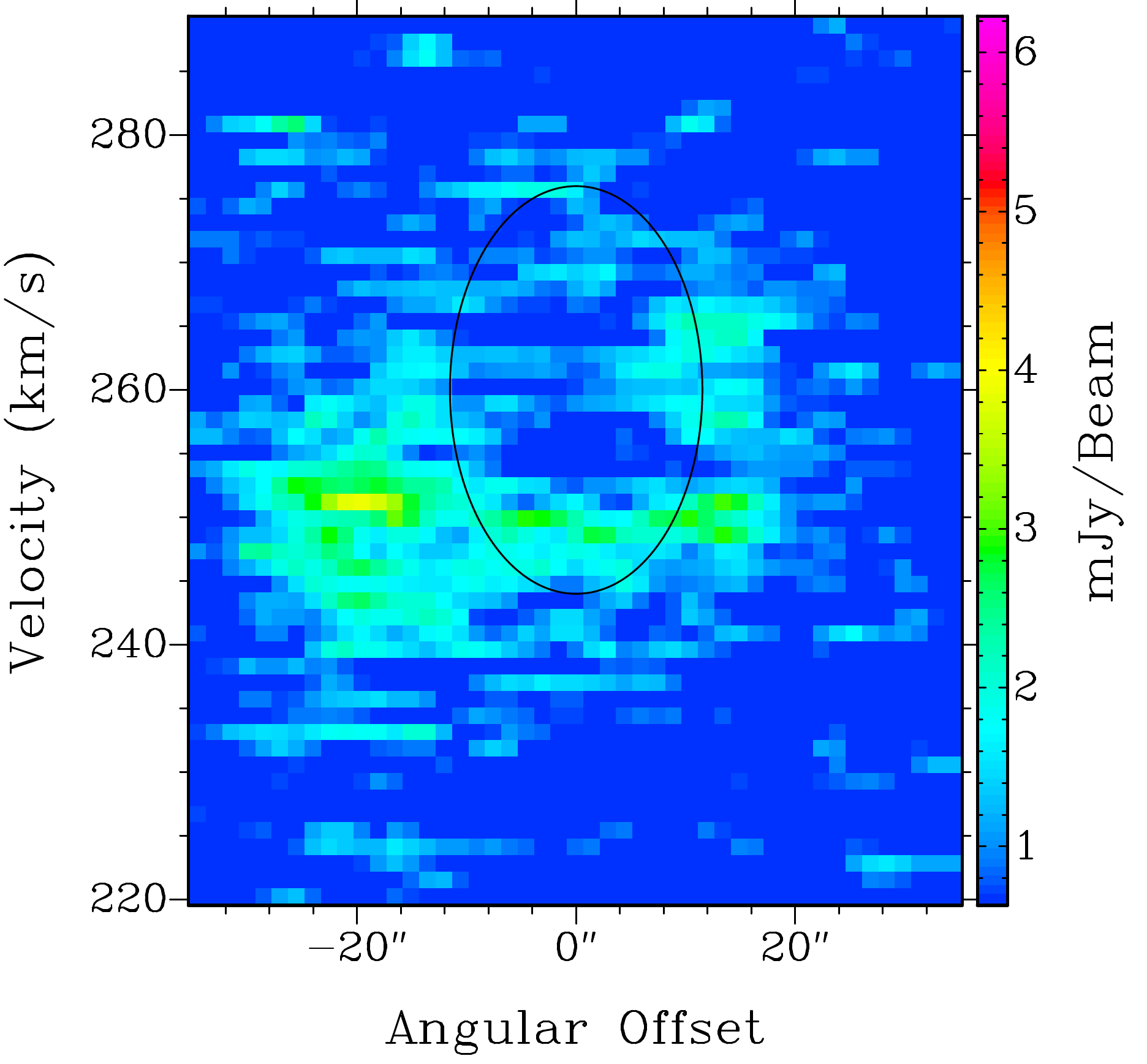}
\epsscale{0.53}
\plotone{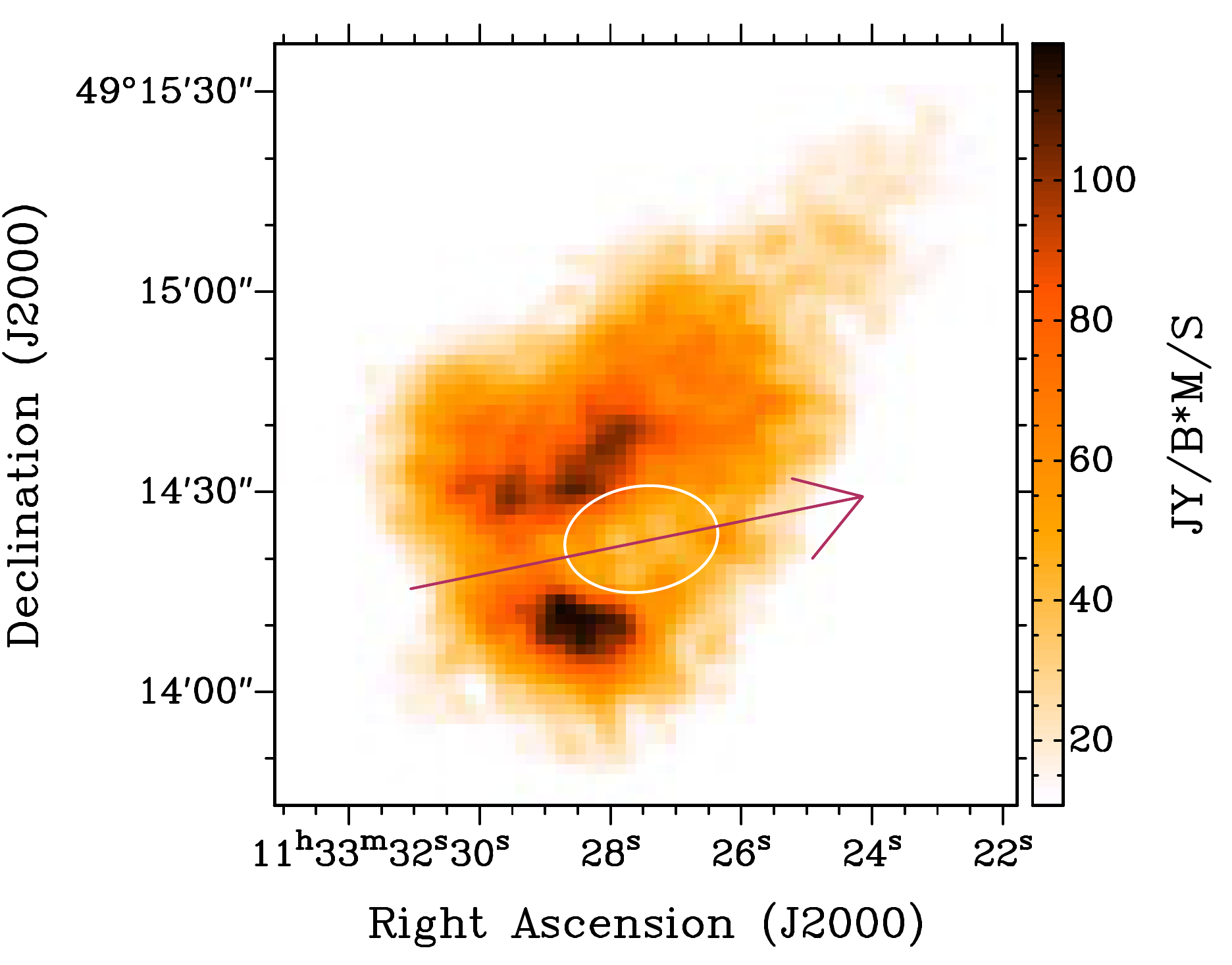}\\
\end{center}
\caption{\textit{Left:} P-V diagram of Mrk 178's potential hole from the VLA natural-weighted cube,  starting at a 1$\sigma$ level (0.64 mJy/beam).  The black ellipse indicates the approximate extent of the hole's parameters in the P-V diagram.  \textit{Right:} The natural-weighted integrated \HI\ map with a red arrow indicating the location of the corresponding slice through the galaxy and pointing in the direction of positive offset. The location of the hole in Mrk 178 is outlined in the white ellipse. \label{m178_hole}}
\end{figure}

In P-V diagrams the emission around an \HI\ hole will create an empty ring or partial ring appearance in the P-V diagram when a hole is present \citep{walter99}.  The left side of Figure~\ref{m178_hole} shows the P-V diagram for Mrk 178's hole.  The black ellipse indicates the parameter space of the hole in the P-V diagram; the hole creates a partial ring defined by the offsets of about \n11.4\arcsec\ and 11.4\arcsec, and a central velocity of \s260 \kms.  The right side of Figure~\ref{m178_hole} is the natural-weighted map of Mrk 178, with the red arrow indicating the location of the slice used for the P-V diagram and the white ellipse indicating the location of Mrk 178's hole. The higher velocity side of the ring may be composed of very tenuous gas or that side of the hole may have blown out of the \HI\ disk.

With the higher velocity side of the \HI\ shell not clearly defined, the estimated expansion rate uncertainty will be high, however, the velocity of the dense edge of the shell can be used to get an estimate of the expansion. The velocity of the center of this hole is found to be \s260 \kms\ and the velocity of the intact side of the \HI\ shell was taken to be \s244 \kms, resulting in an expansion velocity of 16 \kms.  The radius of the hole was taken to be the square root of the product of the major and minor axes of the hole, resulting in a radius of \s180 pc.  A hole of this size, expanding at 16 \kms, would have taken roughly 11 Myr to form.  This calculated age of the hole is a rough estimate, as the expansion rate of the hole may have been much faster when it first formed and the hole looks as though it has blown out of one side of the disk, which means the hole may be older than indicated by its current expansion rate.

 The energy needed to create Mrk 178's hole can be estimated using Equation 26 from \citet{chevalier74}, which calculates the energy from the initial supernova burst:
\begin{equation}\label{e0}
E_{0}=5.3\times10^{-7\ }n_{0}^{1.12}\ v_{sh}^{1.40}\ R^{3.12}
\end{equation}
where $E_{0}$ is the initial energy in units of $10^{50}$ ergs, $n_{0}$ is the initial volumetric density of the gas in atoms cm$^{-3}$, $v_{sh}$ is the velocity of the hole's expansion in \kms, and $R$ is the radius of the hole in pc.  For Mrk 178's hole, $n_{0}$ was taken to be the approximate density of the surrounding gas.  A column density of $7.69\times10^{20}$ cm$^{-2}$, the average density of the surrounding \HI, was used as the initial column density for the hole.  Assuming a scale height of 1740 pc (Pokhrel \et, in prep.), $n_{0}$ is approximately 0.14 cm$^{-3}$.  Using these parameters results in an energy of  \s$3.1\times10^{51}$ ergs.  Assuming that the energy of a supernova explosion is \s$10^{51}$ ergs, it would take approximately 3 supernovae explosions to create a hole of this size.  This is a very small number of supernova explosions that could have easily formed over a period as short as a Myr with Mrk 178's current star formation rate \citep[see Figure 9 of][]{sullivan06}.

The \textit{southern} region of high stellar density in the V-band emission has two bright components centered on 11h 33m 28.7s and 49\degr 14\arcmin 15.9\arcsec\ (see Figure~\ref{m178_star}).  The component located closer to the center of Mrk 178's stellar disk, at 11h 33m 28.5s and 49\degr 14\arcmin 18.3\arcsec\, does look as though it could be in the hole as can be seen in Figure~\ref{m178vla_r}d.   \citet{gonzalez88} calculate the ages of the stars in the V-band's bright southern stellar concentration to be \s9 Myr.  The estimated time needed to create the hole was calculated to roughly be 11 Myr, which, given the uncertainties in the hole age calculation, is roughly in line with the 9 Myr age of the stellar concentration.

Although, interesting on its own, the potential \HI\ hole does not easily explain the rest of the \HI\ morphology and kinematics: the \HI\ is rotating on two separate kinematic major axes (as discussed in Section \ref{m178vlavel}) and the stellar component expands further than the VLA natural-weighted \HI\ (as discussed in  Section \ref{m178vla_morph}). The stellar component that expands further than the \HI\ data is not located near the \HI\ hole and therefore is not likely a result of the hole if it is real.  The two kinematic axes indicate that the gas has been significantly disturbed in the past. 

\subsection{Mrk 178 Has Recently Interacted With Another Galaxy}

Two kinematic major axes and an asymmetric \HI\ distribution (which does not cover the stellar disk) could indicate that Mrk 178 has recently interacted or merged with another galaxy.  There are no known companions close to Mrk 178 and the GBT maps do not show any companion to Mrk 178 at the sensitivity and resolution of the maps.  Therefore, it is unlikely that Mrk 178 has recently had an interaction with a nearby gas-rich companion.  Instead, it is possible that Mrk 178 is interacting with a gas-poor companion or Mrk 178 is the result of a merger still in the process of settling into a regular rotation pattern.  

However, if an interaction or merger has caused a significant asymmetry in the gas morphology of the disk and two kinematic major axes, then why does the outer V-band disk not show any signs of a morphological disturbance such as tidal tails, bridges, or significant asymmetries? The gaseous disk is collisional and thus has a short term memory of past events (on the order of one dynamical period).  The stellar disk is non-collisional and therefore has a longer memory of past mergers than the gaseous disk.  So, if the gaseous disk is still significantly kinematically and morphologically disturbed, then the outer regions of the older stellar disk (reaching \s27 mag/arcsec$^{2}$) would likely still show significant signs of disturbance and yet it appears to be relatively elliptical in shape.  Haro 36 is a BCD with a tidal tail visible in its outer V-band disk which has a limiting surface brightness of \s25.5 mag/arcsec$^{2}$.  Since Haro 36 has a distance of 9.3 Mpc, it is reasonable to assume that Mrk 178, at a limiting surface brightness of 27 mag/arcsec$^{2}$ and a distance of only 3.9 Mpc, would show signs of a tidal disturbance in its outer V-band disk if it existed.  Since there are no signs of tidal disturbances in the outer V-band disk, Mrk 178 is not likely a merger remnant.

\subsection{Mrk 178 Is Experiencing Ram Pressure Stripping}\label{m178_ram_press}

 Mrk 178 could be interacting with intergalactic gas through ram pressure stripping.  The \HI\ intensity map in Figure~\ref{m178vla_na}a has a generally cometary appearance, with a bifucated head of star formation and two gas peaks. Each head (\HI\ peak) is also cometary and pointing in the same direction as the whole galaxy, particularly the northern one.   These \HI\ heads are likely an indication of a subsonic shock front since they are near a sharp \HI\ density edge.  To make this structure as well as clear the gas from the southeastern part of the V-band disk, the galaxy could be moving in the southeast direction at several tens of km s$^{-1}$ into a low density IGM that may be too ionized to see in \HI.  If this motion also had a component toward us, then it could account for the strong velocity perturbation in the south (and the lack of velocity perturbation in the northwest), where all the HI gas redshifted relative to the rest of the galaxy is the gas being stripped from the southeast edge of the galaxy and is moving away from us.  The HI `hole' between the two \HI\ peaks discussed in Section~\ref{m178_hihole} could be a hydrodynamical effect of the streaming intergalactic gas; the intergalactic gas could be moving between and around each \HI\ peak as the galaxy moves through the IGM. Alternatively, the ram pressure from this motion could have made or compressed the head region, promoting rapid star formation there or at the leading shock front.  That star formation could then have made the hole discussed in Section~\ref{m178_hihole}.  This scenario is analogous to that in NGC 1569 \citep{johnson12} where an incoming \HI\ stream apparently compressed the galaxy disk and triggered two super star clusters, which are now causing significant clearing of the peripheral gas.

Some parameters of the interaction can be estimated from the pressure in the head region of Mrk 178 as observed in HI and starlight. From \citet{e89} the pressure in the interstellar medium (assuming comparable stellar and gas disk thicknesses) is approximately: 
\begin{equation}
P\approx \dfrac{\pi}{2}\ G\ \Sigma_{\rm gas}\left(\Sigma_{\rm gas}+\dfrac{\sigma_{g}}{\sigma_{s}}\Sigma_{\rm stars}\right)
\label{pressure}\end{equation}
where $\Sigma_{\rm gas}$ and $\Sigma_{\rm stars}$ are the gaseous and stellar surface densities, $\sigma_{g}$ and $\sigma_{s}$ are the gaseous and stellar velocity dispersions, and G is the gravitational constant.   In the second term of Equation~\ref{pressure}, $\sigma_{g}$ $\sigma_{s}$$^{-1}$ is approximately equal to 1 because the gaseous and stellar velocity dispersions of dwarf irregular galaxies are similar \citep{johnson12,  johnson15}.  The average HI column density in the southern star formation peak is $\sim1\times10^{21}$ cm$^{-2}$\ in Figure~\ref{m178vla_na}a, which is $\Sigma_{\rm gas}=11\;M_\odot$ pc$^{-2}$ or $2.3\times10^{-3}$ g cm$^{-2}$ when corrected for He and heavy elements by multiplying by 1.35. \citet{e12} calculate the properties of a stellar clump in the same region as the star formation peak seen in our Figure 1.  They note that the southern stellar clump of stars (see their Figure 2) has a mass inside the galactocentric radius of the southern stellar clump (390 pc) of $1.3\times10^5\;M_\odot$.  Assuming the mass is evenly spread over a circular region, $\Sigma_{\rm stars}=0.27\;M_\odot$ pc$^{-2}$ or  $5.6\times10^{-3}$ g cm$^{-2}$.  If we assume that this stellar surface density is also the stellar surface density of the interior of the southern clump, then the pressure in the southern clump is approximately $P=5.7\times10^{-13}$ dyne cm$^{-2}$.  Now we can calculate the ambient density of the IGM, $\rho_{\rm IGM}$, assuming that the internal cloud pressure, P, is equal to the ram pressure, $P_{\rm{ram}}=\rho_{\rm IGM}v_{\rm IGM}^2$, where $v_{\rm IGM}$ is Mrk 178's velocity relative to the intergalactic medium.  In order for the IGM to sweep away the southeastern gas and perturb the velocity field on only one side of the galaxy, the velocity of Mrk 178 relative to the IGM has to be comparable to the internal rotational speed of the galaxy, so, the minimum velocity of Mrk 178 relative to the IGM is \s20 \kms.   Therefore, the IGM density that makes the ram pressure, $P_{\rm{ram}}$, equal to the internal cloud pressure is:
\begin{equation}
\begin{gathered} 
\rho_{\rm IGM}=1.4\times10^{-25} \left( \frac{20\ km~s^{-1}}{v_{IGM}} \right)^{2} g\ cm^{-3}\\ 
\rm{or} \\
\rho_{\rm IGM}=0.08 \left( \frac{20\ km~s^{-1}}{v_{IGM}} \right)^{2} atoms\ cm^{-3}.
\end{gathered}\label{density}
\end{equation}
The velocity of Mrk 178 relative to the IGM is likely much higher than 20 \kms\ and closer to 100's of \kms; therefore, at $v_{\rm IGM}$=100 \kms\ $\rho_{\rm IGM}$ would be 0.0032 $cm^{-3}$.  At this density, even if the IGM has a thickness in the line of sight of 5 kpc and assuming that the gas is all neutral and not ionized (the IGM is probably at least partially ionized), the IGM would have a column density of $5\times10^{19}$ or less than 1$\sigma$ in the VLA maps in Figure~\ref{m178vla_na}, meaning that the IGM would be lost in the noise. The GBT maps would be able to pick up this level of emission, however, the cloud is still expected to be partially ionized, dropping the column density of the \HI.  Also, the relative velocity between Mrk 178 and the IGM could be higher than 100 \kms, which would lower the required external column density. Therefore the IGM may not be visible in the GBT maps either.   

The relative motion of Mrk 178 and the IGM could have cleared the gas out of the southern part of the galaxy, produced a cometary appearance to the main clumps at the leading edge of the remaining interstellar gas in addition to the gas in the galaxy as a whole, and produced the large velocity perturbation and dispersion that are observed in the southern region.

\subsection{Mrk 178 Has A Gas Cloud Running Into It}

\citet{sanchez13} suggest that the overall cometary shapes of galaxies, such as that seen prominently in Mrk 178's disk, can be explained by extragalactic gas impacting the disk of the galaxy.  The gas in the south of Mrk 178's disk may have experienced a collision with a cloud in the southeast side of the galaxy, pushing the redshifted side of the galaxy to the west.  This scenario would leave the gas in the northwestern edge of the disk relatively undisturbed.   When the gas cloud impacted Mrk 178's disk, it would have created a dense shock front in the gas as it moved gas from the southeastern edge of the disk to the west.   Therefore, it is possible that the impacting gas cloud is showing up as a region of high density in the south of Mrk 178's disk.   The \HI\ hole discussed in Section~\ref{m178_hihole} could have also been created by the cloud collision.  A gas cloud running into Mrk 178's disk is a situation that is similar to that of ram pressure stripping discussed in Section~\ref{m178_ram_press}.  The main difference between these two situations is that ram pressure stripping is a steady pressure and a gas cloud impacting the disk is a short lived pressure.

The location of the collision would likely be indicated by the morphological peak in \HI\ to the south.  To the west of and around the location of the morphological peak, Mrk 178's velocity dispersions are slightly increased near 11h32m29s and 14\degr15\arcmin\ to about 13-16 \kms\ (the surrounding regions have dispersions of $\lesssim$10 \kms), indicating that the gas has been disturbed in this region.  Assuming that the northwest side of Mrk 178's \HI\ disk has been relatively undisturbed by the impacting gas cloud, we can assume that the southeast side of the \HI\ disk used to rotate with velocities redshifted with respect to Mrk 178's systemic velocity.  If a gas cloud impacted the disk opposite to the rotation, we would expect a large increase in velocity dispersion and we would also expect the redshifted velocities to generally get closer to the systemic velocity of the galaxy as the disk gas gets slowed by the impact.  However, the redshifted gas in the southern end of the disk has a velocity relative to the systemic velocity of 10-16 \kms, which is similar to the blueshifted gas in the northwestern edge of the disk.  This indicates that the gas cloud has struck Mrk 178's disk either co-rotating with it or in a radial direction parallel to the plane, pushing the gas that was in the southeast part of the disk west and away from us relative to the plane of the galaxy.

If the impacting gas cloud had enough energy to move the eastern edge of the disk to the west, then the binding energy of the gas that was pushed west will be approximately equal to the excess energy that has been left behind by the impacting cloud in the southern region of the disk.  We will assume for simplicity that the angular momentum of the system has a relatively small effect on the energies calculated since the forced motion of the gas in the disk is nearly the same as the rotation speed, so the gas in the disk of the galaxy does not have enough time to turn a significant amount.  The binding energy of the gas that was originally in the southeast end of the disk would be about 0.5m$_{se}v^{2}$ where m$_{se}$ is the \HI\ mass of the gas that was in the southeast edge of the galaxy (that now has been pushed west) and v is the rotational velocity that the gas in the southeast end of the disk had before the impact.  Assuming that the galaxy was once symmetric, we can use the mass of the northwest `extension' (see Section~\ref{m178himass}) as the approximate m$_{se}$, $7.5\times10^{5}$ M$_{\sun}$, and the observed velocity of the edge of the disk in the northeast, corrected for inclination \citep[b/a=0.49:][]{hunter04}, to get the rotational velocity of the gas cloud, 18 \kms.  Using these numbers, we estimate that the binding energy of the gas that was in the southeast of the disk is \s$1.2\times10^{8}$ M$_{\sun}$ km$^{2}$ s$^{-2}$ or \s$2.4\times10^{51}$ erg.  

Next we can estimate the excess energy in the southern \HI\ clump: $0.5m_{sc}(\sigma_{sc}^{2}-\sigma_{a}^{2})$  where $m_{sc}$ is the \HI\ mass of the gas in the southern clump of high \HI\ density (now a mixture of the gas originally in the southeastern edge of the disk and the gas cloud that ran into the disk), $\sigma_{sc}$ is the velocity dispersion of the gas in the southern clump of high \HI\ density, and $\sigma_{a}$ is the ambient velocity dispersion.  \textsc{ispec} was used to calculate the mass of the southern clump of high density \HI\ by using a box that contained emission south of the Declination 49\degr14\arcmin18.5\arcsec, resulting in an \HI\ mass of $2.3\times10^{6}$ M$_{\sun}$.  The velocity dispersion of the dense southern \HI\ clump is \s13 \kms\ and the velocity dispersion of the ambient gas is \s9 \kms.  Using these numbers, the excess energy in the southern dense \HI\ clump is $1.0\times10^{8}$ M$_{\sun}$  km$^{2}$ s$^{-2}$ or \s$2.0\times10^{51}$ erg.  This energy is comparable to the estimated binding energy of the gas that was originally in the southeast of the disk. Therefore, it is possible that the southeastern edge of Mrk 178's disk was struck by a gas cloud that pushed the gas west and away from us, and increased the velocity dispersion.  

\section{Results: VII Zw 403}
\subsection{VII Zw 403: Stellar Component}

\begin{figure}[!ht]
\centering
\epsscale{1}
\plottwo{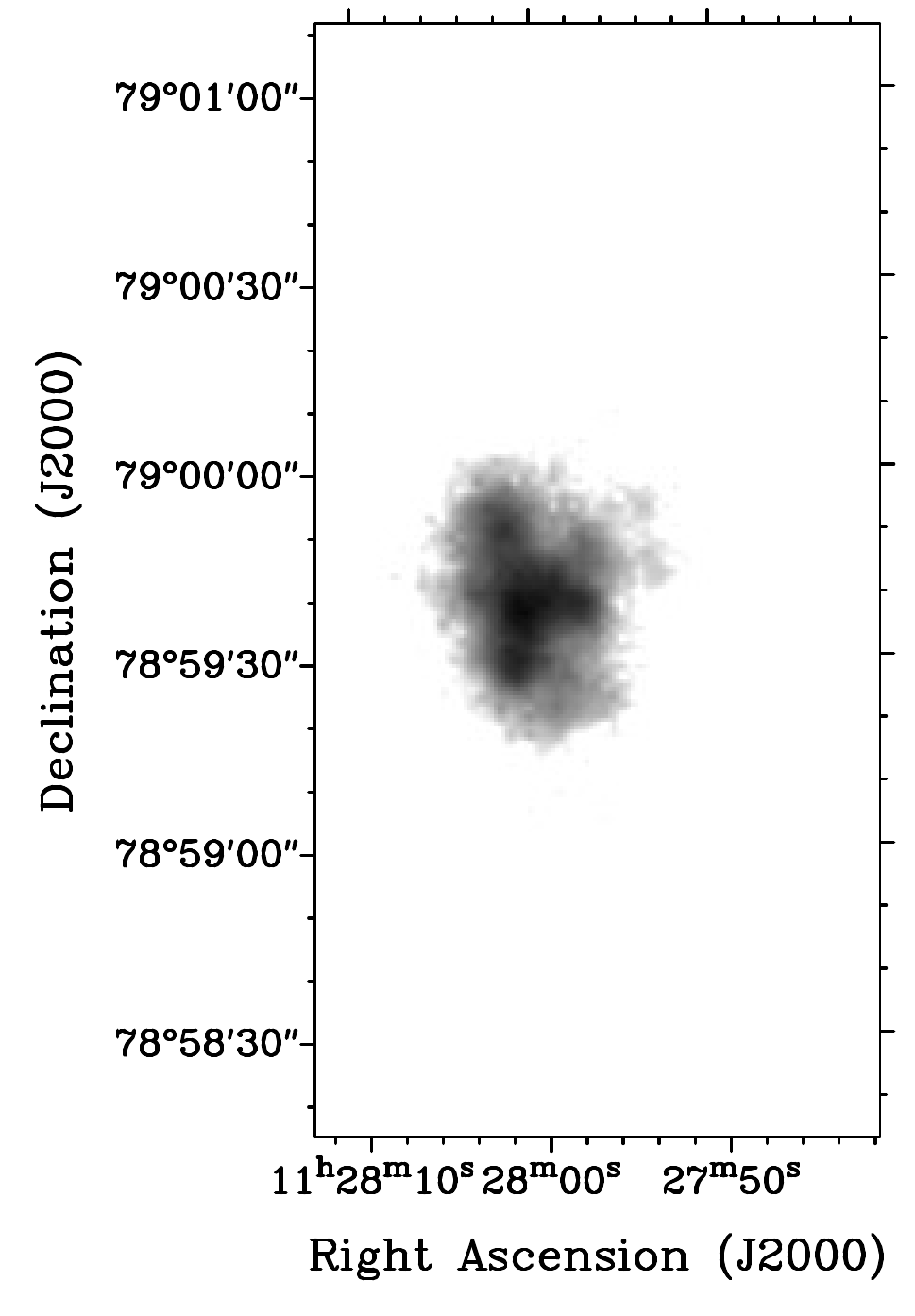}{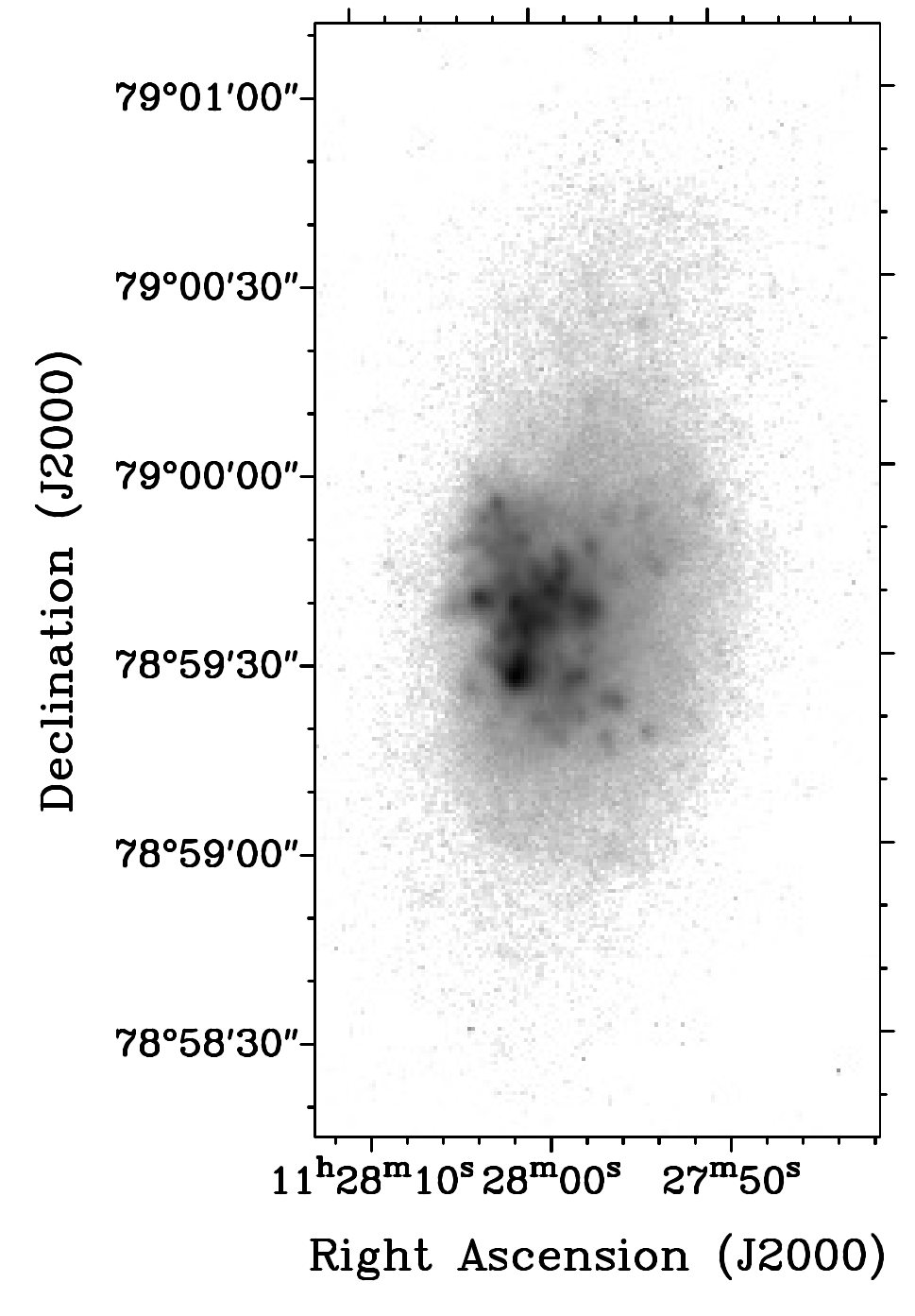}
\caption{VII Zw 403 \ \  \textit{Left:} FUV.;\ \  \textit{Right:} V-band. \label{7zw403_star}}
\end{figure}

VII Zw 403's FUV and V-band data are shown in Figure~\ref{7zw403_star}.  The FUV data were taken with GALEX and the V-band data were taken with the Lowell Observatory 1.8m Perkins Telescope \citep{hunter06, hunter10}.  The FUV and V-band surface brightness limits are \s29.5 and \s27 mag arsec$^{-2}$, respectively \citep{herrmann13}. The FUV morphology is similar to the inner morphology of the V-band.

\subsection{VII Zw 403: VLA \HI\ Morphology}
\begin{figure}[!ht]
\epsscale{0.51}
\plotone{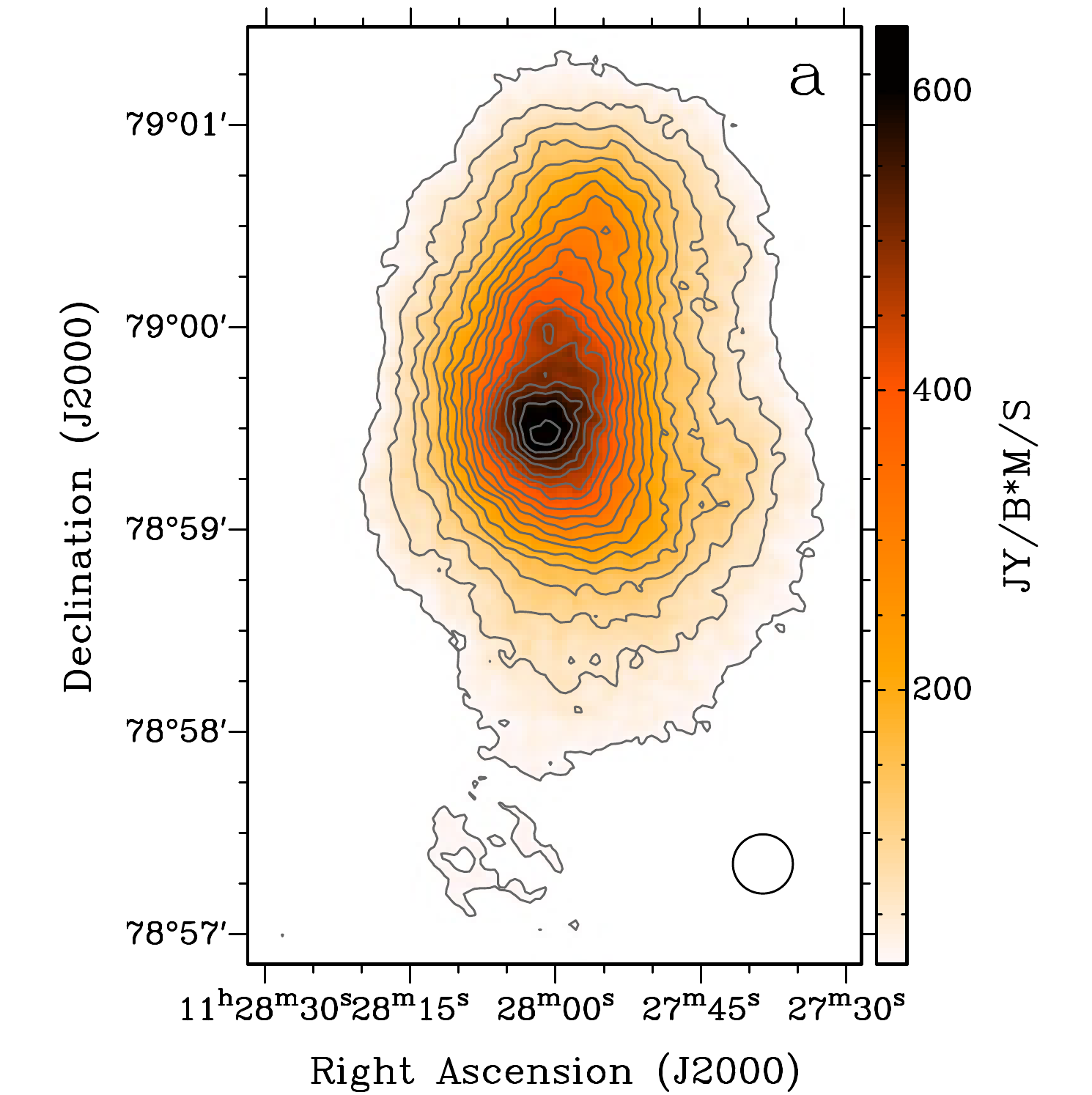}
\epsscale{0.50}
\plotone{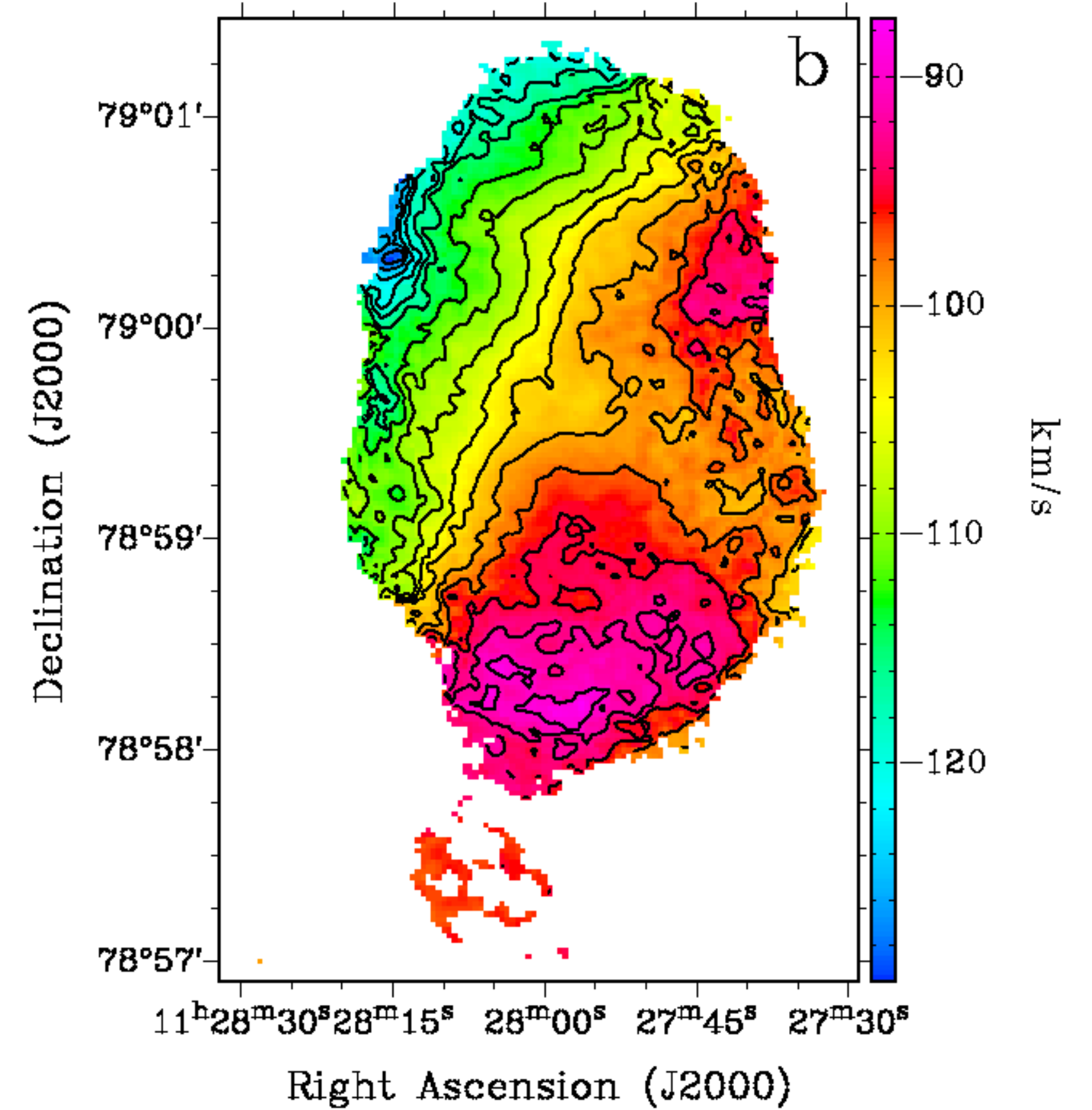}
\epsscale{0.5}
\plotone{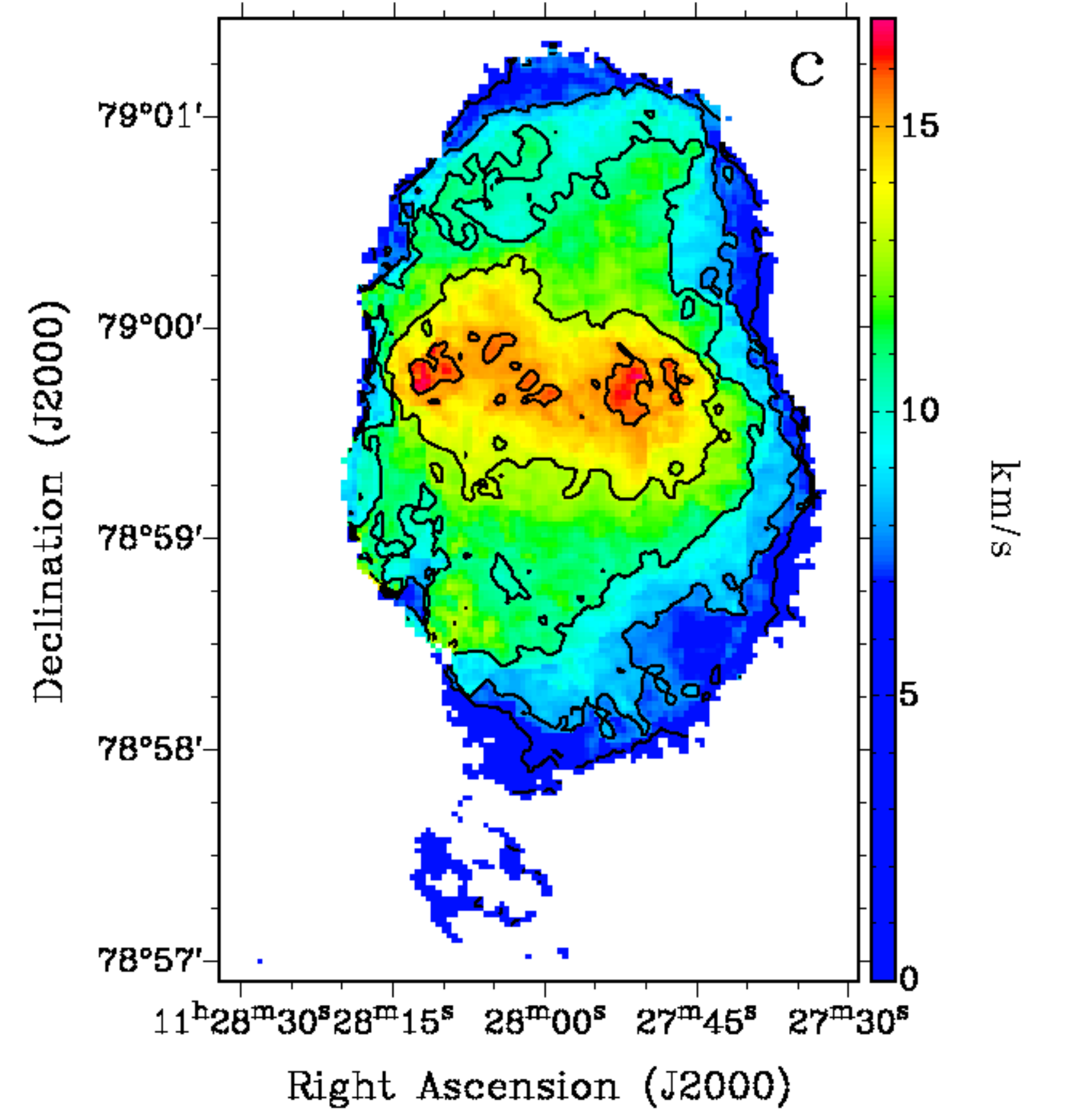}
\caption{VII Zw 403's VLA natural-weighted moment maps. (a): Integrated \HI\ intensity map; contour levels are 1$\sigma\times$(2, 6, 10, 14, 18, 22, 26, 30, 34, 38, 42, 46, 50, 54, 58, 62, 66, 70) where 1$\sigma=3.14\times10^{19}\ \rm{atoms}\ \rm{cm}^{-2}$. The black ellipse in the bottom-right represents the synthesized beam. (b): Intensity-weighted velocity field; contour levels are \n130 \kms\ to \n90 \kms\ separated by 2.5 \kms. (c): Velocity dispersion field; contour levels are 3 \kms\ to 15.5 \kms\ separated by 2.5 \kms. \label{7zw403vla_na}}
\end{figure}

\begin{figure}[!ht]
\epsscale{0.53}
\plotone{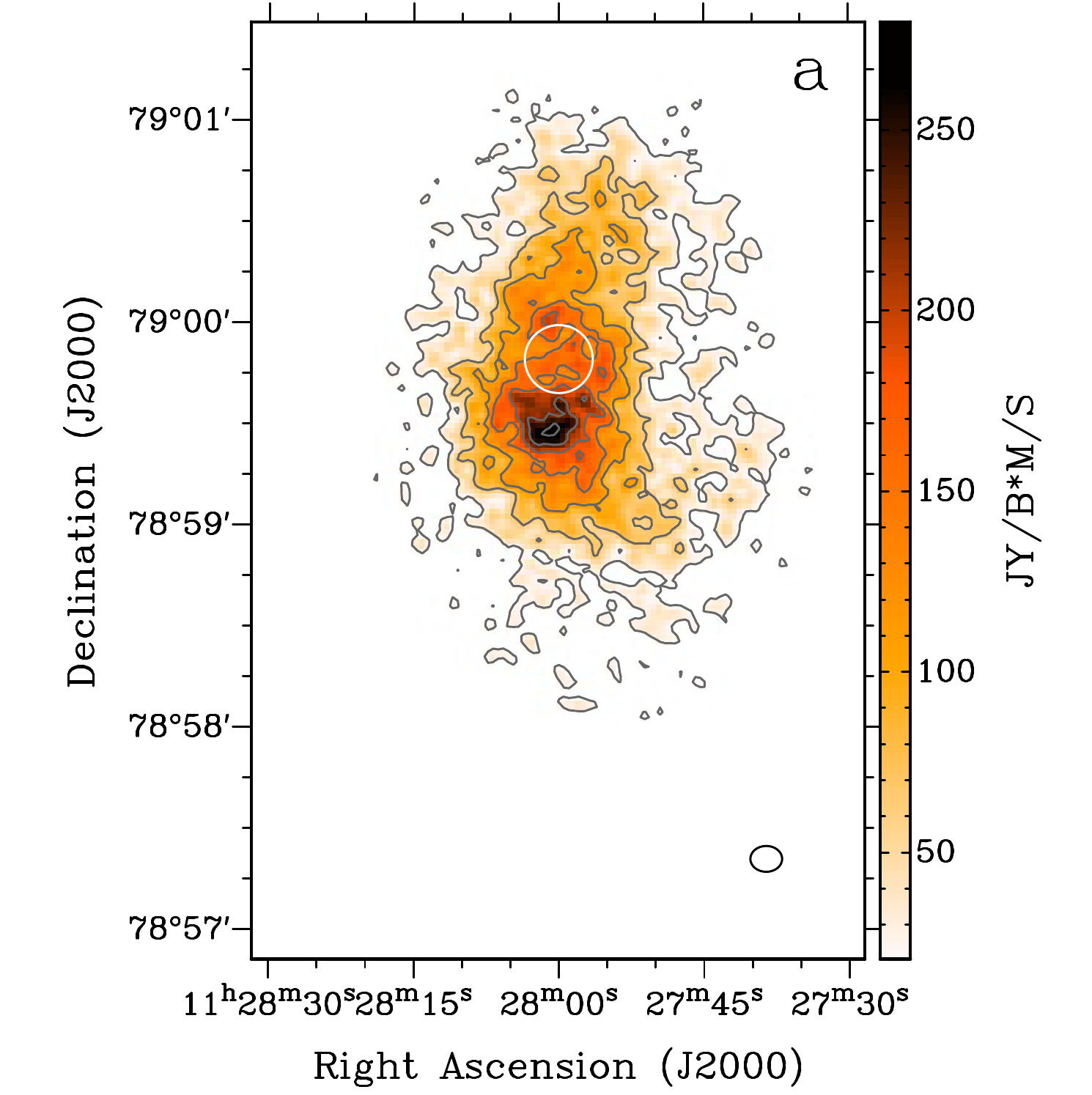}
\epsscale{0.5}
\plotone{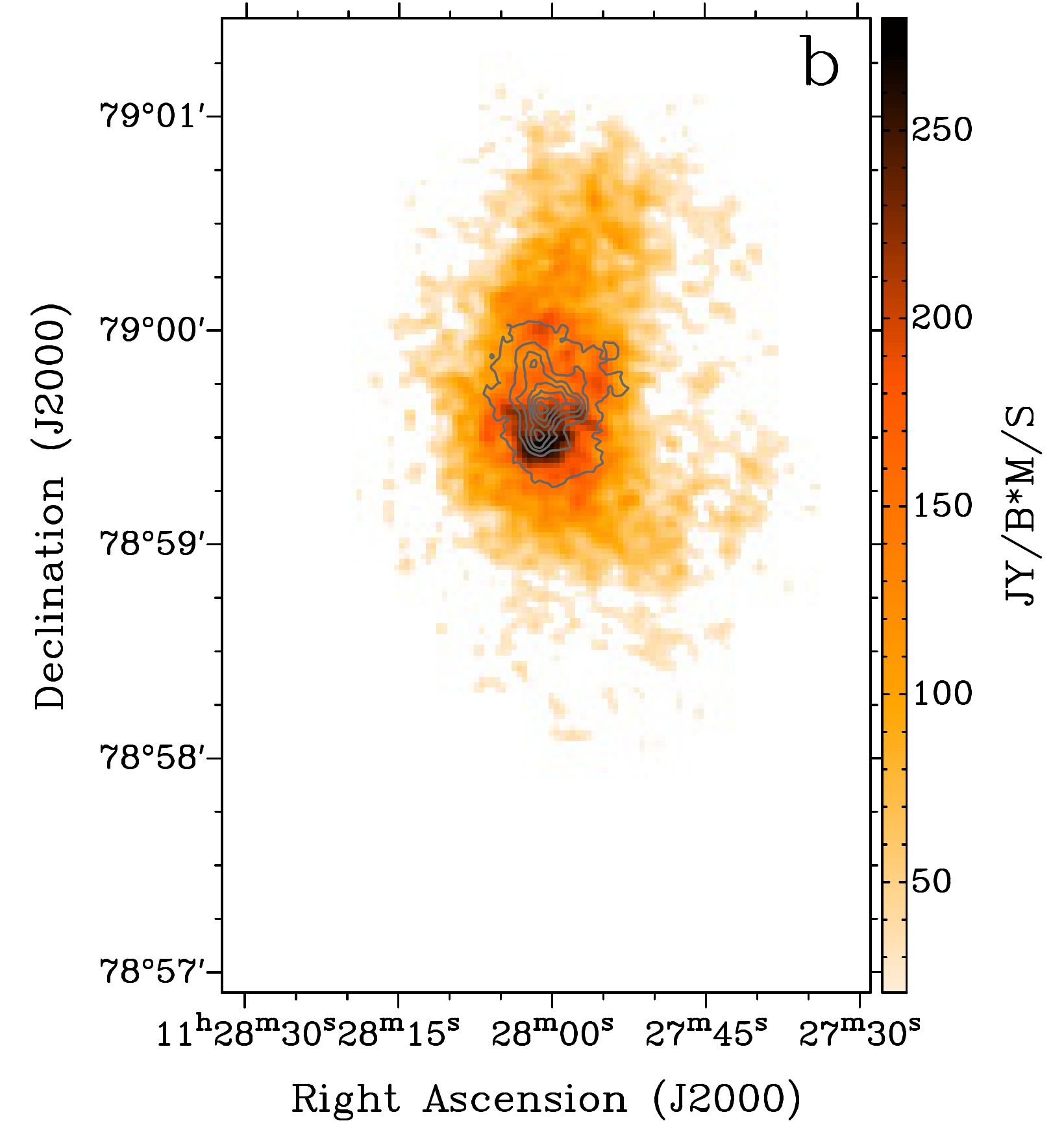}
\epsscale{0.5}
\plotone{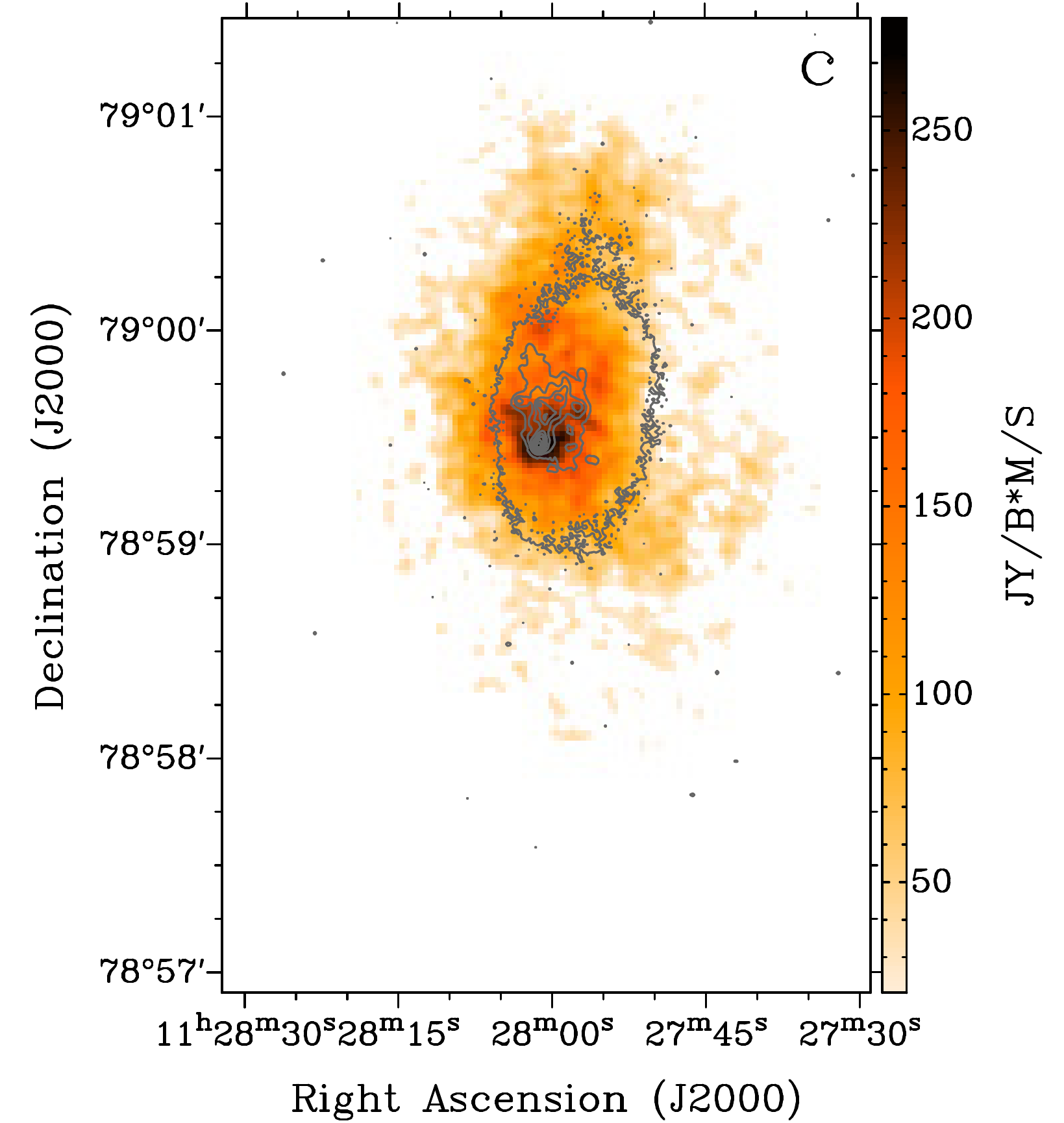}
\caption{VII Zw 403's VLA robust-weighted moment maps. (a): Integrated \HI\ intensity map; contour levels are 1$\sigma\times$(2, 6, 10, 14, 18, 22, 26) where 1$\sigma=1.59\times10^{20}\ \rm{atoms}\ \rm{cm}^{-2}$. The white circle indicates the approximate location of the potential stalled \HI\ hole from \citet{simpson11}. The black ellipse in the bottom-right represents the synthesized beam. (b): Integrated \HI\ intensity map colorscale and FUV contours.  (c): Integrated \HI\ intensity map colorscale and V-band contours. \label{7zw403vla_r}}
\end{figure}

VII Zw 403's natural-weighted integrated \HI\ intensity map as measured by the VLA is shown in Figure~\ref{7zw403vla_na}a.  The \HI\ emission has a morphological major axis in the north-south direction and is centrally peaked.  There is also some detached, tenuous \HI\ emission just to the south of the main disk that may be associated with VII Zw 403.  The robust-weighted integrated \HI\ intensity map in Figure~\ref{7zw403vla_r}a reveals some structure in the inner region of the \HI.  Just to the north of the densest \HI\ region, the fourth contour from the bottom reveals an \HI\ structure that curves toward the east.   This structure was also seen in \citet{simpson11}.  The FUV and V-band stellar contours are both plotted over the colorscale of VII Zw 403's robust-weighted integrated \HI\ intensity map in Figures~\ref{7zw403vla_r}b and \ref{7zw403vla_r}c, respectively.  The highest isophotes from the FUV and V-band data are located on the highest \HI\ column density in projection and extend just north of that.

\subsection{VII Zw 403's Optical Maps: VLA \HI\ Velocity and Velocity Dispersion Field}
The VLA \HI\ velocity field of VII Zw 403 is shown in Figure~\ref{7zw403vla_na}b.  The kinematics of the east side of the galaxy resemble solid body rotation with a major kinematic axis that does not align with the morphological major axis.  The kinematics  of the west side of the galaxy are generally disturbed with some possible organized rotation in the south.  The velocity dispersion field is shown in Figure~\ref{7zw403vla_na}c.  The dispersions reach near 17 \kms\ with the highest dispersions being centrally located.

\subsection{VII Zw 403: GBT \HI\ Morphology And Velocity Field}
\begin{figure}[!ht]
\centering
\epsscale{0.6}
\plotone{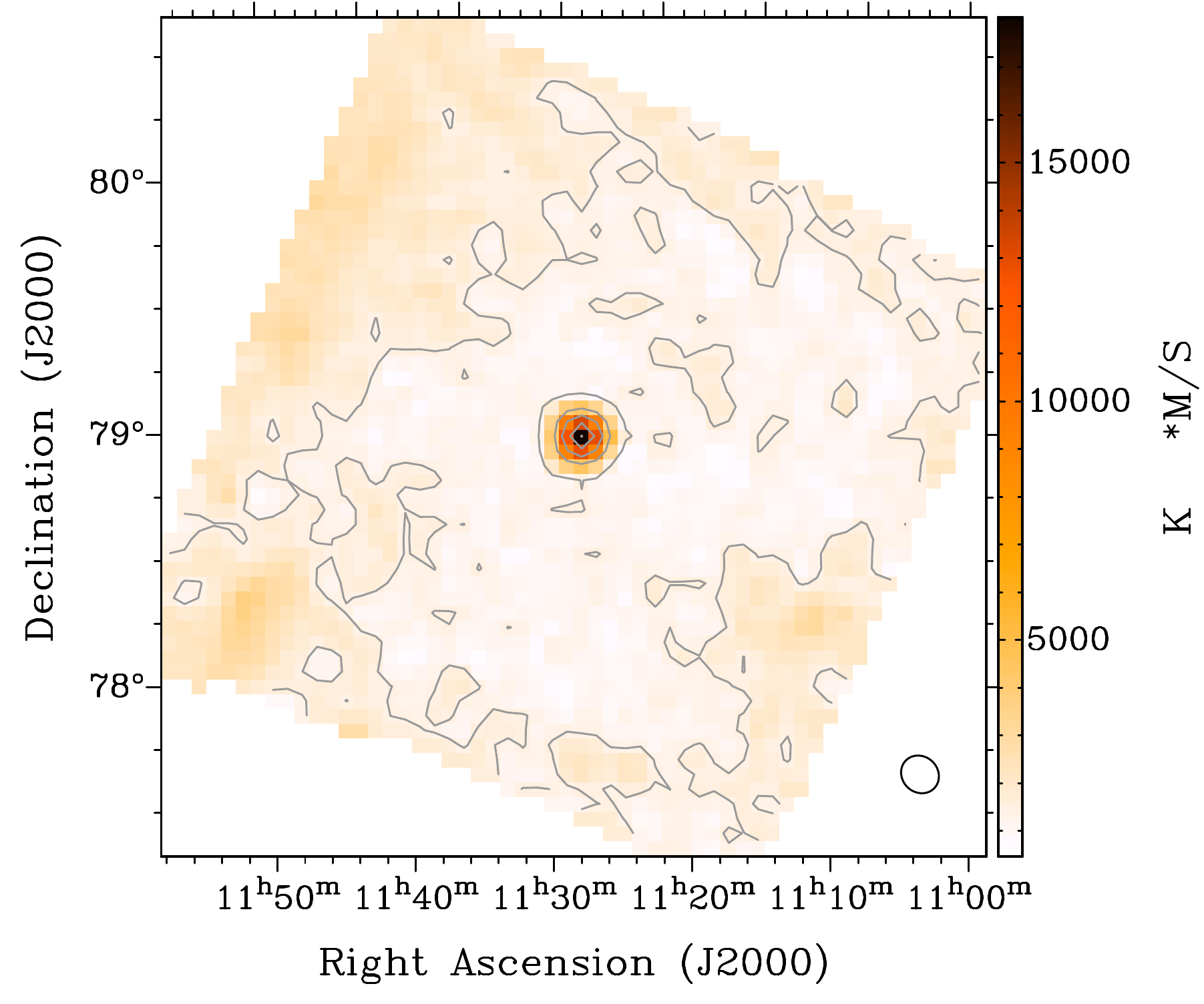}
\caption{VII Zw 403's GBT integrated \HI\ intensity map; contour levels are 1$\sigma\times$(7, 27, 47, 67, 87) where 1$\sigma=1.59\times10^{17}\ \rm{atoms}\ \rm{cm}^{-2}$. The black ellipse represents the GBT beam. \label{7zw403gbt}}
\end{figure}

VII Zw 403's integrated \HI\ intensity map, as measured with the GBT, is shown in the left side of Figure~\ref{7zw403gbt}.  The tenuous emission beyond VII Zw 403's emission (located at the center of the map) is from the Milky Way.  VII Zw 403's velocity range overlaps partially with the velocity range of the Milky Way.  These GBT maps were integrated to allow some of the Milky Way to appear in order to search as many channels as possible  for any extended emission or companions nearby VII Zw 403.  Yet, after using multiple velocity ranges for the integration of the data cube and inspection of individual channels, no companions or extended emission from VII Zw 403 were found before confusion with the Milky Way emission became a problem.  Therefore, VII Zw 403 does not appear to have any extra emission or companions nearby at the sensitivity of this map.  VII Zw 403's GBT \HI\ velocity field was also inspected, however, there was no discernible velocity gradient.

\subsection{VII Zw 403: \HI\ Mass}
VII Zw 403's total \HI\ mass detected in the VLA natural-weighted data is $4.2\times10^{7}$ M$_{\sun}$, while its mass from the GBT data is $5.1\times10^{7}$ M$_{\sun}$.  The Milky Way emission could be contributing to some of the mass measured from the GBT data, however, that is unlikely; the velocity range used in \textsc{ispec} was the same as that used to make the integrated \HI\ intensity map (see Figure~\ref{7zw403gbt}) and the box size in which the flux was summed tightly enclosed the VII Zw 403 emission.  The VLA was able to recover 82\% of the GBT mass, although VII Zw 403's GBT mass should be higher since some channels that contain emission from VII Zw 403 were excluded from the mass measurement to avoid confusion with the Milky Way.

\section{Discussion: VII Zw 403}\label{7zwdisc}
\begin{figure}[!ht]
\epsscale{0.75}
\centering
\plotone{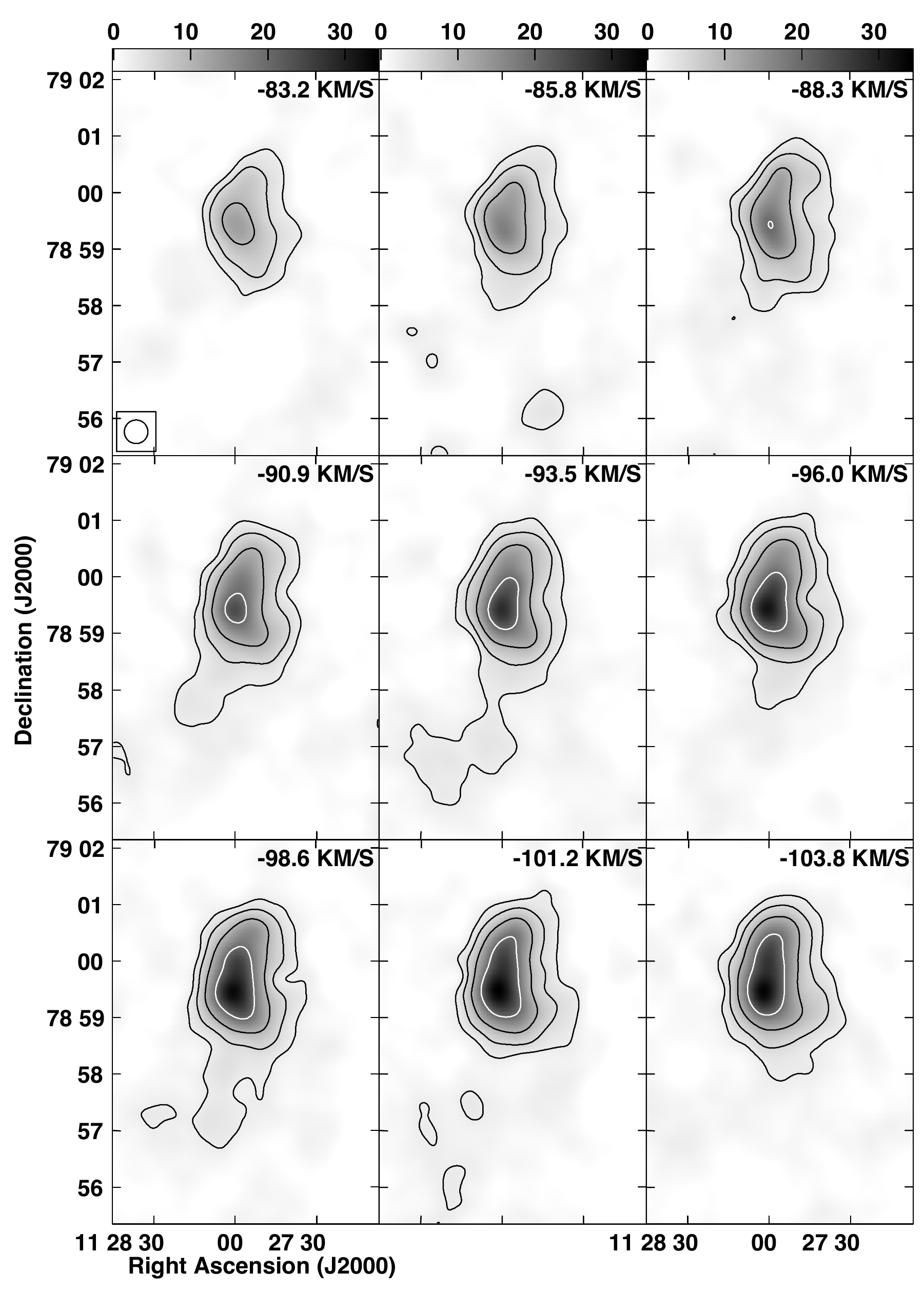}
\caption{VII Zw 403's $25\arcsec\times25\arcsec$ convolved VLA channel maps; contour levels are 1$\sigma\times$(2.5, 5, 10, 20), where 1$\sigma=1\ \rm{mJy}\ \rm{beam}^{-1}$. The synthesized beam is represented by an ellipse in the top left panel.  The colorscales on top of the panels are given in units of mJy/beam. \label{7zw403vla_cvl}}
\end{figure}

\begin{figure}[!ht]
\epsscale{0.8}
\centering
\plotone{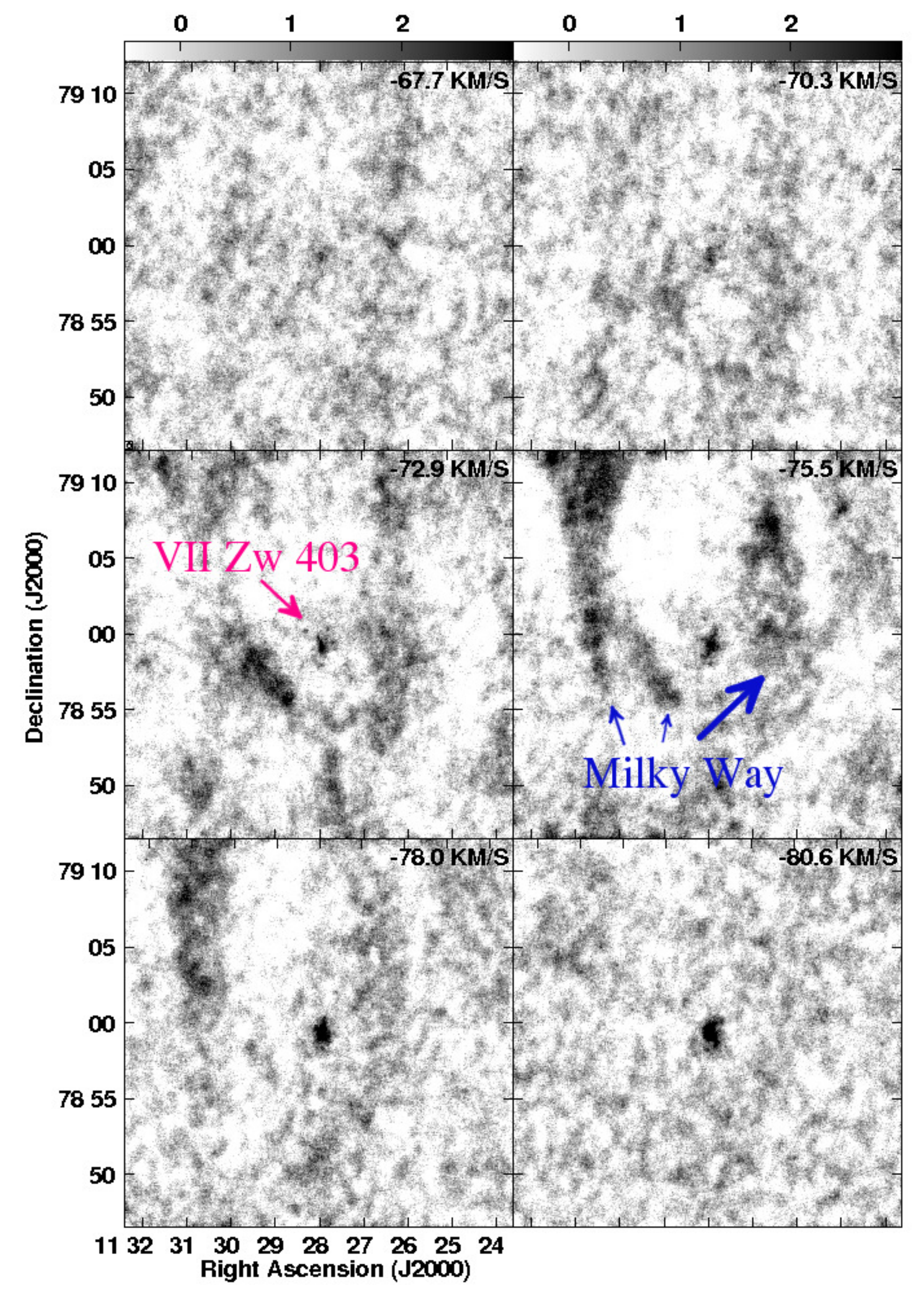}
\caption{VII Zw 403's channel maps from the natural-weighted VLA data cube where some overlap of VII Zw 403's and the Milky Way's velocity range occurs. The synthesized beam is represented by an ellipse in the top left panel.  The colorscales on top of the panels are given in units of mJy/beam. \label{7zw403vla+mw}}
\end{figure}

\begin{figure}[!ht]
\epsscale{1}
\centering
\plotone{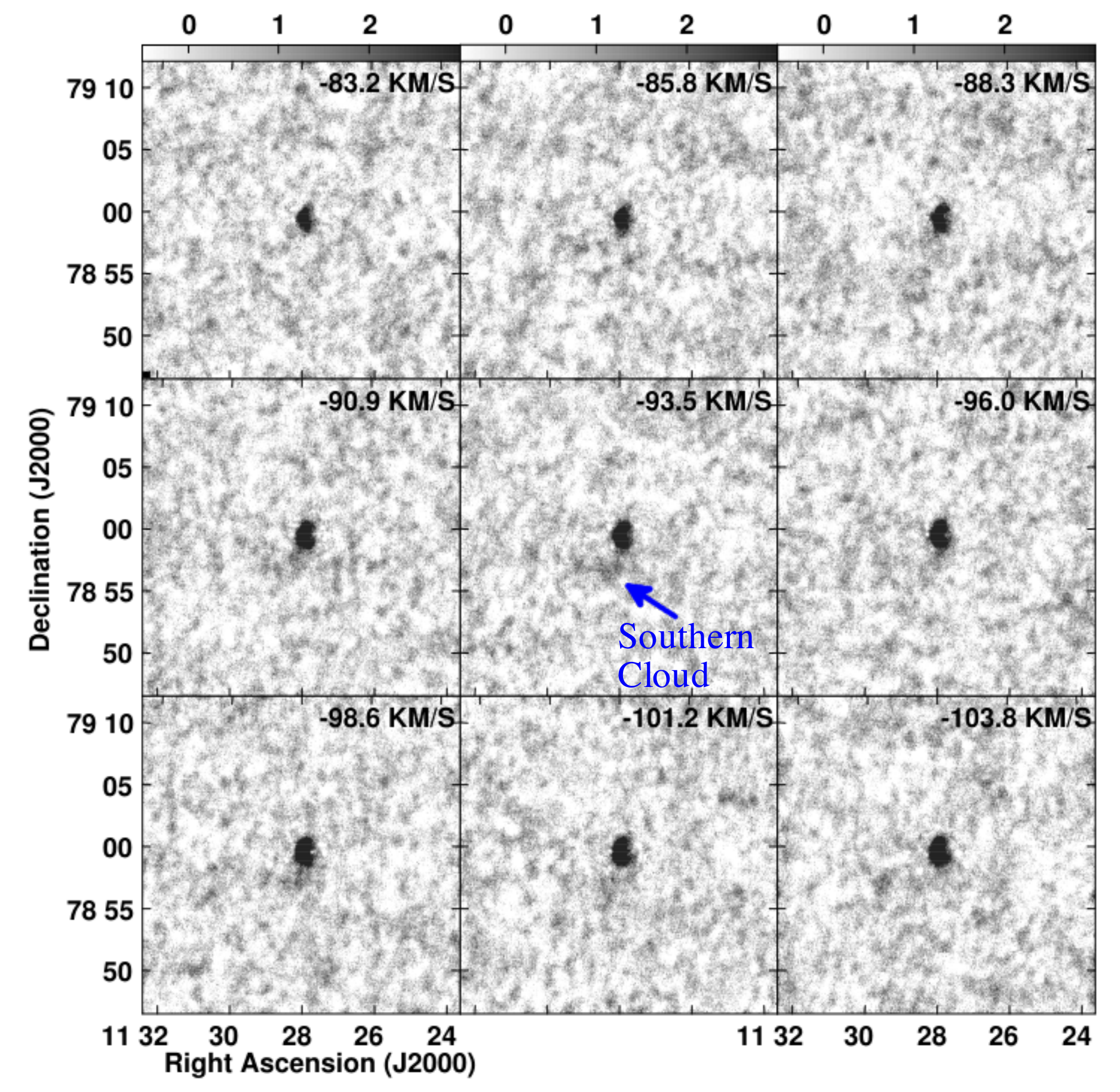}
\caption{The channel maps from the natural-weighted VLA data cube from which the extra cloud emission south of VII Zw 403's main body comes. The synthesized beam is represented by an ellipse in the top left panel.  The colorscales on top of the panels are given in units of mJy/beam. \label{7zw403tail_channels}}
\end{figure}

The most noticeable morphological peculiarity in VII Zw 403's VLA data is the detached gas cloud to the south of the disk in the natural-weighted integrated \HI\ intensity map (Figure~\ref{7zw403vla_na}).  Because this feature appears in more than three consecutive channels in the 25\arcsec$\times$25\arcsec convolved natural-weighted data cube, as can be seen in Figure~\ref{7zw403vla_cvl}, it is unlikely to be noise in the map and is therefore considered real emission. However, Figure~\ref{7zw403vla+mw} shows that there is overlap in velocity between VII Zw 403 and HI in the Milky Way. To make sure that the southern detached cloud is not Milky Way emission in the foreground of VII Zw 403, the channels that contained the southern detached cloud were checked for Milky Way emission (see Figure~\ref{7zw403tail_channels}); none was found. We are therefore confident that the cloud is a real feature connected with VII Zw 403.

\begin{figure}[!ht]
\epsscale{0.51}
\begin{center}
\plotone{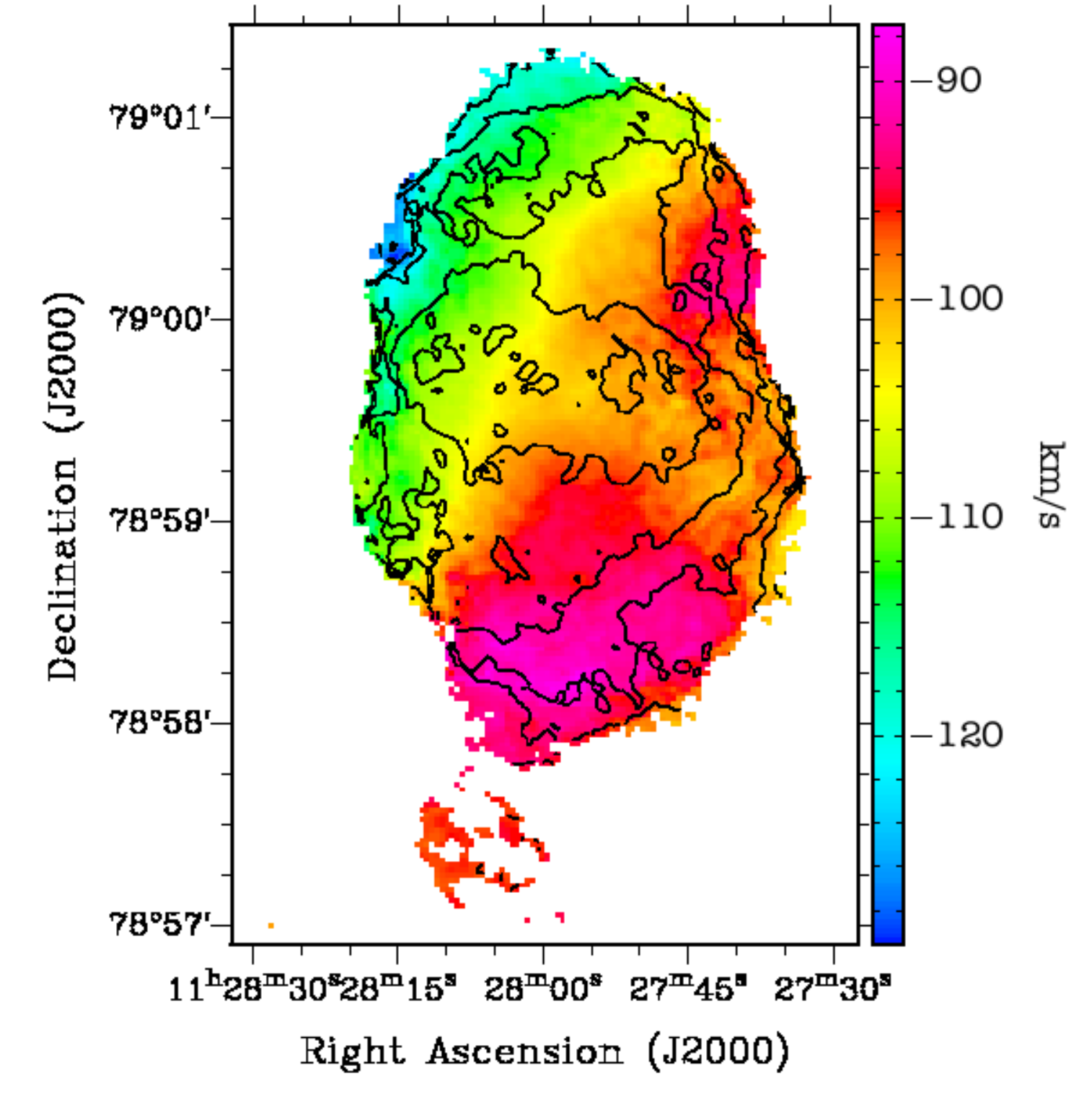}
\end{center}
\caption{The colorscale of VII Zw 403's VLA \HI\ velocity field and contours from the VLA \HI\ velocity dispersion map from Figure~\ref{7zw403vla_na}.  \label{7zw403_x1_lininup}}
\end{figure}

Kinematically, the east side of the galaxy has rotation that resembles solid body rotation,  while the velocity field on the west side shows a break in the isovelocity contours from northeast to southwest.  Strikingly, the velocity dispersions in the natural-weighted data show higher values along this break.  The alignment of the two features can be seen in Figure~\ref{7zw403_x1_lininup}, where the velocity dispersion field contours have been plotted over the colorscale of the velocity field.  

\begin{figure}[!ht]
\epsscale{0.65}
\begin{center}
\plotone{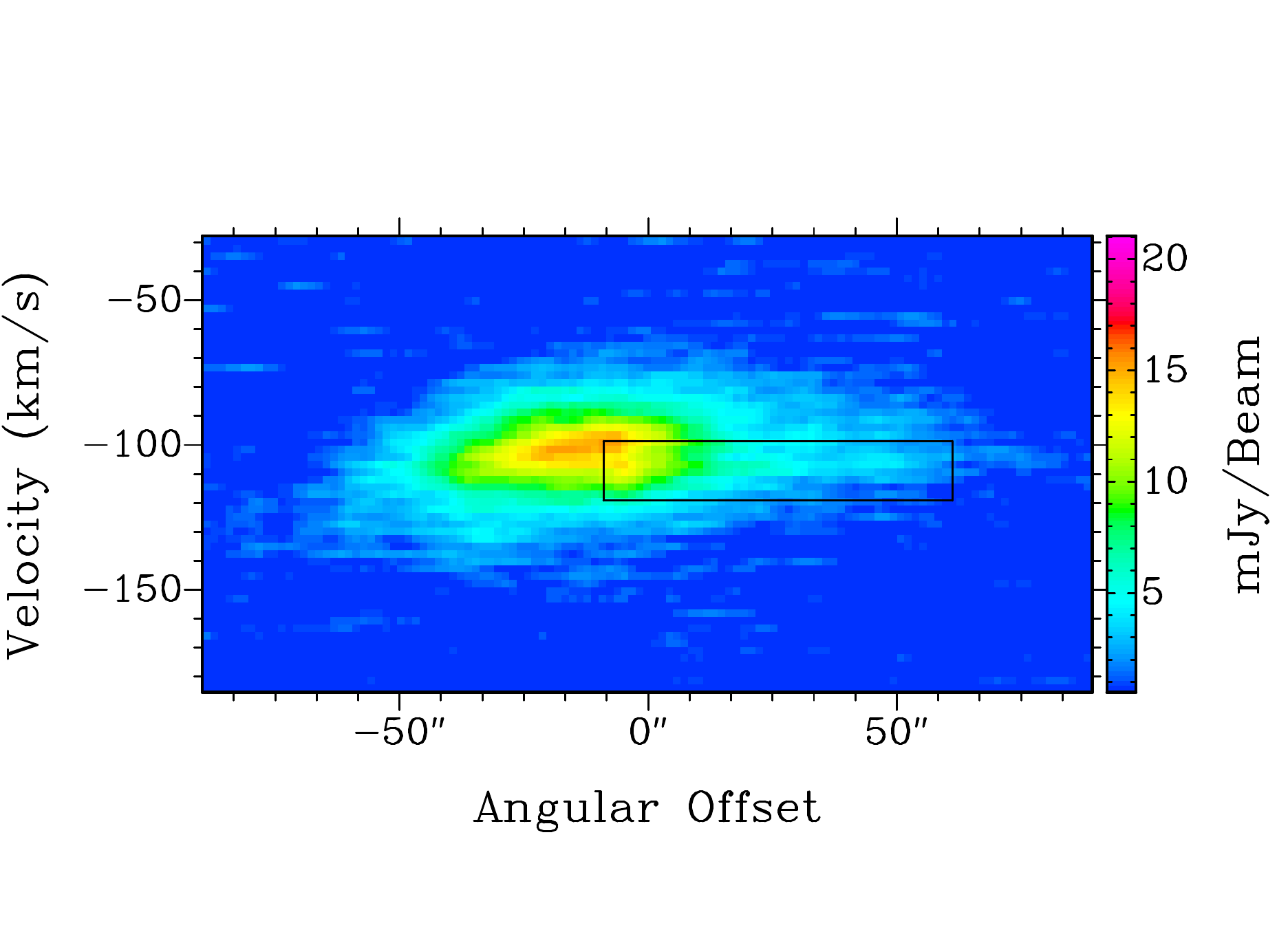}
\epsscale{0.44}
\plotone{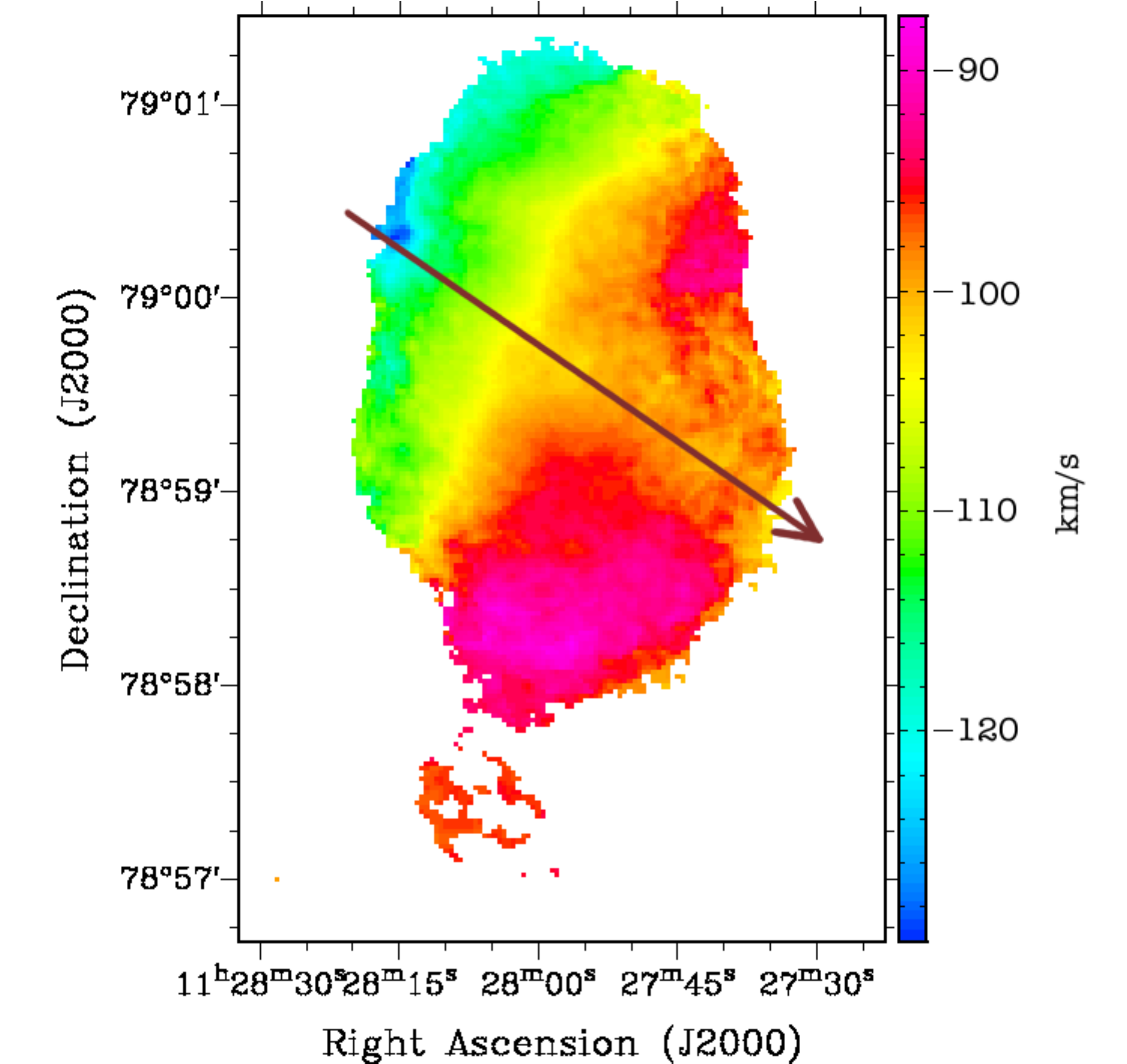}
\end{center}
\caption{\textit{Left:} P-V diagram of VII Zw 403's western velocity disturbance from the VLA natural-weighted data cube, starting at a 1$\sigma$ level (0.60 mJy/beam).  The black box indicates the emission from the potential external gas cloud. \textit{Right:}  VII Zw 403's natural-weighted \HI\ velocity field with a red arrow indicating the location of the corresponding slice through the galaxy and pointing in the direction of positive offset.  \label{7zw403_pv_veldist}}
\end{figure}

A P-V diagram through the western velocity disturbance using the natural-weighted data cube is shown in Figure~\ref{7zw403_pv_veldist}.  The gas goes from blueshifted to redshifted velocities in a solid body manner as the P-V diagram moves towards positive offsets.  However, in the P-V diagram there is a thin horizontal streak of high brightness from \n8\arcsec\ to 58\arcsec\ and \n110 to \n100 \kms\ as outlined in the black box.  Most of the rest of the tenuous gas at this angular offset range appears to be above \n100 \kms, consistent with the rest of the gas on the west side rotating in a solid body fashion with the east side of the disk.  It is possible that this density enhancement is a external gas cloud in the line of sight that is disturbing the velocity field of VII Zw 403.

\begin{figure}[!ht]
\epsscale{1}
\begin{center}
\plottwo{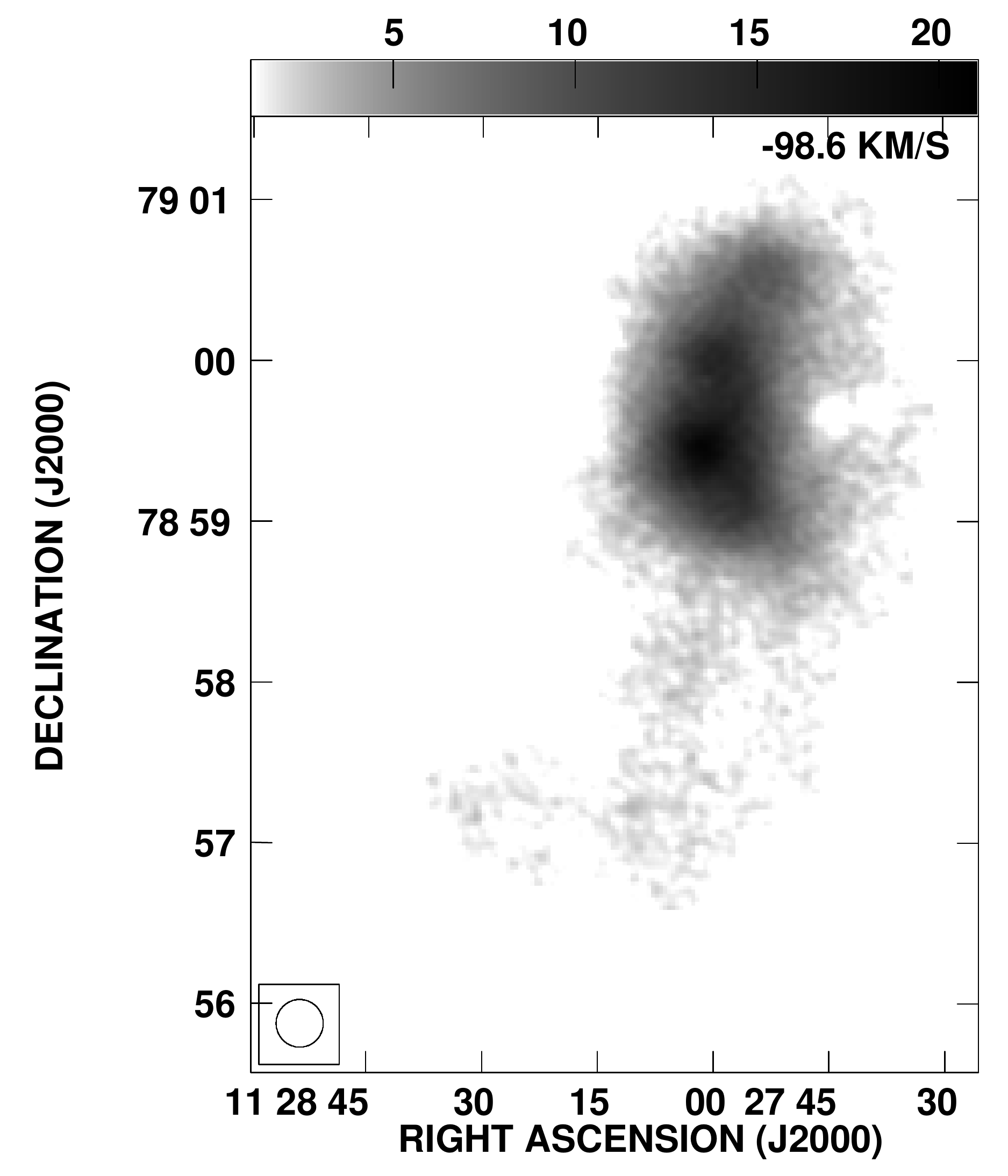}{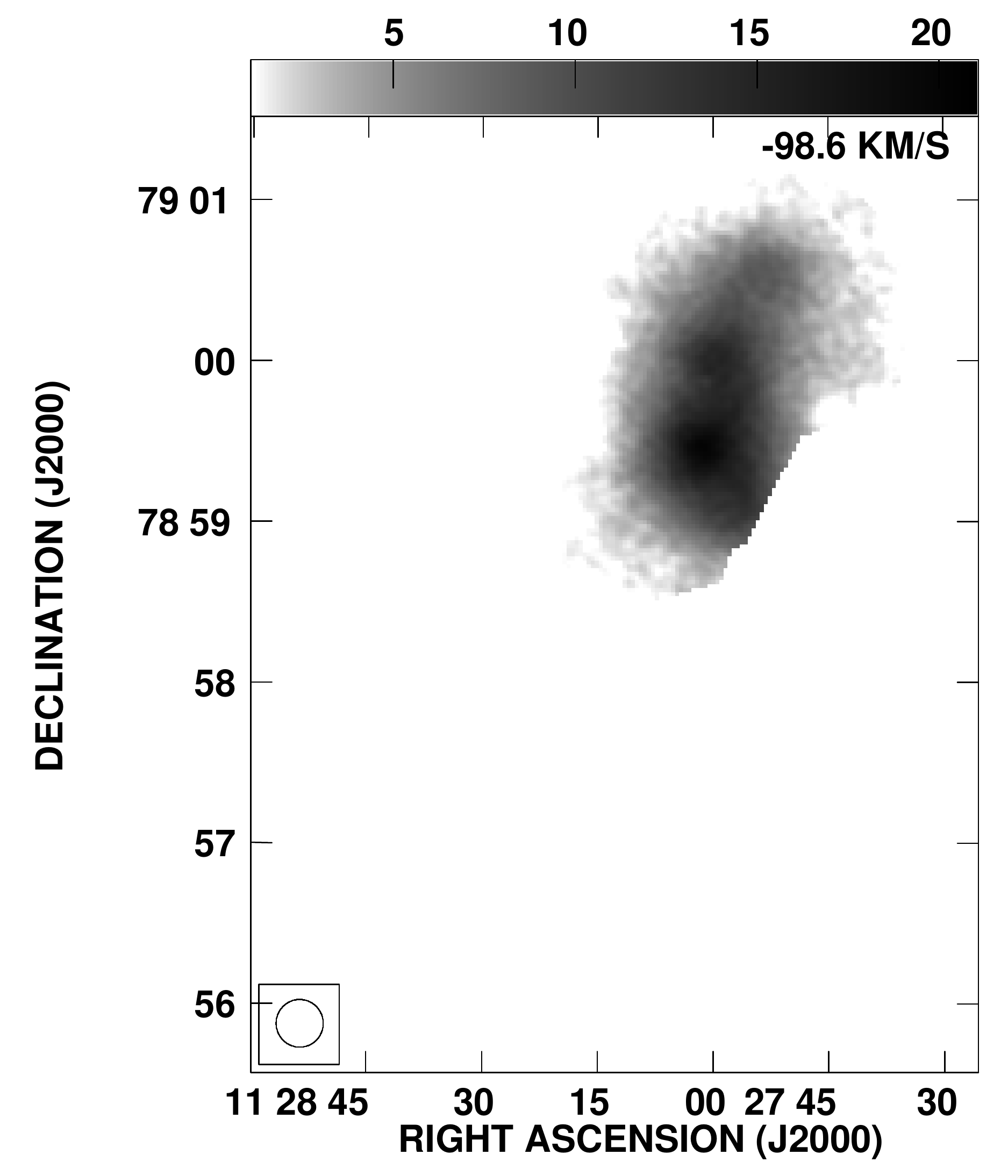}
\end{center}
\caption{\textit{Left:} A channel map from VII Zw 403's VLA natural-weighted data cube showing the data \textit{prior} to blanking out the emission thought to be associated with a gas cloud in the line of sight of VII Zw 403's disk.    \textit{Right:} The same channel from VII Zw 403's VLA natural-weighted data cube showing the data \textit{after} blanking out the emission thought to be associated with a gas cloud in the line of sight of VII Zw 403's disk.  The colorscales on top of the panels are given in units of mJy/beam. \label{7zw403_blank_example}}
\end{figure}

Since the gas at anomalous velocities appears at higher negative velocities than the rest of the gas in the disk in the same line of sight (see Figure~\ref{7zw403_pv_veldist}) and since the velocity disturbance is highly directional (northeast to southwest), the emission associated with it morphologically extends from the southwestern edge of the disk when individual velocity channels are viewed (as in the left image of Figure~\ref{7zw403_blank_example}). Both to remove this anomalous gas component from the line of sight and also examine whether it could be an external gas cloud, the AIPS task \textsc{blank} was used to manually mark-out the regions containing the emission. This was done twice: once to blank out the emission from the potential cloud, resulting in a data cube just containing the disk emission; and a second time using the first blanked data cube as a mask to retain just the cloud emission. A map of an unblanked channel and a map of the same channel with the potential cloud emission blanked are shown in Figure~\ref{7zw403_blank_example} as an example.

\begin{figure}[!ht]
\epsscale{1.03}
\begin{center}
\plottwo{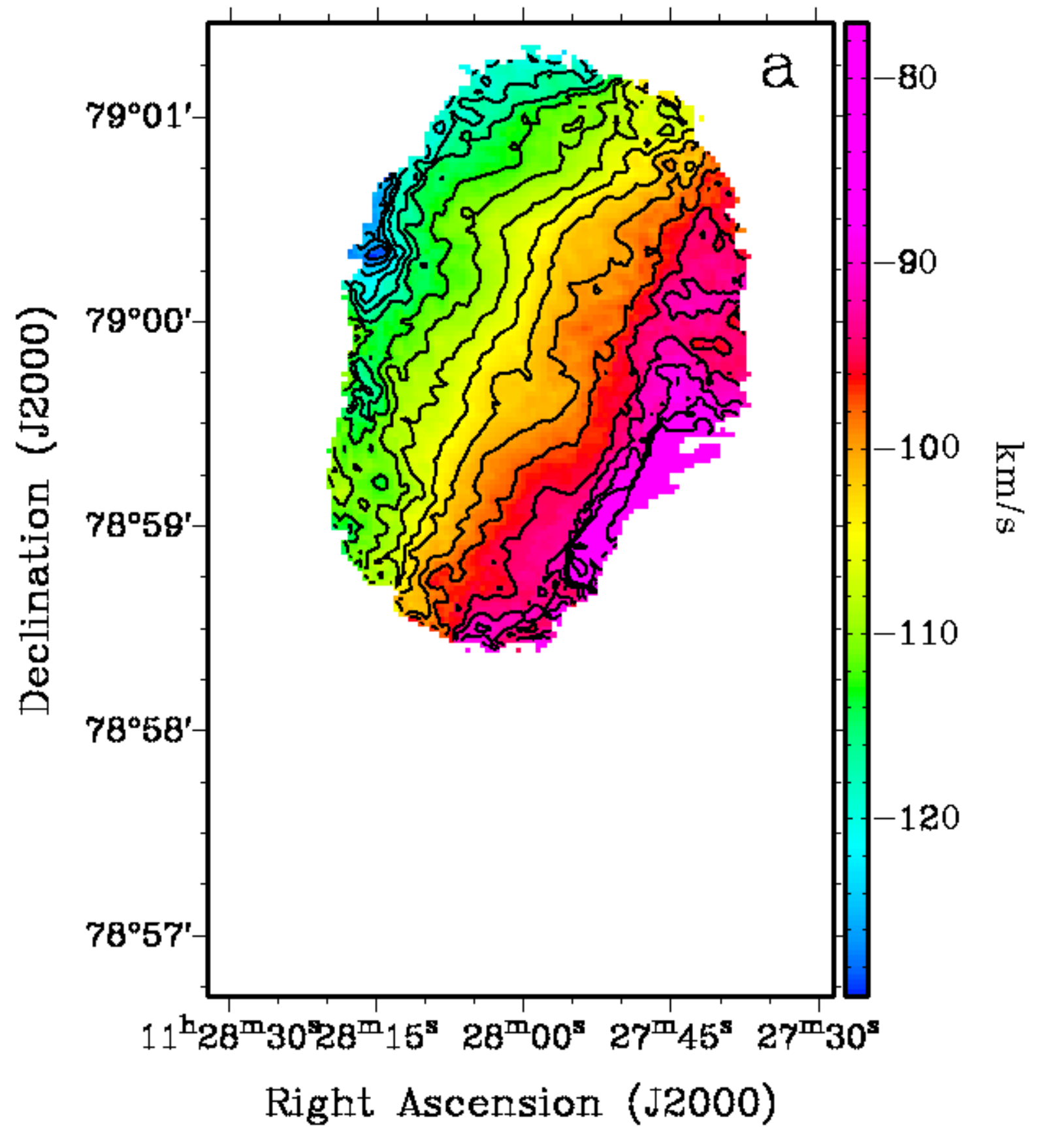}{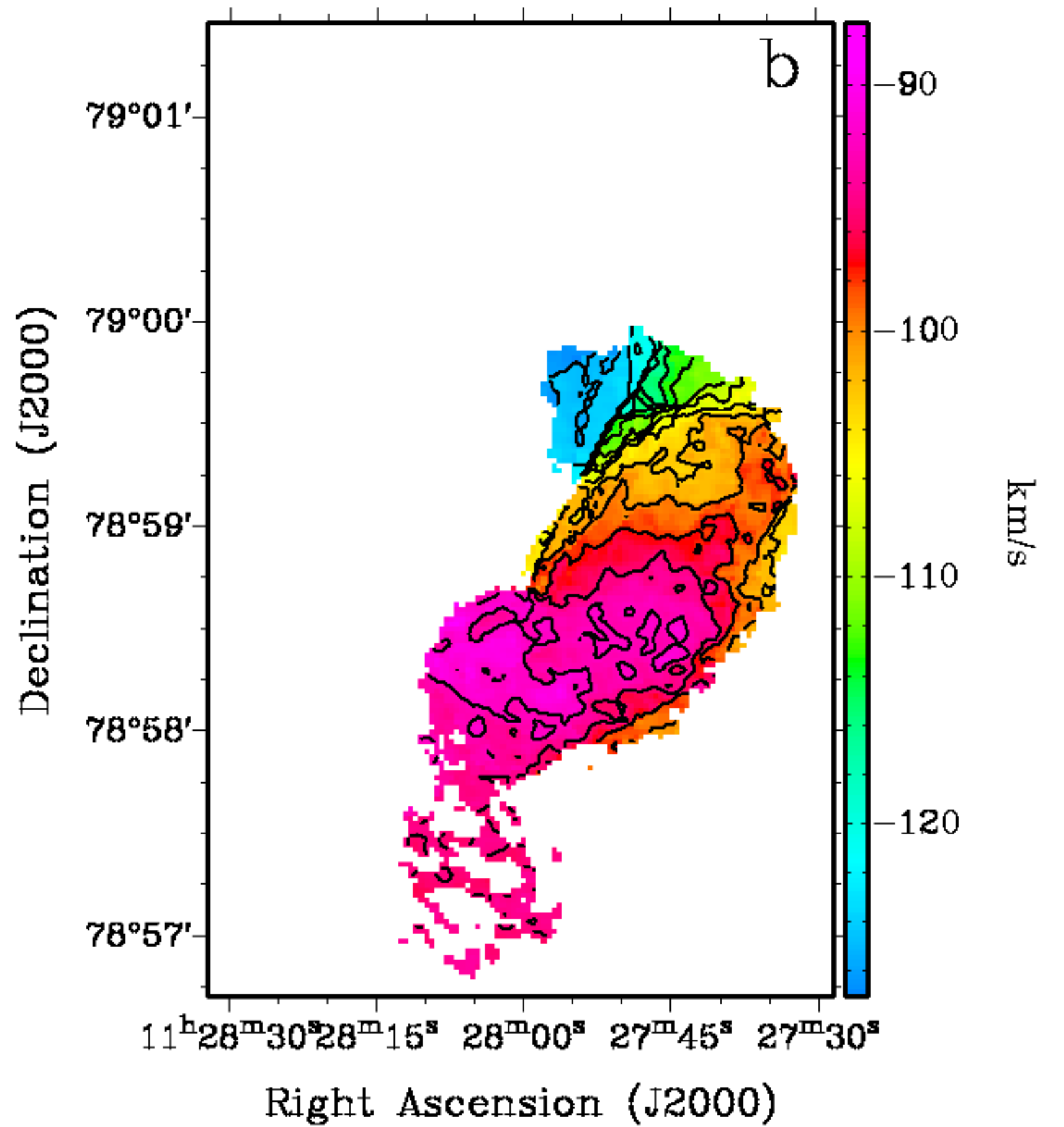}\\
\plottwo{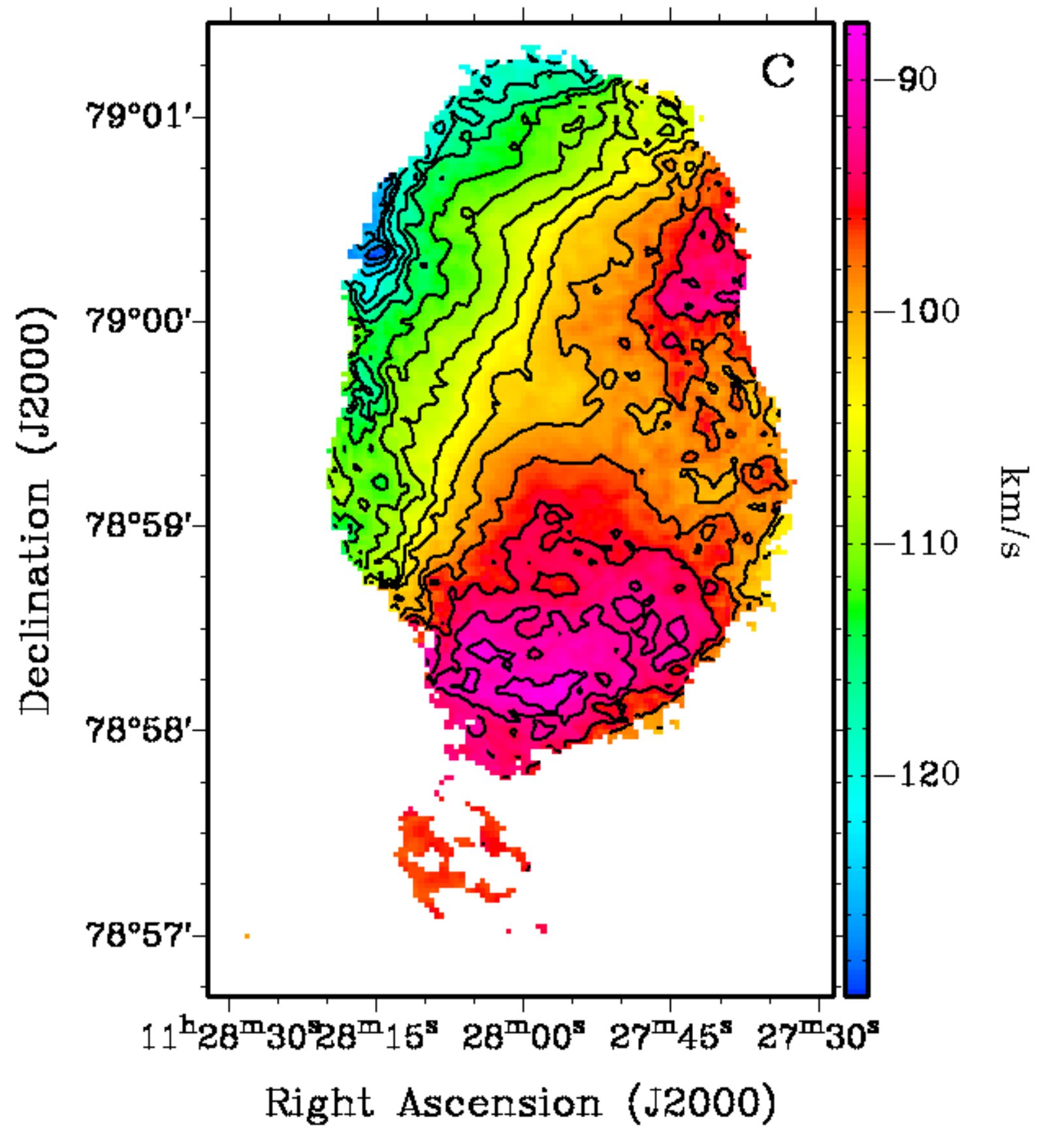}{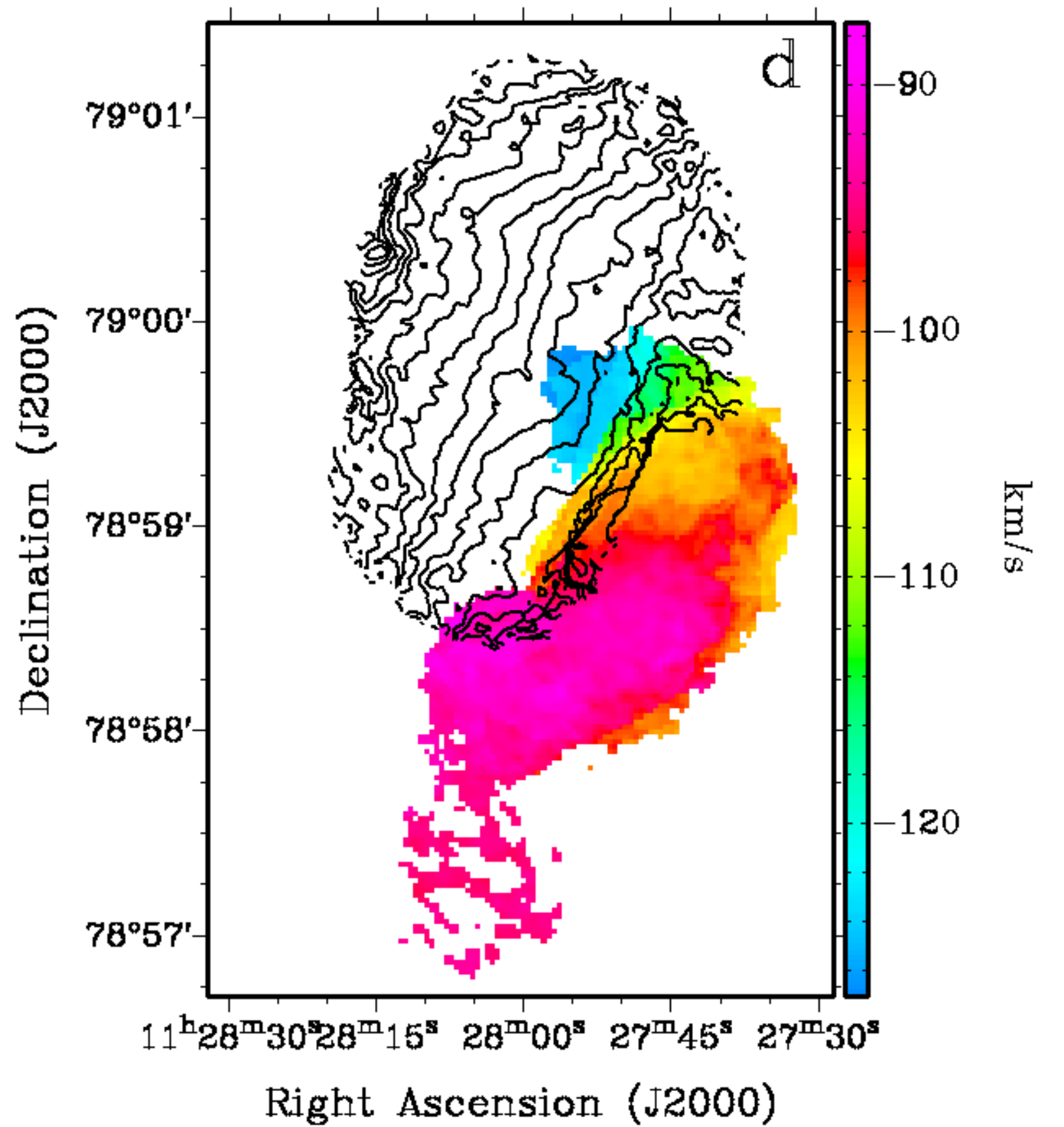}
\end{center}
\caption{VII Zw 403's VLA data: (a): Intensity-weighted velocity field of VII Zw 403 without the emission from foreground gas cloud.  The contours are \n127.5 \kms\ to \n85.0 \kms\ in intervals of 2.5 \kms. (b): The intensity-weighted velocity field of the foreground gas cloud.  The contours are \n125.0 \kms\ to \n92.5 \kms\ in intervals of 2.5 \kms.  (c): The original velocity field map from Figure~\ref{7zw403vla_na} for comparison. (d): The contours from the velocity field in the upper left and colorscale of the velocity field in the upper right.  \label{7zw403_chim}}
\end{figure}

The results of blanking are shown in Figure~\ref{7zw403_chim}.  Figure~\ref{7zw403_chim}a is the velocity field of the galaxy without most of the emission from the gas cloud, Figure~\ref{7zw403_chim}b is the velocity field of the gas cloud, and Figure~\ref{7zw403_chim}d compares the two by showing the contours of the first and the colorscale of the second.  Figure~\ref{7zw403_chim}c is the original velocity field from Figure~\ref{7zw403vla_na} for comparison.  With most of the gas cloud emission removed, the velocity field on the west side of the galaxy now generally follows the solid body trend seen on the east side of the galaxy.  The emission from the gas cloud also has a generally smooth transition in velocities.  The mass of the gas cloud in the line of sight of VII Zw 403 is \s$7.5\times10^{6}$ M$_{\sun}$ or 18\% of the total VLA \HI\ mass measured.  The projected length of the cloud at the distance of VII Zw 403 is \s4.5 kpc when the curves of the cloud are included.  The V-band image of VII Zw 403 (a cropped version is shown in Figure~\ref{7zw403_star}) extends to the new gas cloud presented in Figure~\ref{7zw403_chim}, however, it does not show any emission in that region above the limiting magnitude of \s27 mag arsec$^{-2}$.

\begin{figure}[!ht]
\epsscale{1.1}
\begin{center}
\plottwo{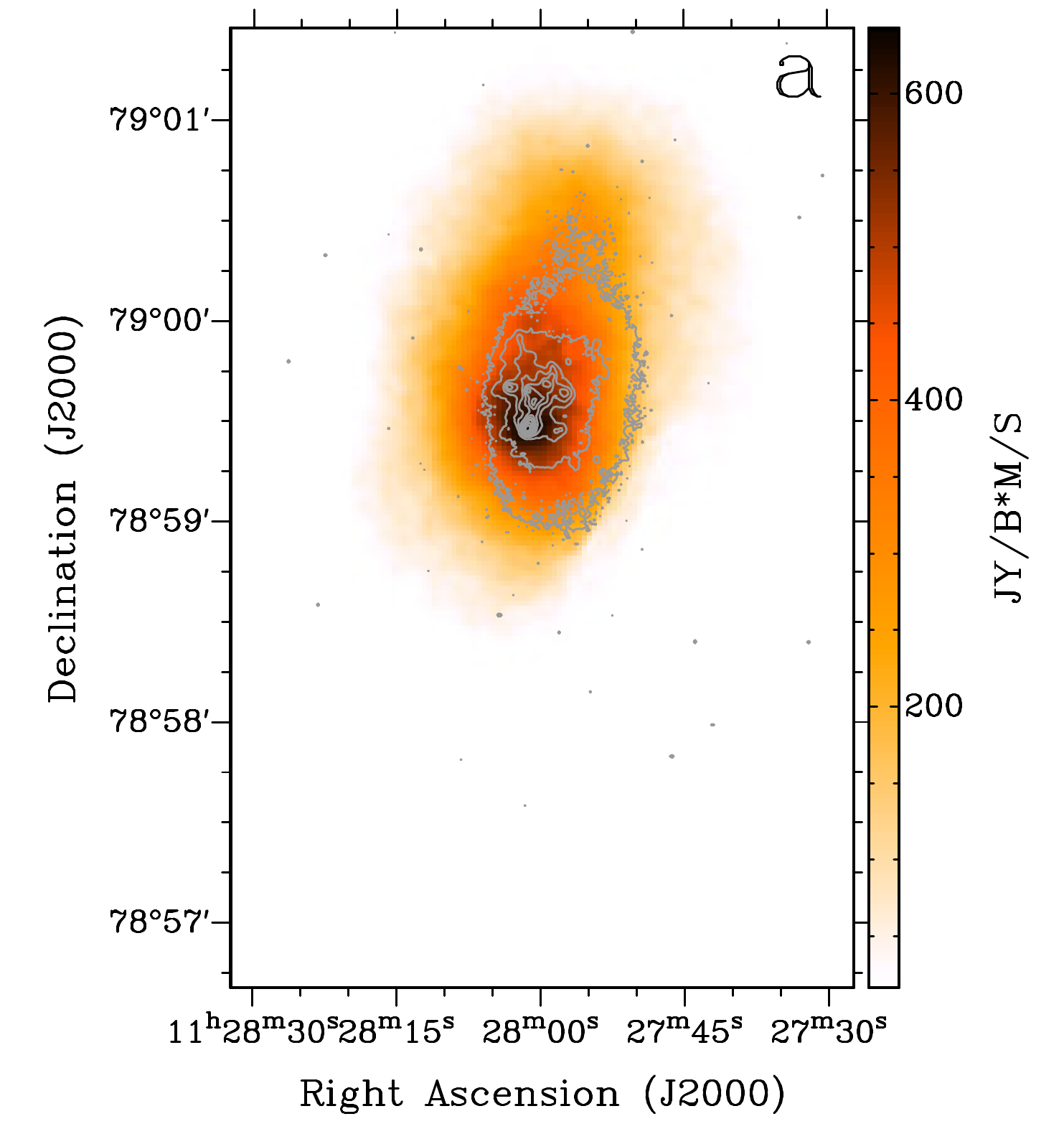}{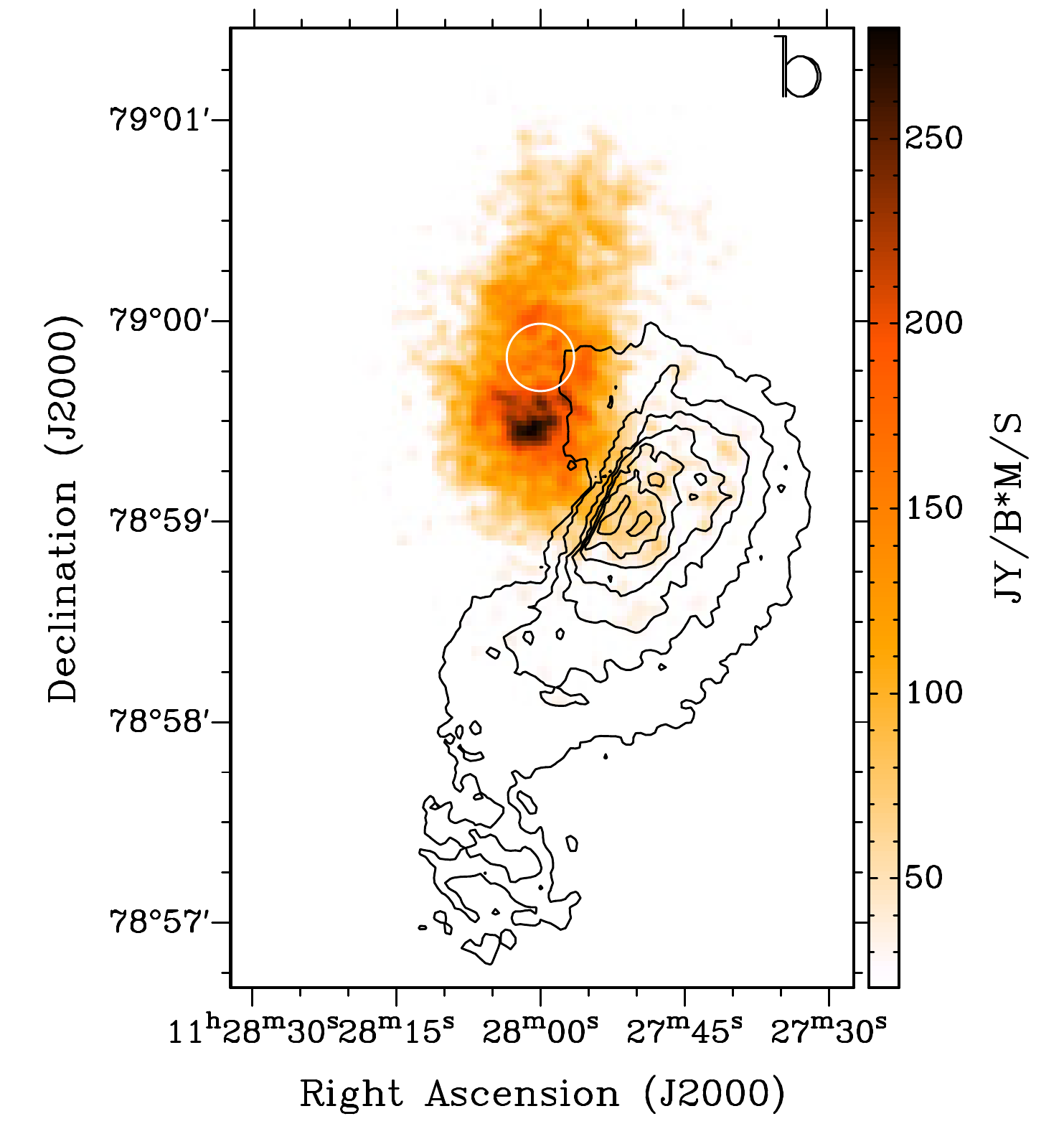}
\epsscale{0.55}
\plotone{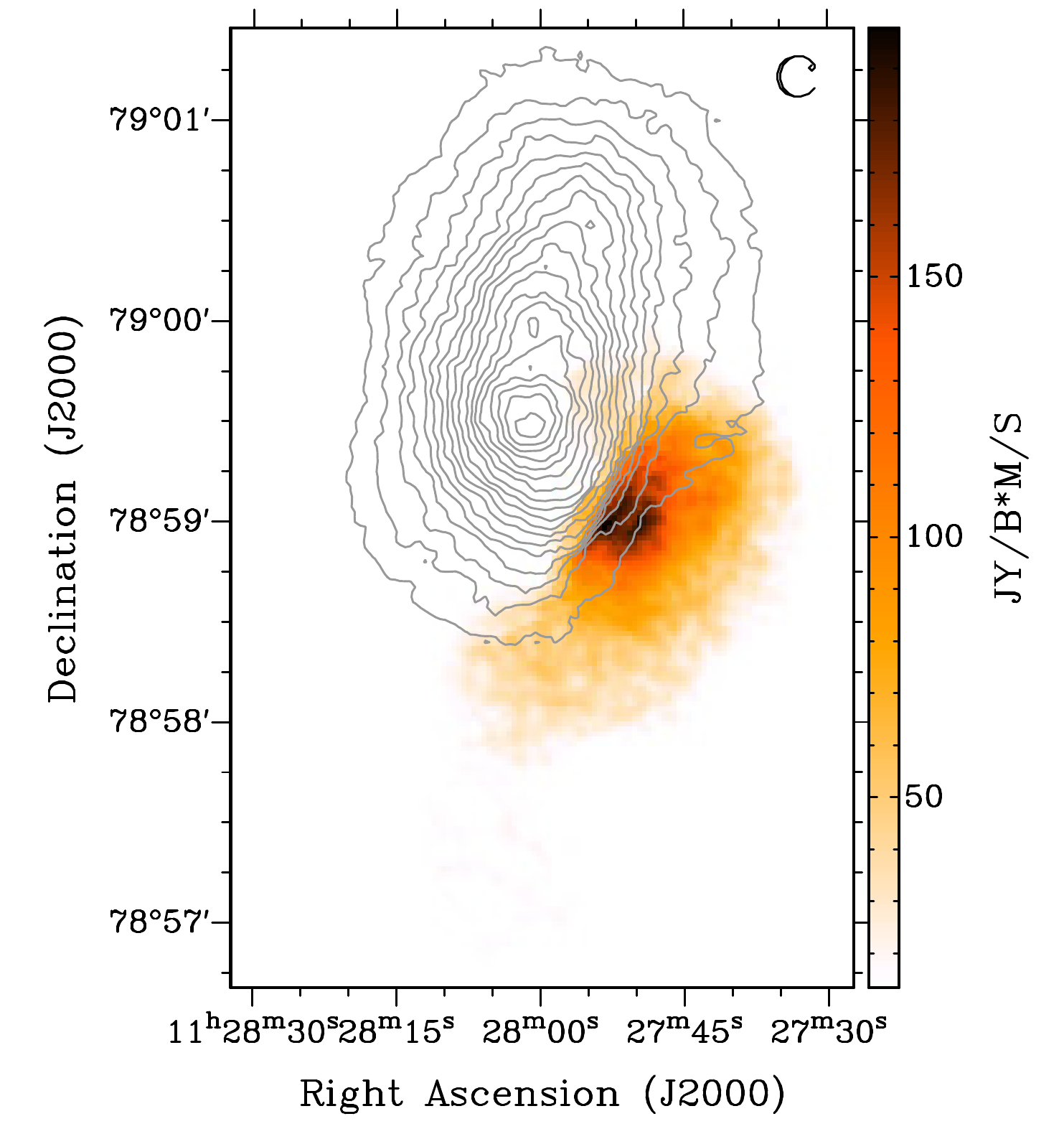}
\end{center}
\caption{VII Zw 403's VLA data: (a): The colorscale of the integrated \HI\ intensity map of the disk without the foreground gas cloud with contours of the V-band data.  (b): The colorscale of VII Zw 403's robust-weighted VLA integrated \HI\ intensity map overlaid with the  \HI\ intensity map contours of the foreground gas cloud. The white circle represents the location of a potential stalled hole in VII Zw 403, as suggested by \citet{simpson11}.  (c):  The colorscale of the integrated \HI\ intensity map of the foreground gas cloud with contours of the integrated \HI\ intensity map of the rest of the disk. 
\label{7zw403_foregroundhi}}
\end{figure}

The remaining velocity field of VII Zw 403's disk shows that the disk does not appear to have a kinematic major axis that is aligned with the morphological major axis of the \textit{gaseous or stellar} disk. The PA of the kinematic major axis is roughly 235\degr\ (estimated by eye), while the stellar morphological major axis is \s169.2\degr\ \citep{hunter06}.  The stellar morphological axis matches the \HI\ morphological disk major axis well without the gas cloud, as can be seen in Figure~\ref{7zw403_foregroundhi}a.  The misalignment of the stellar morphological and \HI\ kinematic major axes indicates that the gaseous disk of VII Zw 403 is disturbed, possibly from past gas consumption \citep[as suggested by][]{simpson11}, a past interaction, or a past dwarf-dwarf merger.   It is also possible that VII Zw 403 is highly elongated or bar-like along the line of sight, resulting in an offset kinematic major axis.

Below we discuss two main possible explanations for the curved gas cloud in Figure~\ref{7zw403_chim}b.  

\subsection{Gas Expelled From A Hole}\label{7zw403_hole}
The northern edge of the gas cloud in Figure~\ref{7zw403_chim}b spatially lines up well with a potentially stalled hole.  \citet{simpson11} suggested that the \HI\ structure just to the north of the the densest \HI\ region, denoted by a white circle in Figure~\ref{7zw403vla_r}a, may be a stalled hole that has broken out of the disk.  The alignment of these two features can be seen in Figure~\ref{7zw403_foregroundhi}b, where the robust-weighted \HI\ intensity map colorscale has been plotted with the contours of the \HI\ intensity map of the gas cloud in Figure~\ref{7zw403_chim}b and the location of the potential \HI\ hole has been denoted by a white circle.  If a hole has broken through the disk, then that hole could be ejecting material into the line of sight of the western side of the disk.  The ejected material could then be distorting the velocity field and creating the higher velocity dispersions seen in the galaxy.  The velocity dispersions could also in this case be the result of disturbed gas on both the west and the east side of the disk since the hole could have been expanding into the other side of the disk.  The gas cloud in Figure~\ref{7zw403_chim}b does have a large mass at 18\% of the total mass for VII Zw 403, but, according to dwarf galaxy models in \citet{elbadry16}, dwarf galaxies can have most of their gas expelled, at least momentarily, beyond the stellar disk due to large amounts of stellar feedback (see their Figure 2).  Therefore, this cloud is not necessarily too large to be material ejected from stellar feedback.  However, it should be noted that the data cubes were searched for \HI\ holes as part of the LITTLE THINGS project and no holes passed the hole quality checks outlined in \citet{bagetakos11} in VII Zw 403 (Pokhrel \et, in prep.) including the stalled hole suggested by \citet{simpson11}.  However, in the interest of following up with every potential explanation for VII Zw 403's distorted velocity field, the possibility of a stalled hole ejecting material in the line of sight of the western side of the disk is explored below.  

The velocities of the gas cloud in Figure~\ref{7zw403_chim}b indicate that the most recently ejected material (closest in projection to the potential hole) is blueshifted with respect to the systemic velocity of VII Zw 403.  This indicates that the material would be in the foreground of the disk (being pushed towards us).  The hole would eject material nearly perpendicular to the disk into the foreground of the west side of the galaxy, orienting the disk so that the east side is closest to us and the west side is furthest away.   Such an orientation is perpendicular to that expected from the velocity contours.  

The rough estimates made by \citet{simpson11} using the equations in \citet{mccray87} show that this cavity would have to be made by 2800 stars with masses greater than 7 M$_{\sun}$.   \citet{simpson11} did not see evidence for this very large number of stars and suggested that the cavity may have instead been made over a long period of time or was made through consumption when the star formation rate was higher around 600-800 Myr ago. If the cavity is a hole that broke through the disk roughly 600-800 Myr ago and the foreground gas cloud is gas that has been expelled from the disk by a now stalled hole, then the hole would require an outflow velocity of only \s8 \kms\ to have moved the gas \s4.5 kpc.  This is a reasonable rough estimate for an outflow velocity.  However, when looking at the velocities of the gas cloud, the tip nearest to the potentially stalled hole has a velocity of about \n125 \kms\ and the disk has a velocity of about \n100 \kms\ at the same location.  This is a velocity difference of 25 \kms, which is three times larger than our calculated outflow velocity.  With an outflow velocity this high, the approximate time that the gas would require to move 4.5 kpc would be \s180 Myr, when the SFR of the disk was lower \citep{lynds98}.  The velocity field of the underlying disk does still appear disturbed in Figure~\ref{7zw403_chim}a, so it is likely that the gas cloud extends further to the east to a different velocity.  Yet, it would be expected to continue the velocity trend seen in the rest of the gas cloud since there is a clear gradient of higher negative velocities towards the northeast tip of the gas cloud.  Therefore, the gas cloud's velocity is unlikely to get closer to the velocity of the disk near the potentially stalled hole.

\begin{figure}[!ht]
\epsscale{1.11}
\begin{center}
\plottwo{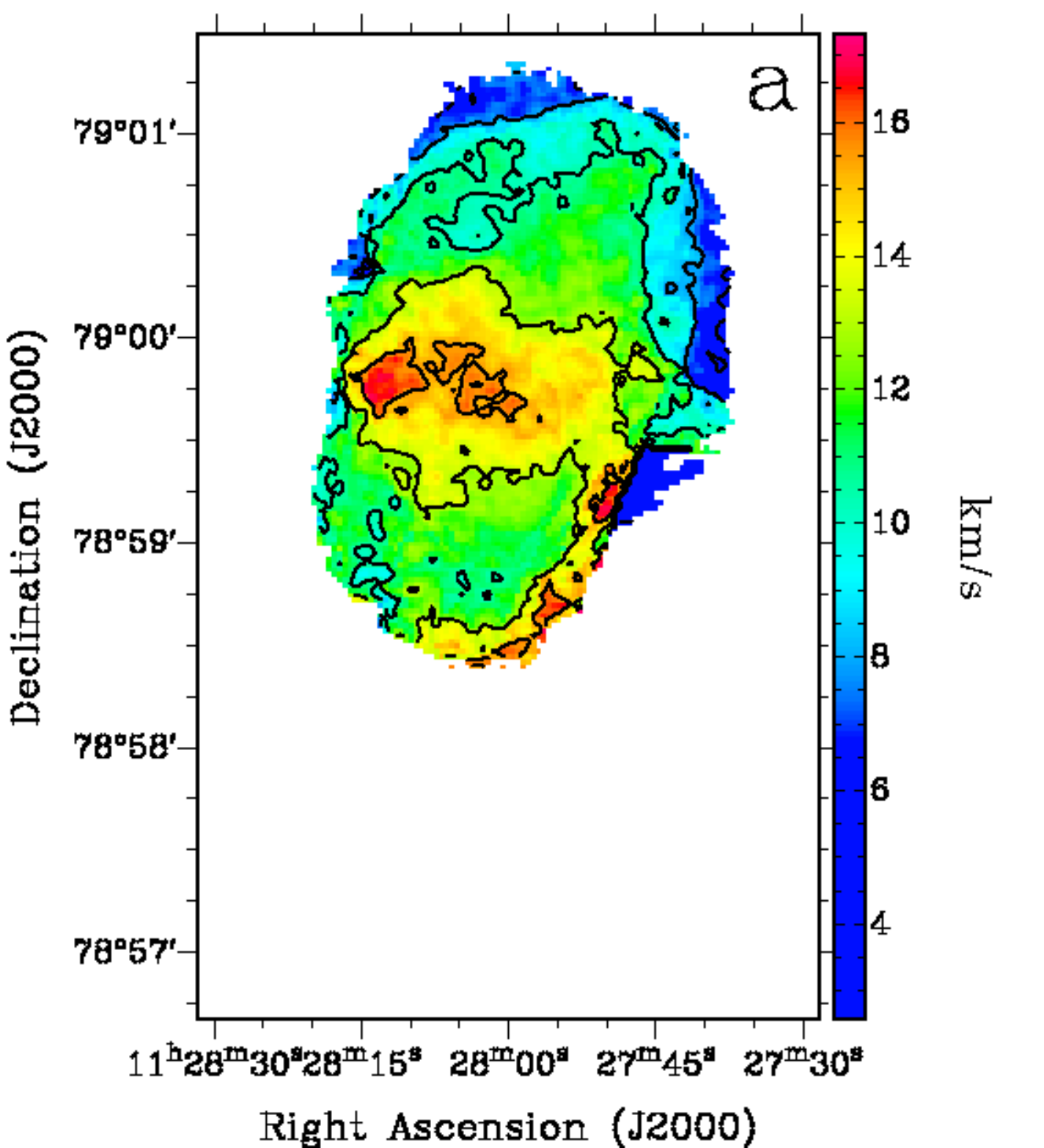}{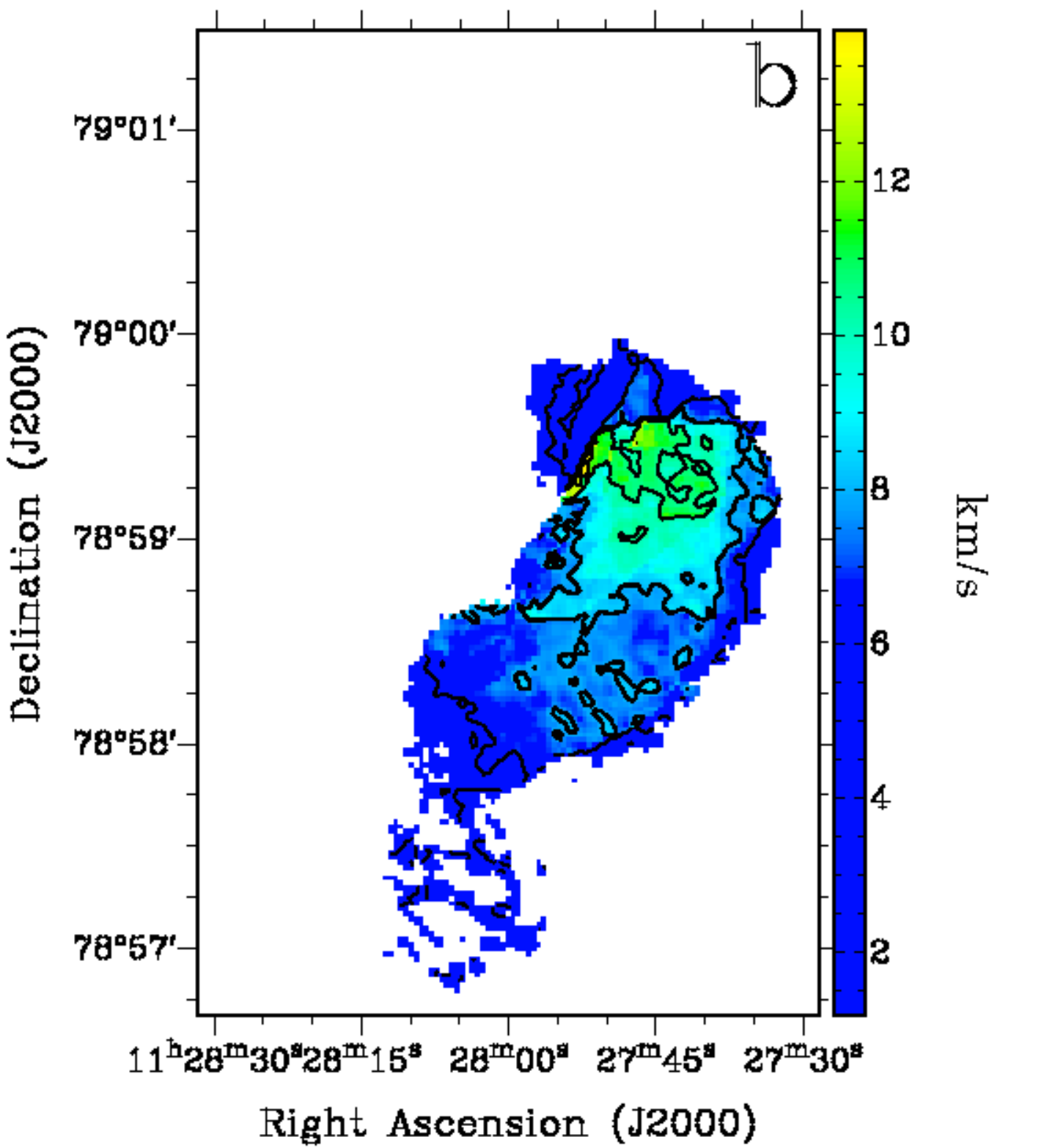}\\
\end{center}
\caption{VII Zw 403's VLA data: The left side is the main disk from Figure~\ref{7zw403_chim}a  and the right side is a map of the velocity dispersions of the gas cloud from Figure~\ref{7zw403_chim}b.  \label{7zw403_cloudx2}}
\end{figure}

The \HI\ velocity dispersions of VII Zw 403's disk without the gas cloud and the velocity dispersions of the gas cloud alone are shown in Figure~\ref{7zw403_cloudx2}.  In Figure~\ref{7zw403_cloudx2}a of the underlying disk velocity dispersions we note that the higher velocity dispersions seen in Figure~\ref{7zw403vla_na}d are still there with the exception of some decrease where the cloud is in the line of sight.  However, the southwest edge of the disk has higher velocity dispersions than in Figure~\ref{7zw403vla_na}d, this may be due to the removal of too much gas from the edge of the disk, resulting in the edge of the disk having velocities in Figure~\ref{7zw403_chim} that are not the true velocity values.   In Figure~\ref{7zw403_cloudx2}b the velocity dispersions of the gas cloud are generally \s10 \kms\ or less.   Such low velocity dispersions are inconsistent with turbulent gas that has been ejected from the disk from a hole with outflow velocities of \s25 \kms.   For example, dwarf irregular galaxy NGC 4861 has three outflow regions with \HI\ expansion velocities of 25 \kms\ and velocity dispersions of \s20 \kms\ \citep{eymeren09}. The low velocity dispersions in VII Zw 403's gas cloud are instead consistent with cold gas in the line of sight of VII Zw 403.   Considering all of the evidence against outflow from an \HI\ hole, this gas cloud has probably not been expelled by supernova explosions and stellar winds.

\subsection{An External Gas Filament}
Another possible explanation for the gas cloud in Figure~\ref{7zw403_chim}b is an external cloud of gas (potentially IGM) that is in the line of sight of VII Zw 403.  An external gas cloud would explain VII Zw 403's high \HI\ mass-to-light ratio relative to other BCDs \citep{simpson11}.  With a mass of \s$7.5\times10^{6}\ M_{\sun}$, the gas cloud is a reasonable size to be a starless gas cloud near the disk with IGM origins \citep{sancisi08}.  

The integrated \HI\ intensity map of the external gas cloud is shown in Figure~\ref{7zw403_foregroundhi}c in colorscale with the contours of the remaining disk's integrated \HI\ intensity map.  The cloud has an \HI\ column density peak near a declination of 78\degr\ 59\arcmin.  It is likely that some of the southwestern edge of the disk was included in the \HI\ map of the line-of-sight gas cloud during the process of using \textsc{blank}, which could explain the entirety of the cloud's peak in \HI.  This can be seen in Figure \ref{7zw403_chim}a as the southwestern edge of the disk looks as though it may be missing, but it is also possible that the \HI\ peak in the colorscale of Figure~\ref{7zw403_foregroundhi}c is, at least in part, the \HI\ peak of a gas cloud being consumed by VII Zw 403.  

It is possible that some of the external gas cloud has not been removed from the main \HI\ disk in Figures~\ref{7zw403_chim} and \ref{7zw403_foregroundhi} due to the nature of the manual identification of the cloud or due to overlapping velocities in the disk and gas cloud.  For example, some of the structure that \citet{simpson11} labeled as a potentially stalled hole denoted by the white circle in Figure~\ref{7zw403_foregroundhi}b can still be seen in the contours of the remaining disk in Figure~\ref{7zw403_foregroundhi}c as a small second \HI\ peak to the north of the dense \HI\ region.  The structure inside and on the north and west edges of the white circle in Figure~\ref{7zw403vla_r}a lines up well with the external gas cloud's northern edge and could actually be resolved structure of the external gas cloud.   

 The velocities of the gas cloud, in this case, cannot tell us which side of the disk (foreground/background) the cloud is on.  If the northeastern edge of the cloud is closest to the disk, then the gas cloud would be in the background of the disk falling towards it and potentially impacting it.   If the southeastern edge of the cloud is closest to the disk, the cloud is then falling away from us and towards the disk in the foreground of the disk. 

The velocity dispersions are nearly constant along the east-west axis in Figure~\ref{7zw403vla_na}.  This implies that if there is a foreground/background gas cloud extending across VII Zw 403's kinematic major axis, then that cloud must have a velocity gradient similar to that of VII Zw 403's disk (with offset velocities in the same line of sight).  Assuming that the velocity gradient of the gas cloud seen in Figure~\ref{7zw403_chim} continues into the line of sight of the disk, then it seems plausible that the velocity dispersion could stay nearly constant through the kinematic major axis of VII Zw 403.  Also, if we assume that the cloud is falling into VII Zw 403, then the gas cloud would acquire about the same velocity gradient as VII Zw 403's disk.  

It is possible that the velocity dispersions on the east side of the galaxy are from the external gas cloud running into the disk and pushing gas in the center of the disk from the west, which in turn, is pushing on gas on the east side of the disk.  If this is the case, then the northeastern end of the external gas cloud would be impacting the disk and the gas cloud would be falling towards us and towards the disk from behind the disk.

The location of the star forming region may also have been affected by the external gas cloud running into the disk over some time.  The densest star forming region, seen in Figure~\ref{7zw403_star}, is offset to the east of the center of the main V-band disk.  When the gas cloud impacted the disk, it would push gas from behind the disk towards us and to the east.  This may have caused star formation in the compressed gas that is located in the east of the disk and extending out towards us.  If this is the case, then the new stars would visually be in the line of sight of the far side of the stellar disk since they are in the foreground and presumably not at the edge of the disk.  The V-band image in Figure~\ref{7zw403_star} thus shows that the west side of the disk would be the near side of the disk.  

As the external gas cloud continues to be consumed by VII Zw 403, it could trigger another burst of star formation in the galaxy.  The cloud has a mass of \s$7.5\times10^{6}\ M_{\sun}$ and \citet{verbeke14} suggest that a mass of $\ge10^{7}\ M_{\sun}$ is required for a burst of star formation.  Considering the uncertainties associated with distance calculations, the gas cloud in the line of sight of VII Zw 403 is potentially large enough to trigger another burst of star formation in VII Zw 403 if it strikes the disk retrograde to its rotation as discussed in \citet{verbeke14}.

The gas cloud may also be the remnants of a past merger.  However, without tidal arms or double central cores, there is not much evidence for something as dramatic as a merger in VII Zw 403.

\section{Results: NGC 3738}
\subsection{Stellar Component}

\begin{figure}[!ht]
\centering
\epsscale{1.1}
\plottwo{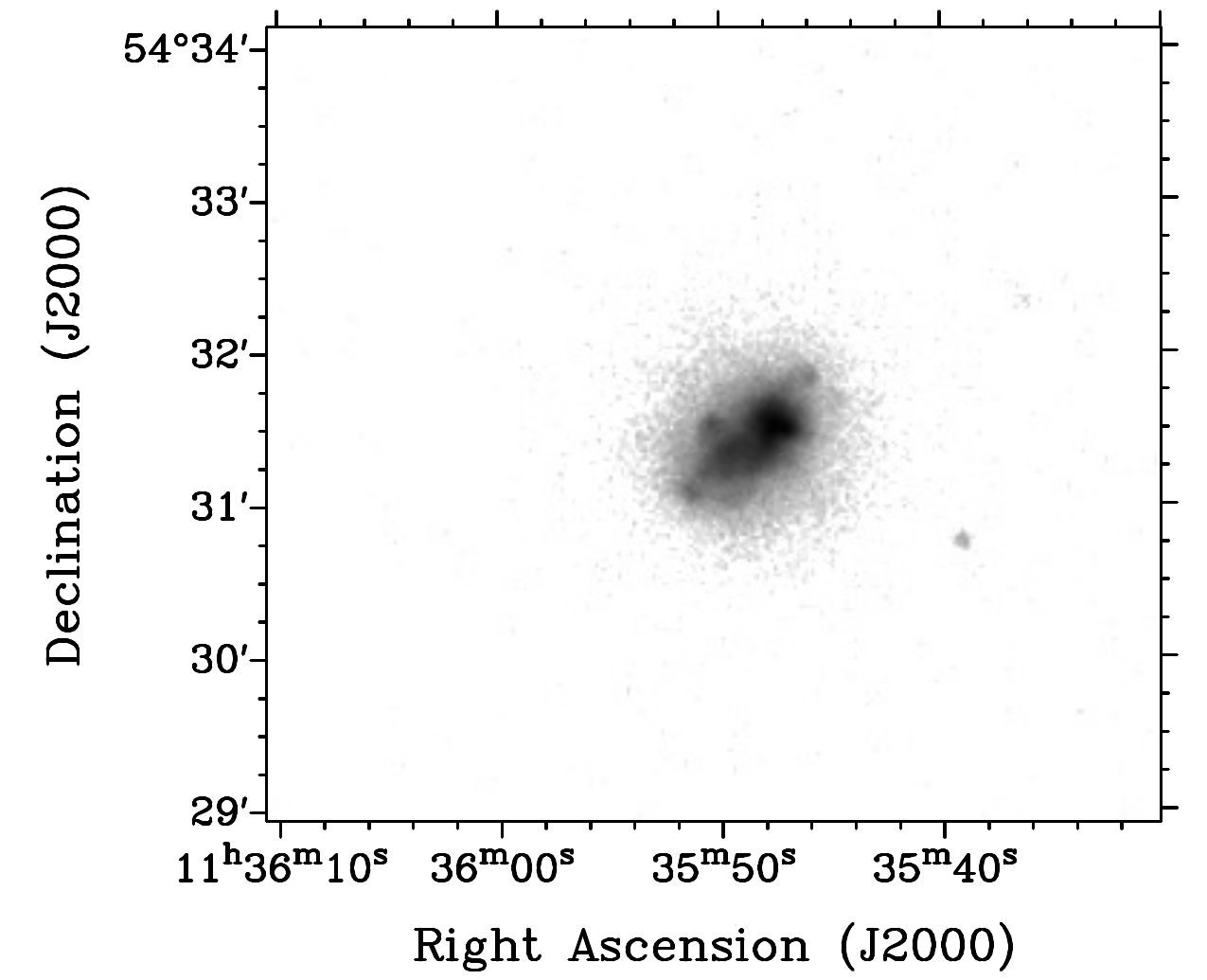}{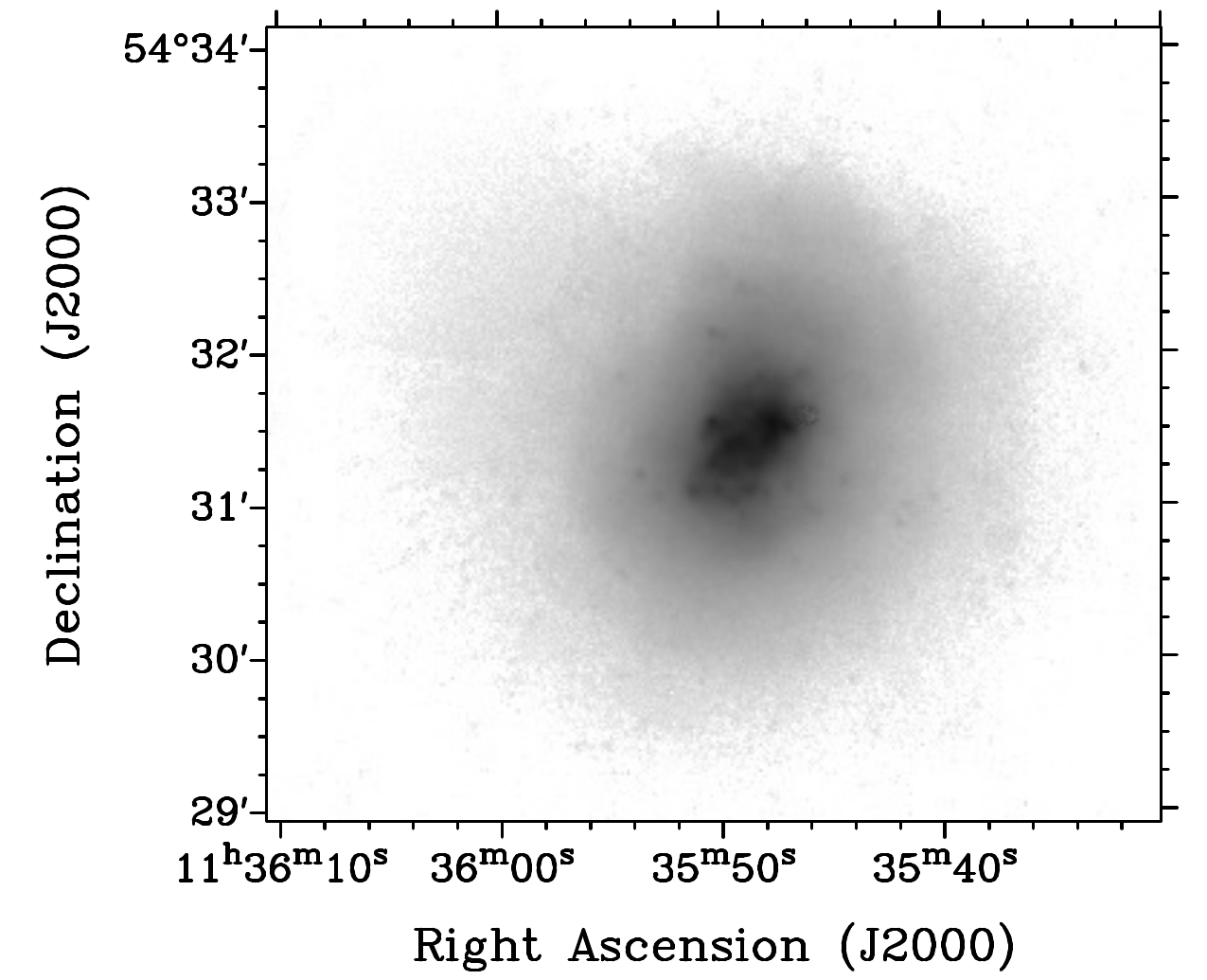}
\caption{NGC 3738's optical maps \ \  \textit{Left:} FUV.;\ \  \textit{Right:} V-band. \label{n3738_star}}
\end{figure}

The FUV and V-band data for NGC 3738 are shown in Figure~\ref{n3738_star}.   The V-band data were taken with the 1.1 m Hall Telescope at Lowell Observatory and the FUV were taken with GALEX \citep{hunter06, hunter10}.  The FUV and V-band surface brightness limits are \s30 and \s27 mag arcsec$^{-2}$, respectively \citep{herrmann13}.  The FUV and V-band data have very similar morphologies, with the  V-band disk being more extended than the FUV disk.  

\subsection{VLA \HI\ Morphology}
\begin{figure}[!ht]
\epsscale{0.505}
\plotone{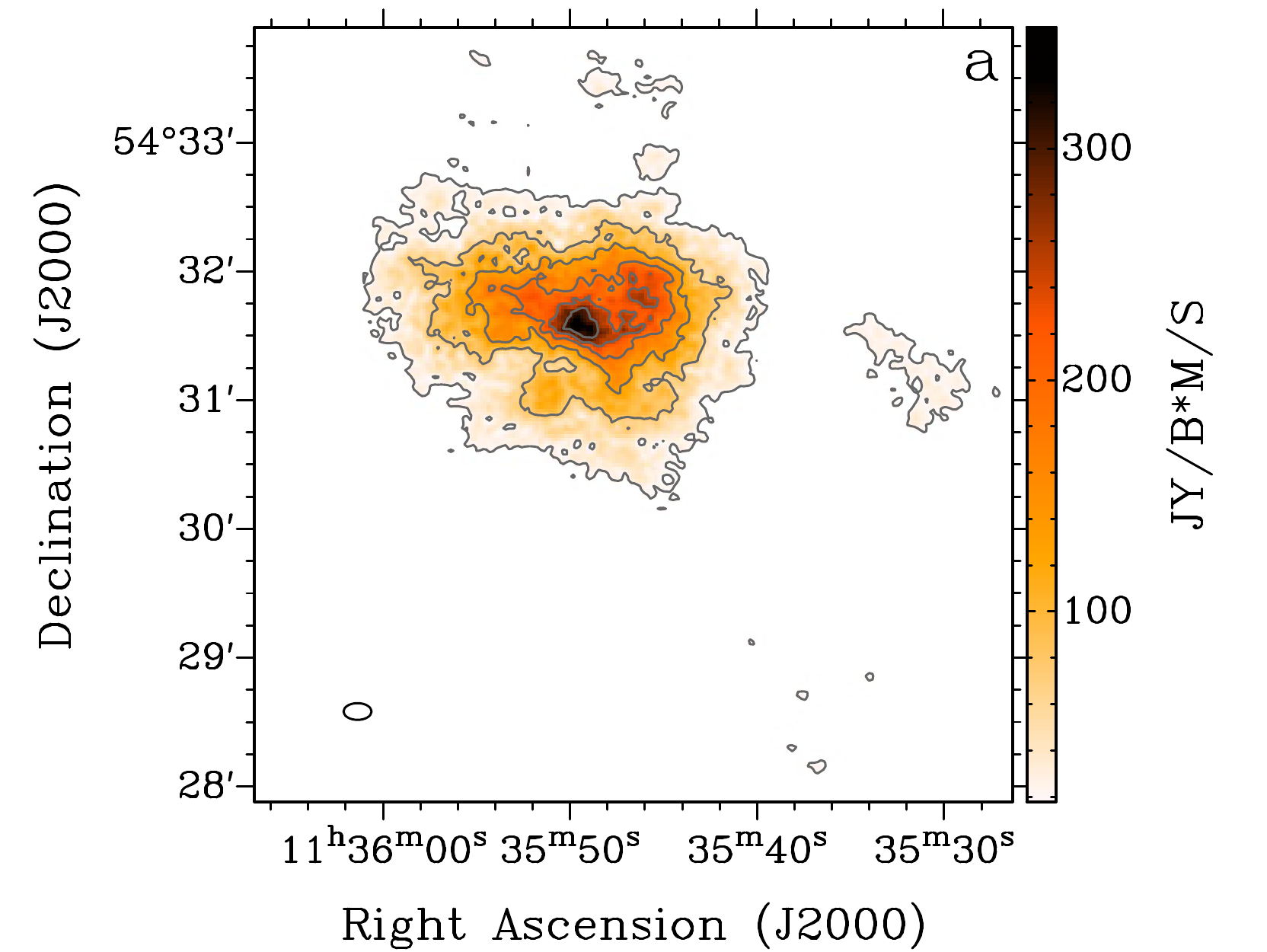}
\epsscale{0.525}
\plotone{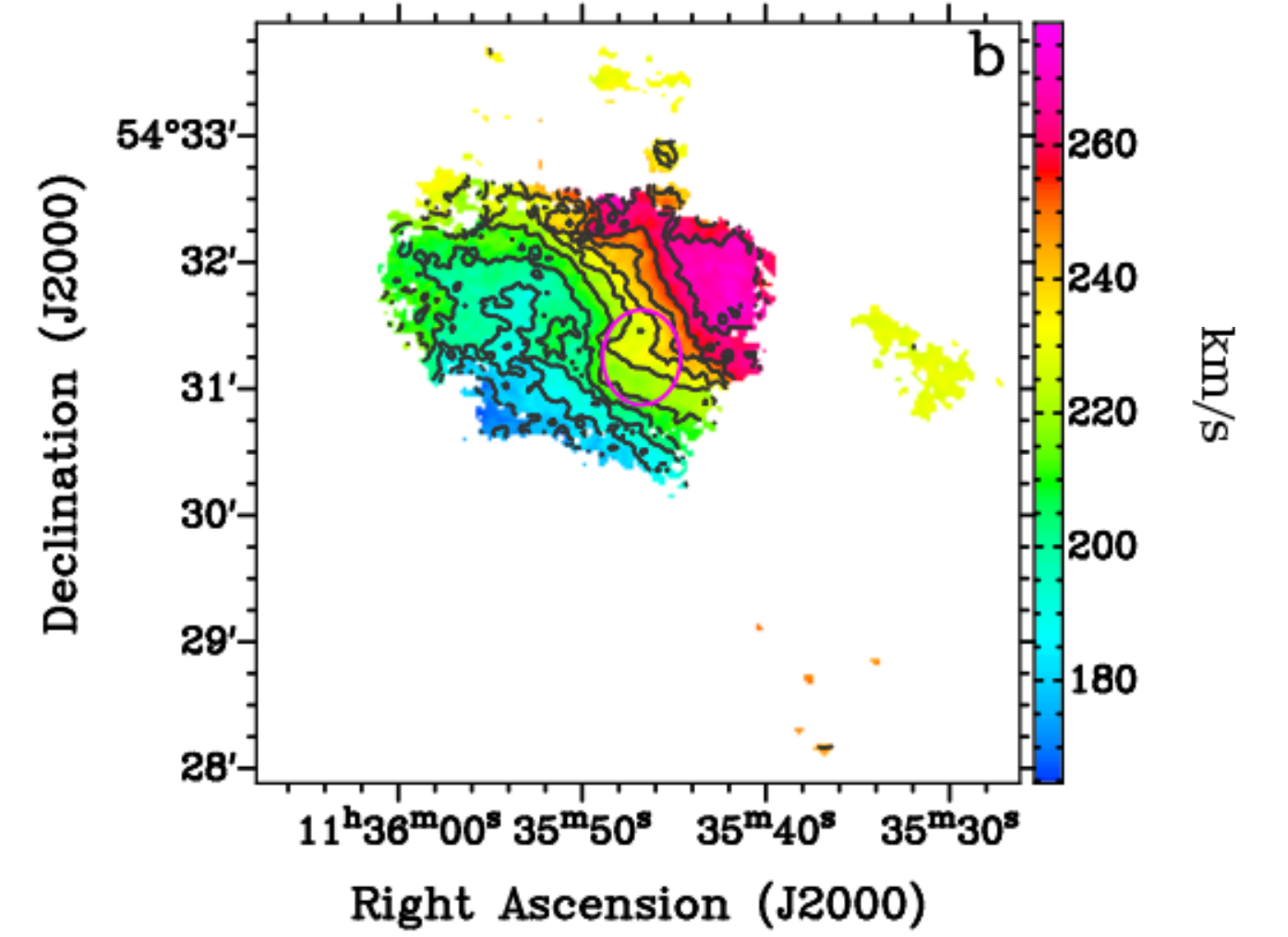}
\epsscale{0.495}
\plotone{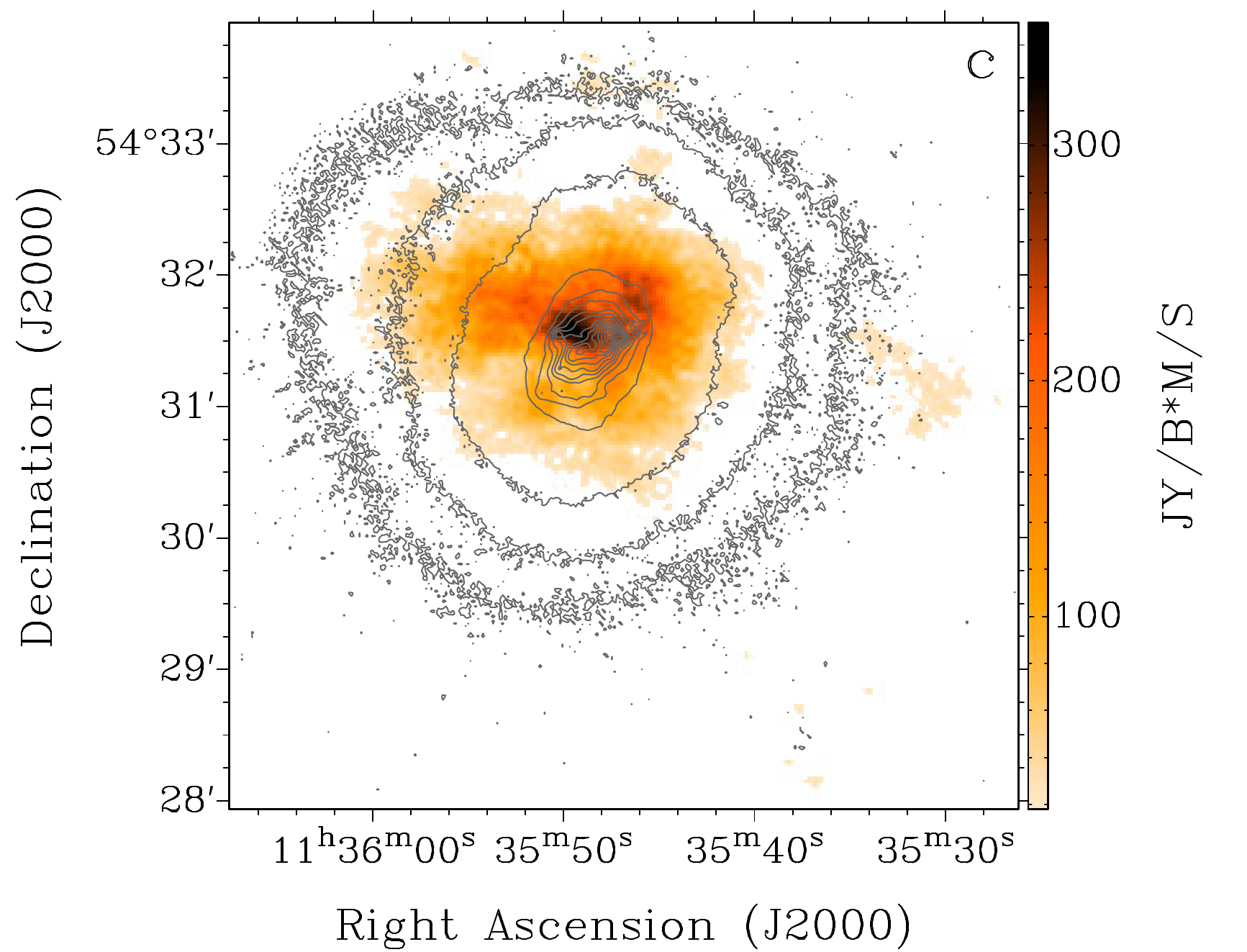}
\epsscale{0.525}
\plotone{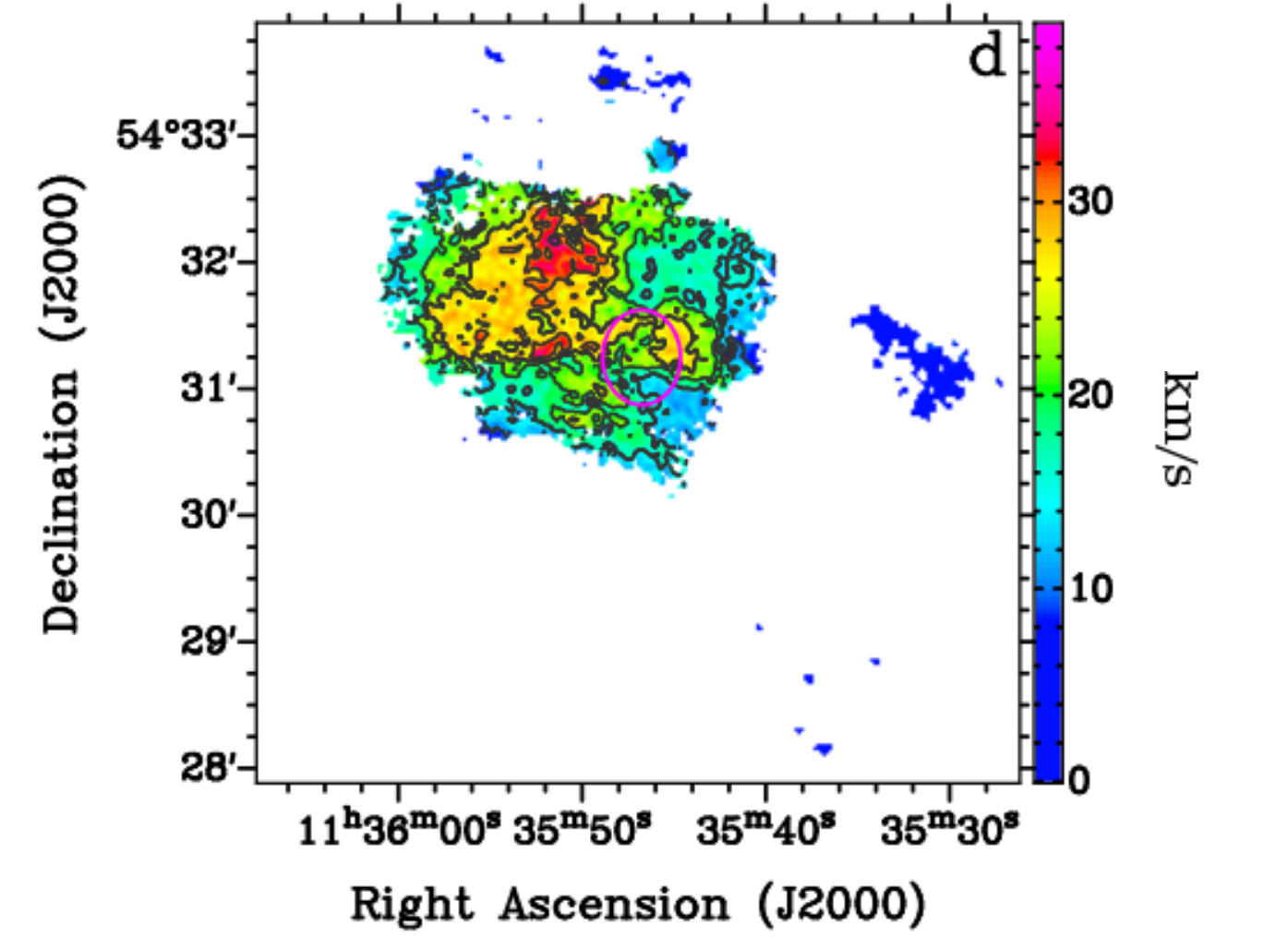}
\caption{NGC 3738's VLA natural-weighted moment maps. (a): Integrated \HI\ intensity map; contour levels are 1$\sigma\times$(2, 8, 14, 20, 26, 32) where 1$\sigma=1.01\times10^{20}\ \rm{atoms}\ \rm{cm}^{-2}$. The black ellipse in the bottom-left represents the synthesized beam. (b): Intensity-weighted velocity field; contour levels are 165 \kms\ to 275 \kms\ separated by 10 \kms. The magenta ellipse represents the approximate location of the kinematically distinct cloud discussed in Section~\ref{disc_n3738}. (c): The colorscale of Figure~\ref{n3738vla_na}a and the V-band contours.  (d): Velocity dispersion field; contour levels are 5 \kms\ to 35 \kms\ separated by 5 \kms. \label{n3738vla_na}}
\end{figure}

\begin{figure}[!ht]
\epsscale{0.515}
\plotone{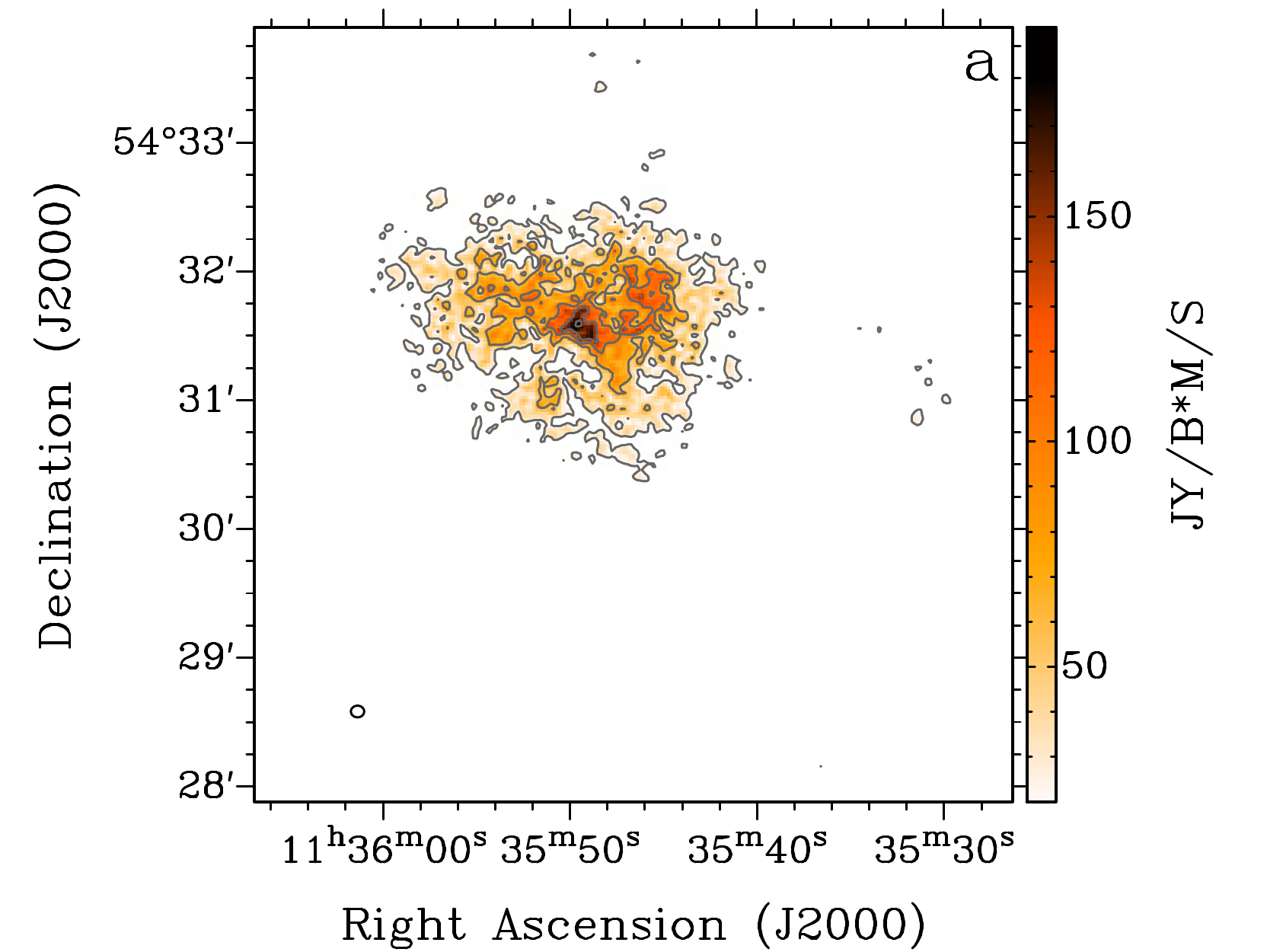}
\epsscale{0.48}
\plotone{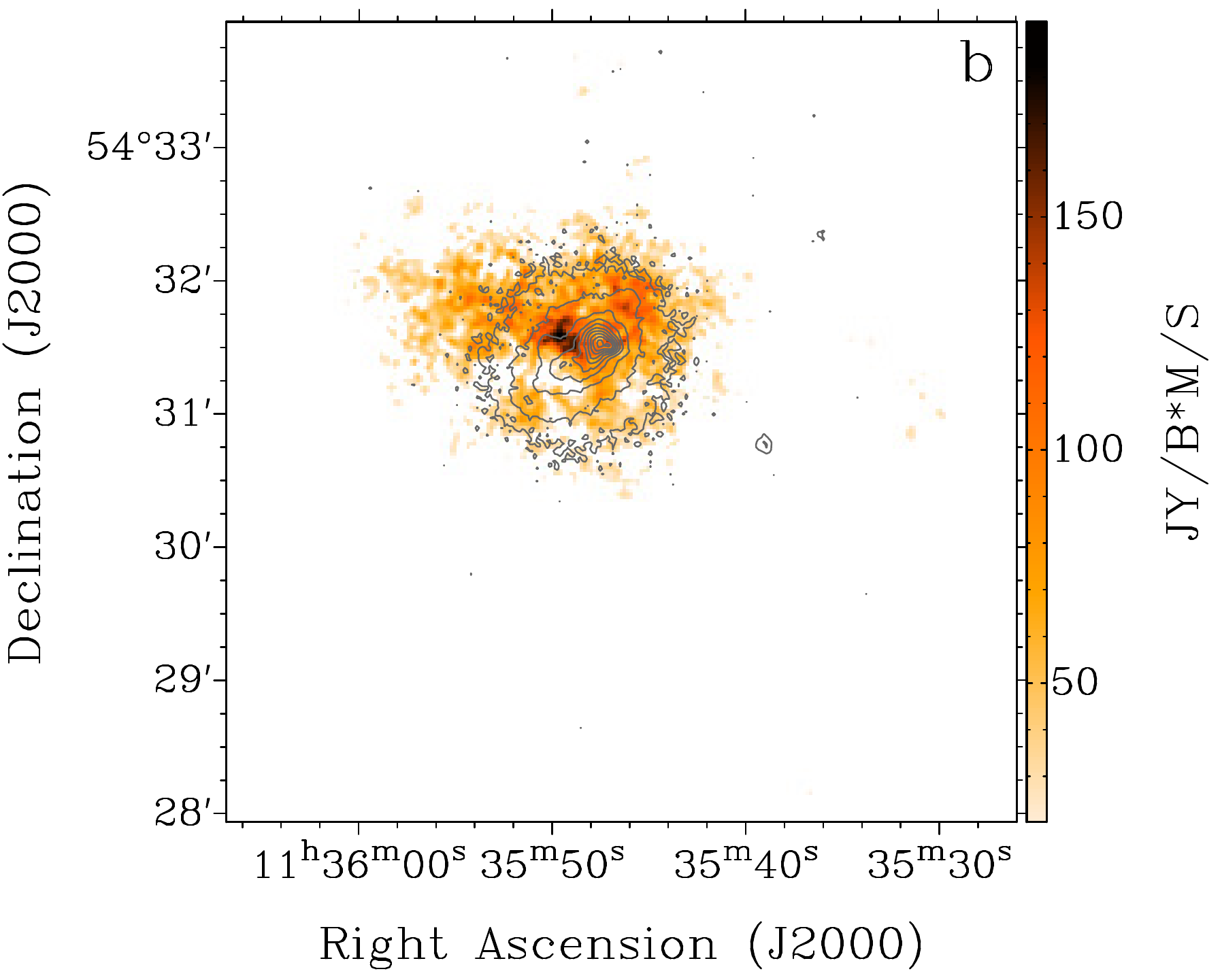}
\epsscale{0.48}
\plotone{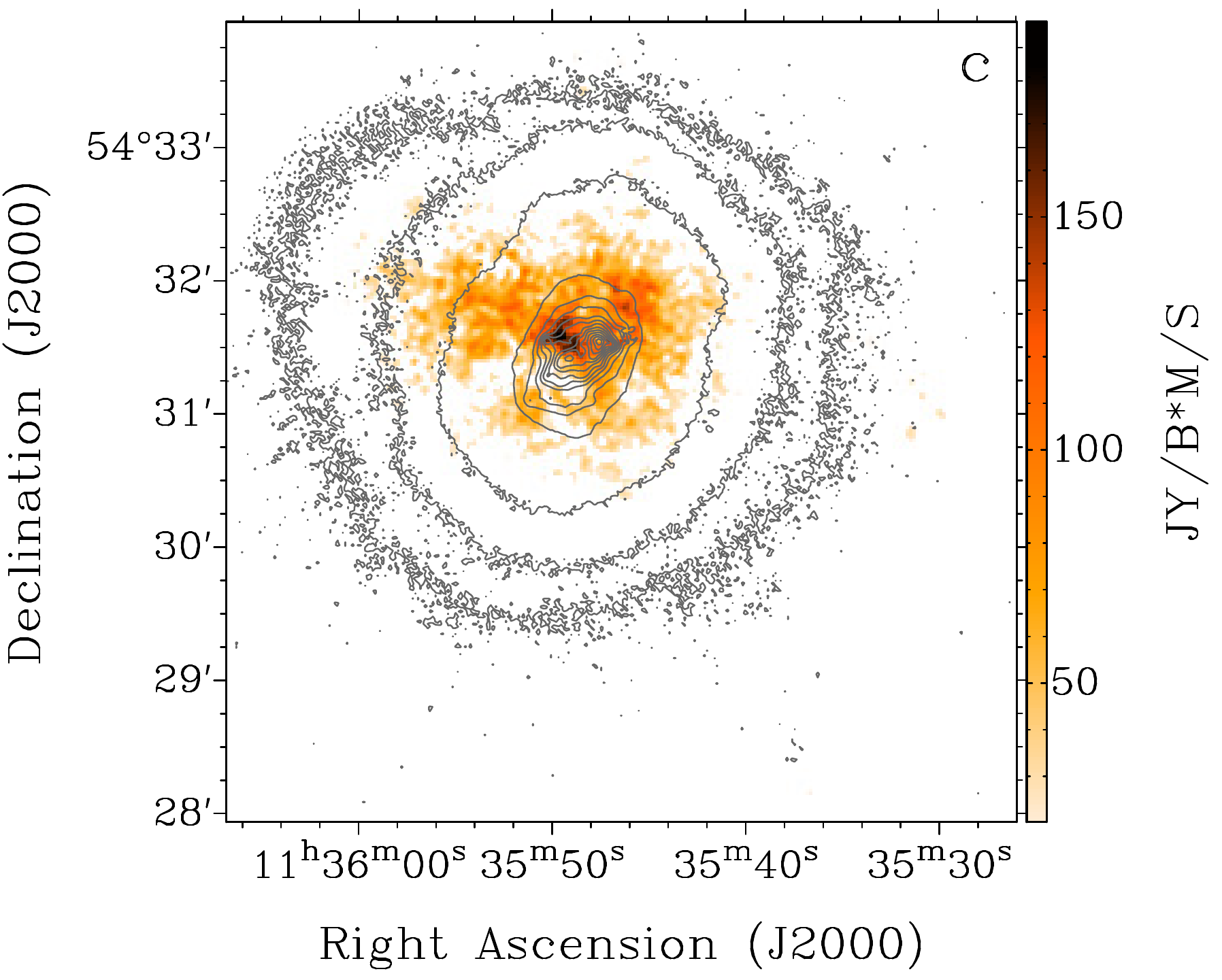}
\caption{NGC 3738's VLA robust-weighted moment maps. (a): Integrated \HI\ intensity map; contour levels are 1$\sigma\times$(2, 6, 10, 14, 18) where 1$\sigma=3.27\times10^{20}\ \rm{atoms}\ \rm{cm}^{-2}$. The black ellipse in the bottom-left represents the synthesized beam.  (b): Integrated \HI\ intensity map colorscale and FUV contours.  (c): Integrated \HI\ intensity map colorscale and V-band contours.  \label{n3738vla_r}}
\end{figure}

NGC 3738's natural-weighted integrated \HI\ intensity map as measured by the VLA is shown in Figure~\ref{n3738vla_na}a.  There are several regions of emission surrounding the disk.  The robust-weighted map is shown in Figure~\ref{n3738vla_r}a.  Figure~\ref{n3738vla_r}b and \ref{n3738vla_r}c show the colorscale of the robust-weighted, integrated \HI\ intensity map and contours of the FUV and V-band data, respectively.  The V-band data stretch beyond the gaseous disk even in the natural-weighted map Figure~\ref{n3738vla_na}c.  

\subsection{VLA \HI\ Velocity And Velocity Dispersion Field}
\begin{figure}[!ht]
\begin{center}
\epsscale{.65}
\plotone{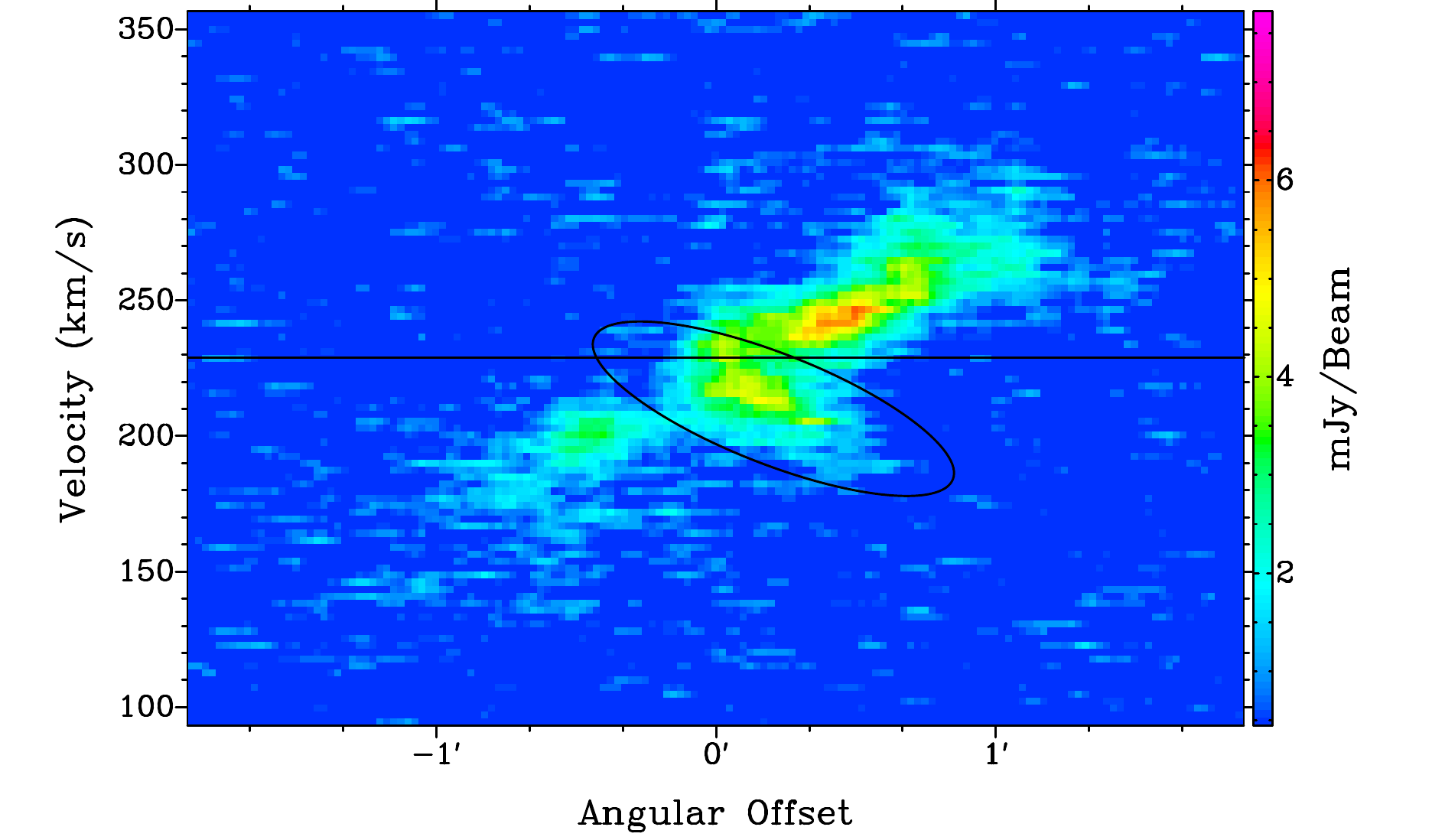}
\epsscale{.5}
\plotone{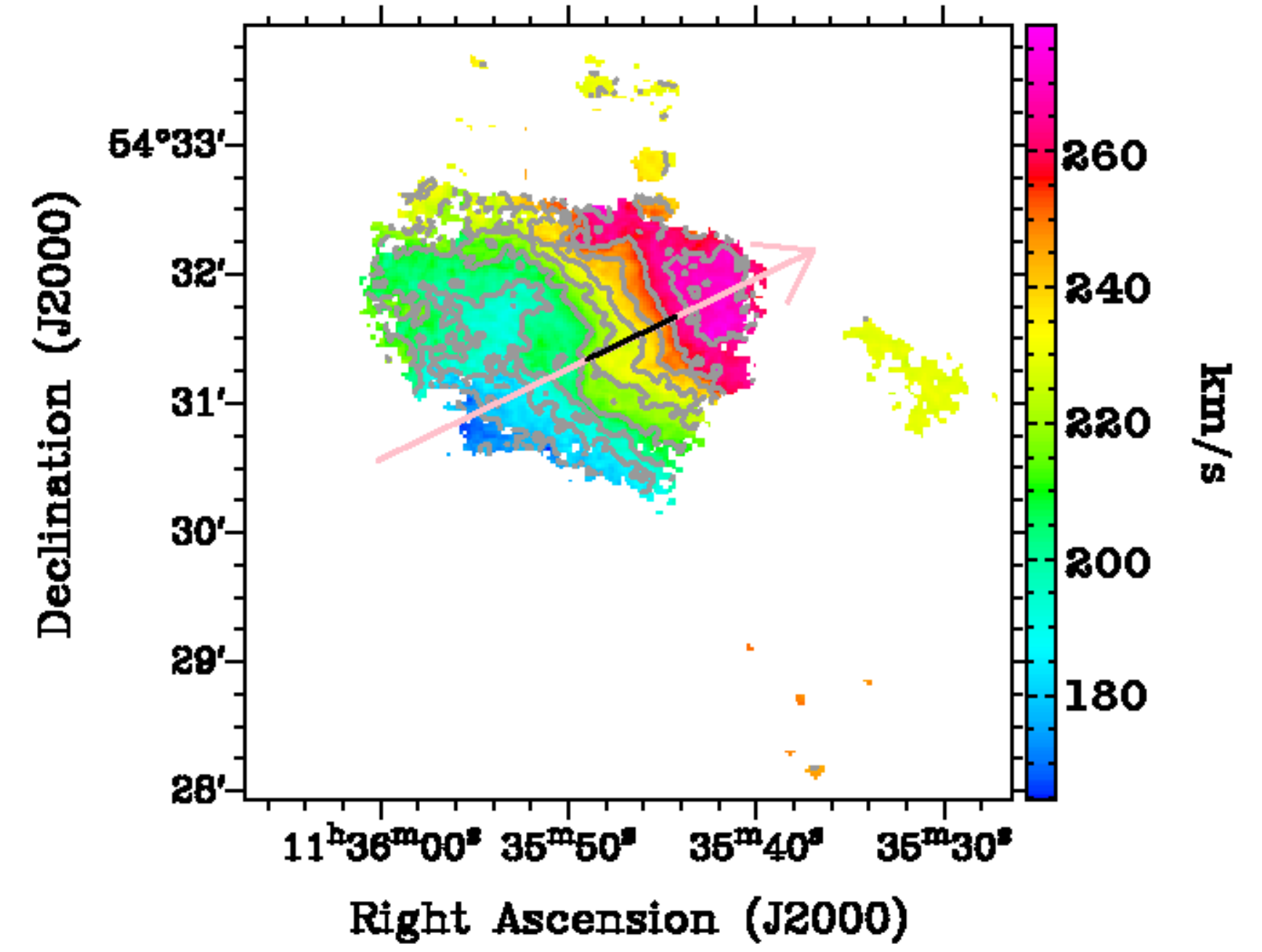}
\end{center}
\caption{NGC 3738: \textit{Left:}  P-V diagram of \HI\ major axis from the natural-weighted VLA data cubes, starting at a 1$\sigma$ level (0.82 mJy/beam). A kinematically distinct gas cloud is located with an ellipse and the systemic velocity of NGC 3738 is denoted by a horizontal line.  \textit{Right:} The natural-weighted \HI\ velocity map with a pink arrow indicating the location of the corresponding slice through the galaxy and pointing in the direction of positive offset. The location of the kinematically distinct gas is indicated on the pink arrow by a black line.\label{n3738_pv}}
\end{figure}

The intensity-weighted \HI\ velocity field is shown in Figure~\ref{n3738vla_na}b.  The disk is participating in near solid body rotation, with some small kinks in the isovelocity contours as can be seen in Figure~\ref{n3738_pv}.  The velocities of the separate regions of emission that surround the disk are all near the systemic velocity of the galaxy of 229 \kms\ indicating that they could be associated with NGC 3738.  The intensity-weighted FWHM of the \HI\ line profiles is shown in Figure~\ref{n3738vla_na}d.  Velocity dispersions reach up to \s35 \kms\ and are above \s20 \kms\ throughout most of the disk.

\subsection{GBT \HI\ Morphology And Velocity Field}
\begin{figure}[!ht]
\epsscale{1.15}
\plottwo{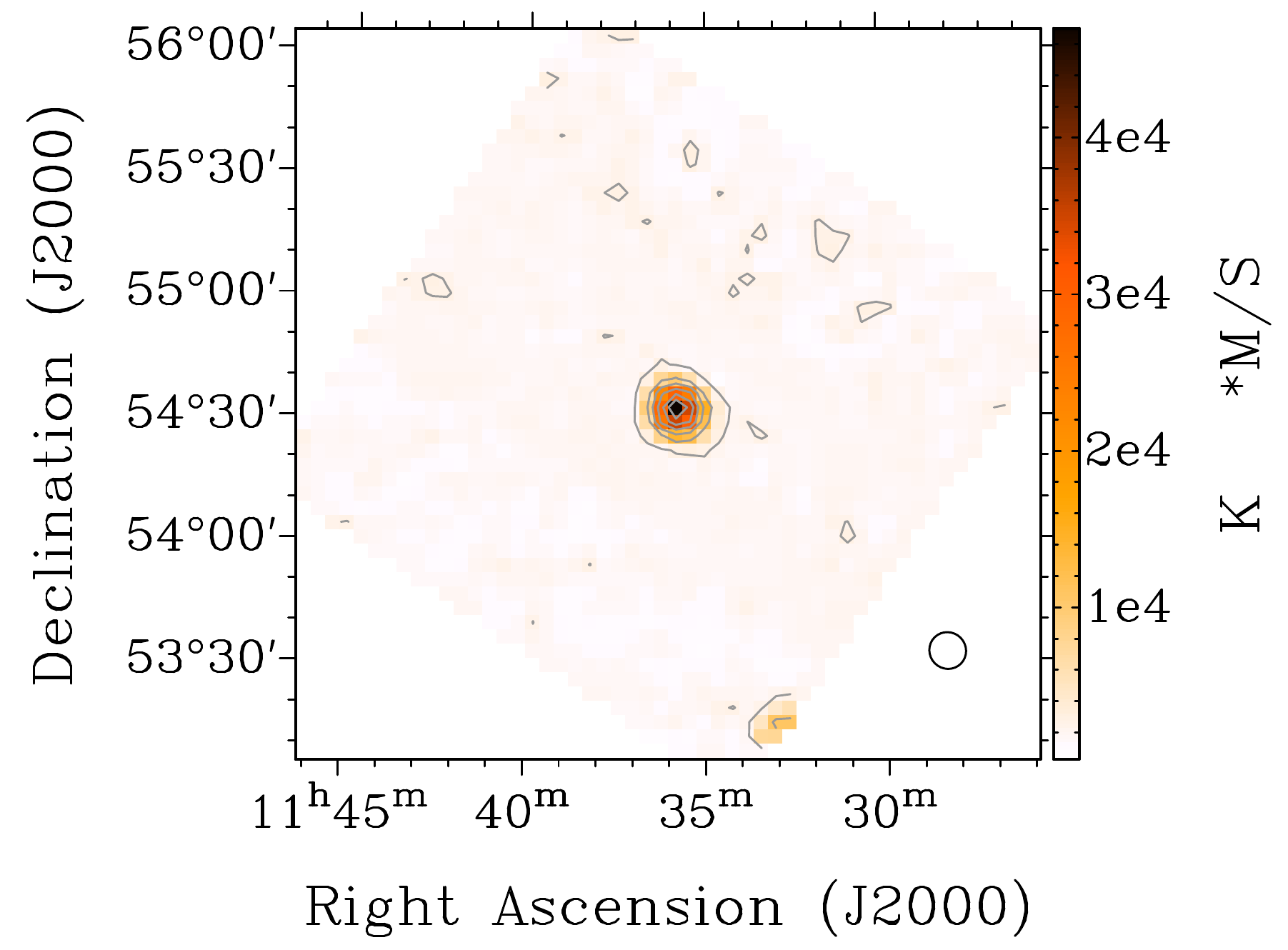}{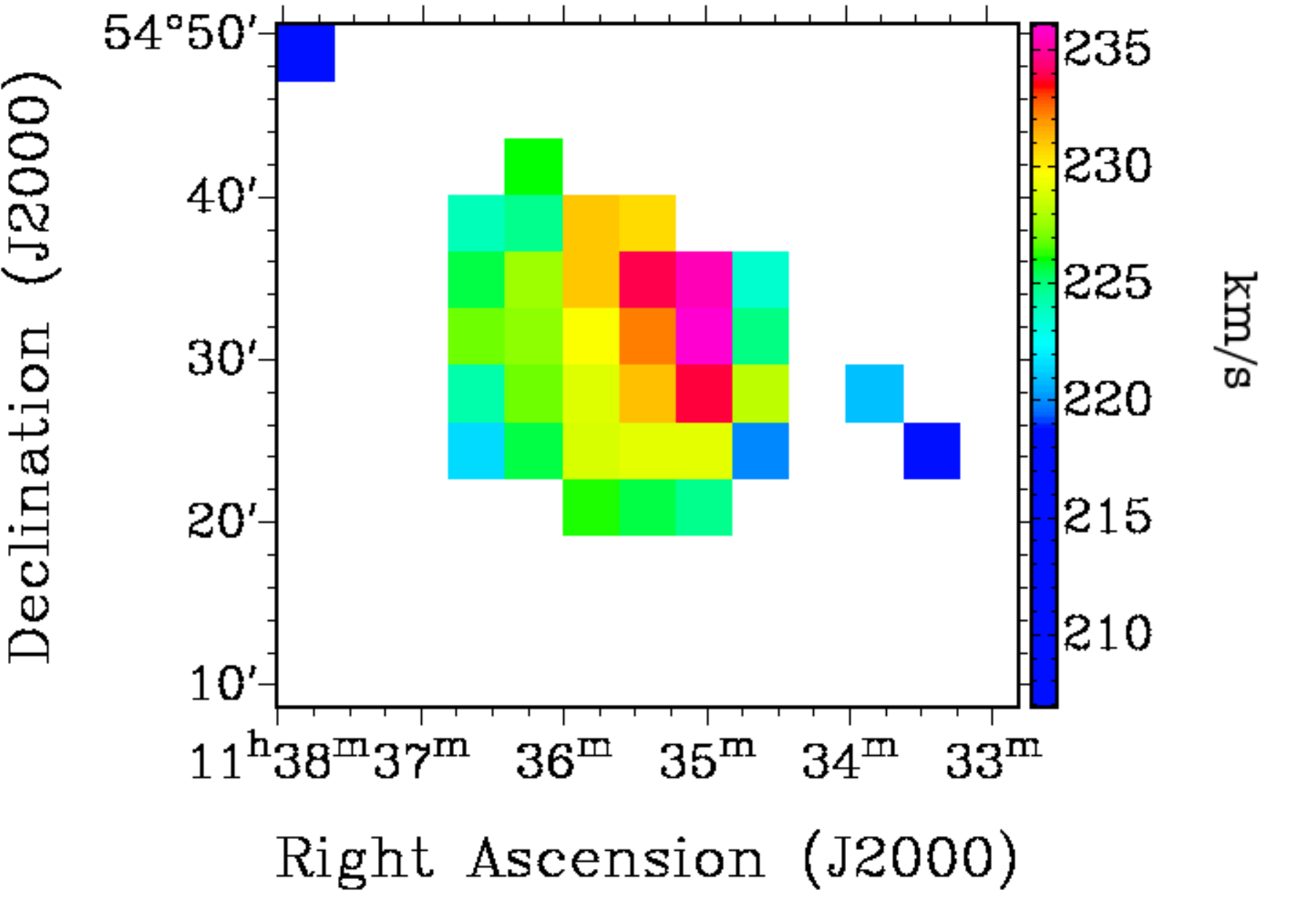}
\caption{NGC 3738's GBT moment maps. \textit{Left:} Integrated \HI\ intensity map; contour levels are 1$\sigma\times$(7, 27, 47,  67,  87, 107) where 1$\sigma=2.73\times10^{17}\ \rm{atoms}\ \rm{cm}^{-2}$.  The black ellipse represents the GBT beam. \textit{Right:} A close up of NGC 3738's GBT intensity-weighted velocity field. \label{n3738gbt}}
\end{figure}

NGC 3738's integrated \HI\ intensity map, as measured with the GBT, is shown in the left side of Figure~\ref{n3738gbt}.   NGC 3738 is at the center of the map surrounded by some noise and part of a galaxy to the south.  A close up of NGC 3738's GBT \HI\ velocity field is shown in the right side of Figure~\ref{n3738gbt}.  There is a gradient evident in NGC 3738's emission from bottom left to top right which matches the gradient direction seen in the VLA maps.  NGC 3738 does not appear to have any companions at the sensitivity of the GBT \HI\ map.

\subsection{\HI\ Mass}
The total \HI\ mass of NGC 3738 measured from the VLA emission, including the separate regions of emission, is $9.5\times10^{7}$ M$_{\sun}$, and the total \HI\ mass measured from the GBT emission for NGC 3738 is $1.7\times10^{8}$ M$_{\sun}$.  The VLA was able to recover 56\% of the GBT mass.  The masses of the individual regions of gas external to the disk in NGC 3738's VLA data add up to $1.3\times10^{7}$ M$_{\sun}$ or 14\% of the total VLA mass.

\section{Discussion: NGC 3738}\label{disc_n3738}
NGC 3738 is morphologically and kinematically disturbed.   In the VLA data, NGC 3738 has high velocity dispersions in its \HI\ disk, several gas clouds around its \HI\ disk, and it does not appear to have an extended, tenuous outer \HI\ disk like those seen in other BCDs \citep{thuan81}.  The \HI\ mass measurements from the GBT maps, however, indicate that there may be a significant amount of tenuous \HI\ surrounding the disk.   Additionally, the stellar morphological major axis, as measured by \citet{hunter06}, is \s179.6\degr, which is offset from the \HI\ kinematic major axis, measured to be \s115\degr\ (estimated by eye), however, the inner isophotes of the V-band image do appear to be more closely aligned with the kinematic major axis of NGC 3738 (see Figures \ref{n3738_star}b and \ref{n3738vla_na}c).  

\begin{figure}[!ht]
\epsscale{0.8}
\centering
\plotone{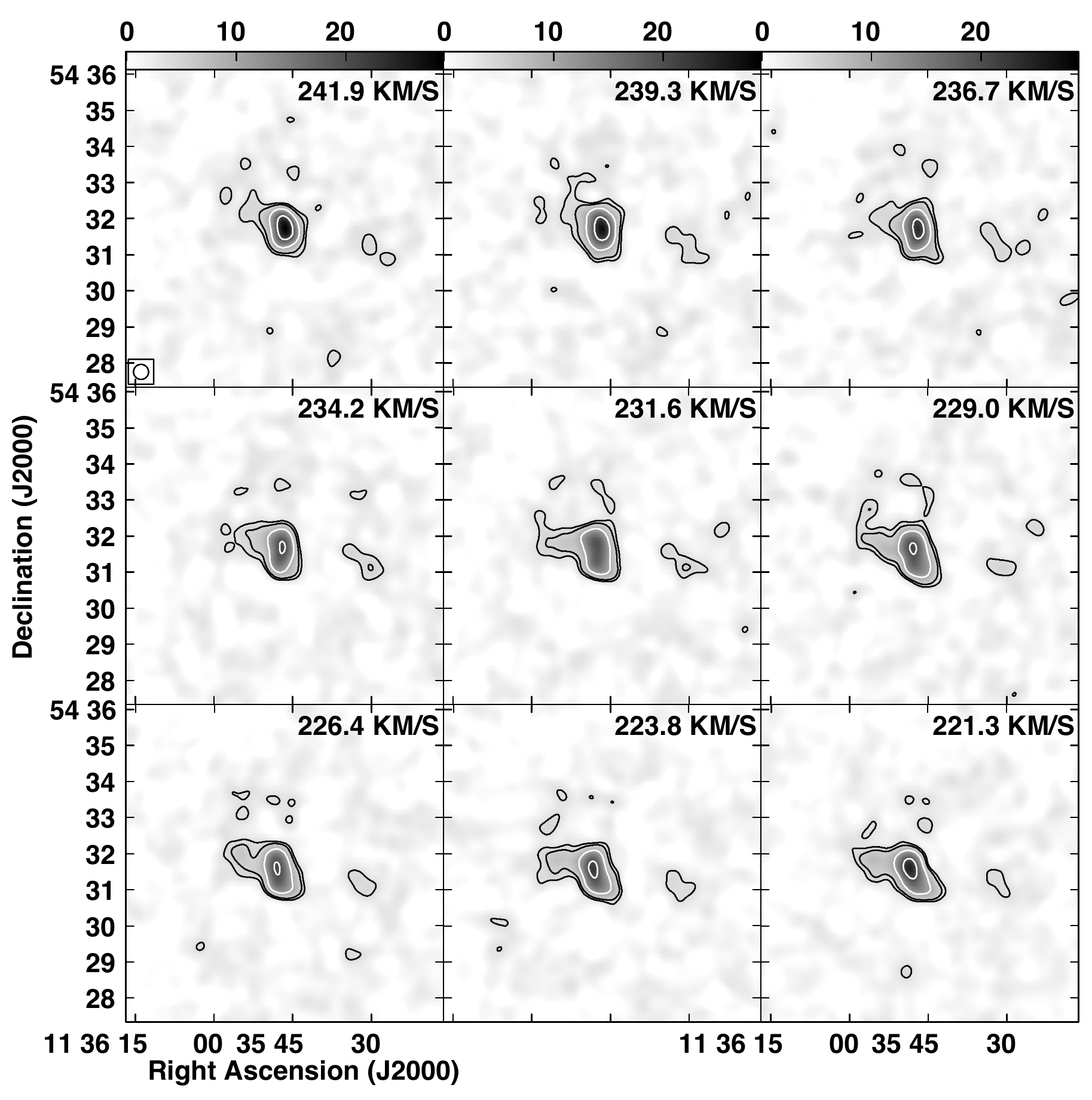}
\caption{NGC 3738's $25\arcsec\times25\arcsec$ convolved VLA natural-weighted channel maps; contour levels are 1$\sigma\times$(3.5, 5, 10, 20), where 1$\sigma=1\ \rm{mJy}\ \rm{beam}^{-1}$. The synthesized beam is represented by an ellipse in the top left panel.  The colorscales on top of the panels are given in units of mJy/beam. \label{n3738_cvl25}}
\end{figure}

The small regions of \HI\ emission around the disk of NGC 3738 in the natural-weighted VLA \HI\ maps could be noise in the natural-weighted maps.  As discussed in Section~\ref{7zwdisc}, the 25\arcsec$\times$25\arcsec\ convolved cube can be checked to see if these \HI\ features are real (appear in three consecutive channels or more).  In Figure~\ref{n3738_cvl25}, the cloud features to the north and west of the NGC 3738's main \HI\ body are visible in more than three consecutive channels each and therefore are likely real.  

With several regions of separate emission surrounding the disk, it is possible that the high dispersions are due to a gas cloud(s) in the line of sight of NGC 3738's disk.  P-V diagrams were made to search the disk for kinematically distinct gas clouds.  Figure~\ref{n3738_pv} is a P-V diagram of a slice that generally traces the kinematic major axis that shows the only kinematically distinct gas cloud that was found in the line of sight of the \HI\ disk. The overall trend seen in the gas through this slice is an increase in velocity moving towards positive offsets.  However, at an angular offset range of \n10\arcsec\ to 30\arcsec\ there is gas that has a decreasing velocity from \s220 \kms\ to 190 \kms\ (the emission circled in the P-V diagram and approximately located on the black segment of the pink slice in Figure~\ref{n3738_pv}).  This kinematically distinct gas cloud could be a foreground/background gas cloud or gas in the disk.  It has a large velocity range of \s30 \kms\ and is just  below the velocities of the surrounding separate regions of emission also seen in the VLA map.  The approximate location and extent of the cloud is also outlined with a magenta ellipse in Figures~\ref{n3738vla_na}b and d. This kinematically distinct cloud is likely causing the distortion of the isovelocity contours seen in the same region and some of the high velocity dispersions seen in the disk of the galaxy. The kinematically distinct gas in Figure~\ref{n3738_pv} cannot, however, account for the high velocity dispersions seen in the northeast side of the galaxy.  

There are no tidal tails or bridges that are apparent in the \HI\ maps of NGC 3738; however, it is possible that NGC 3738 is an advanced merger.  If NGC 3738 is an advanced merger, then the tidal tails in the \HI\ and stellar disk may have had enough time to dissipate or strong tidal tails may not have formed depending on the initial trajectories of the two merged galaxies \citep{toomre72}.  Most of the \HI\ disk at least would have had time to get back onto a regular rotation pattern, as seen in Figure~\ref{n3738vla_na}.  Also, mergers can result in efficient streaming of \HI\ towards the center of the galaxy \citep{bekki08}, causing a central starburst and, perhaps in NGC 3738's case, leaving behind a tenuous outer \HI\ pool that is detected by the GBT and not the VLA.  Some of the tenuous \HI\ detected by the GBT and not the VLA may also be remnants left behind by the progenitor galaxies as they merged.  The gas clouds surrounding the \HI\ disk in Figure~\ref{n3738vla_na} could also be material that was thrown outside of the main disk during the merger.  These gas clouds could then re-accrete back onto the galaxy later.  Although it is possible that NGC 3738 is an advanced merger, the lack of obvious tidal tails in either the stellar or gaseous disk means that we do not have evidence to state with certainty that NGC 3738 is an advanced merger.

Unique features of NGC 3738, such as the gas clouds to the west and north of the main \HI\ disk, an optical disk that extends beyond the main \HI\ disk in the VLA natural-weighted intensity maps, and high \HI\ velocity dispersions (see Figure~\ref{n3738vla_na}), could be an indication that ram pressure stripping is taking place. Like Mrk 178, NGC 3738 belongs to the Canes Venetici I group of galaxies, therefore, it may be being stripped by tenuous, ionized gas (see Section~\ref{m178_ram_press}).  However, unlike Mrk 178, the V-band emission of NGC 3738 stretches beyond the \HI\ disk on all sides of the disk.  Therefore the IGM would need to uniformly sweep out the gas from the edges of the disk, meaning that NGC 3738 would be moving through the IGM nearly face on.  Ram pressure stripping may also have left NGC 3738 with a tenuous outer \HI\ pool which is being detected by the GBT, but may be too tenuous to be detected in the VLA maps, explaining why the GBT maps are resulting in nearly twice the \HI\ mass as the VLA.

It is also possible that NGC 3738's gas has been pushed into its halo by the outflow winds created by a burst of star formation, resulting in the \HI\ emission to the west and the north of the main \HI\ disk. Three holes were found in NGC 3738's \HI\ disk that met the quality selection criteria outlined in \citet{bagetakos11} and have quality values (discussed in Section~\ref{m178_hihole}) of 6-7. All three holes are of the same type: they appear to have blown out both sides of the disk (Pokhrel \et, in prep.).  Thus, we are unable to get any accurate estimates of expansion velocities and other properties of the holes. Feedback can explain not only the \HI\ holes and emission outside of the main disk, but also the stellar disk.  Models presented in \citet{elbadry16} show the stellar disk can expand during periods of strong feedback along with the gaseous disk (see their Figure 2).  Therefore, the outer regions of NGC 3738's gaseous disk may be such low density from the expansion that they do not appear to cover NGC 3738's expanded stellar disk in the natural-weighted \HI\ maps.  The regions of \HI\ emission to the north and west of the main \HI\ disk could then reaccrete onto the disk at later times, resulting in another period of increased star formation activity.

\section{Comparison to other dwarf galaxies}
\begin{figure}[!ht]
\centering
\epsscale{1.11}
\plottwo{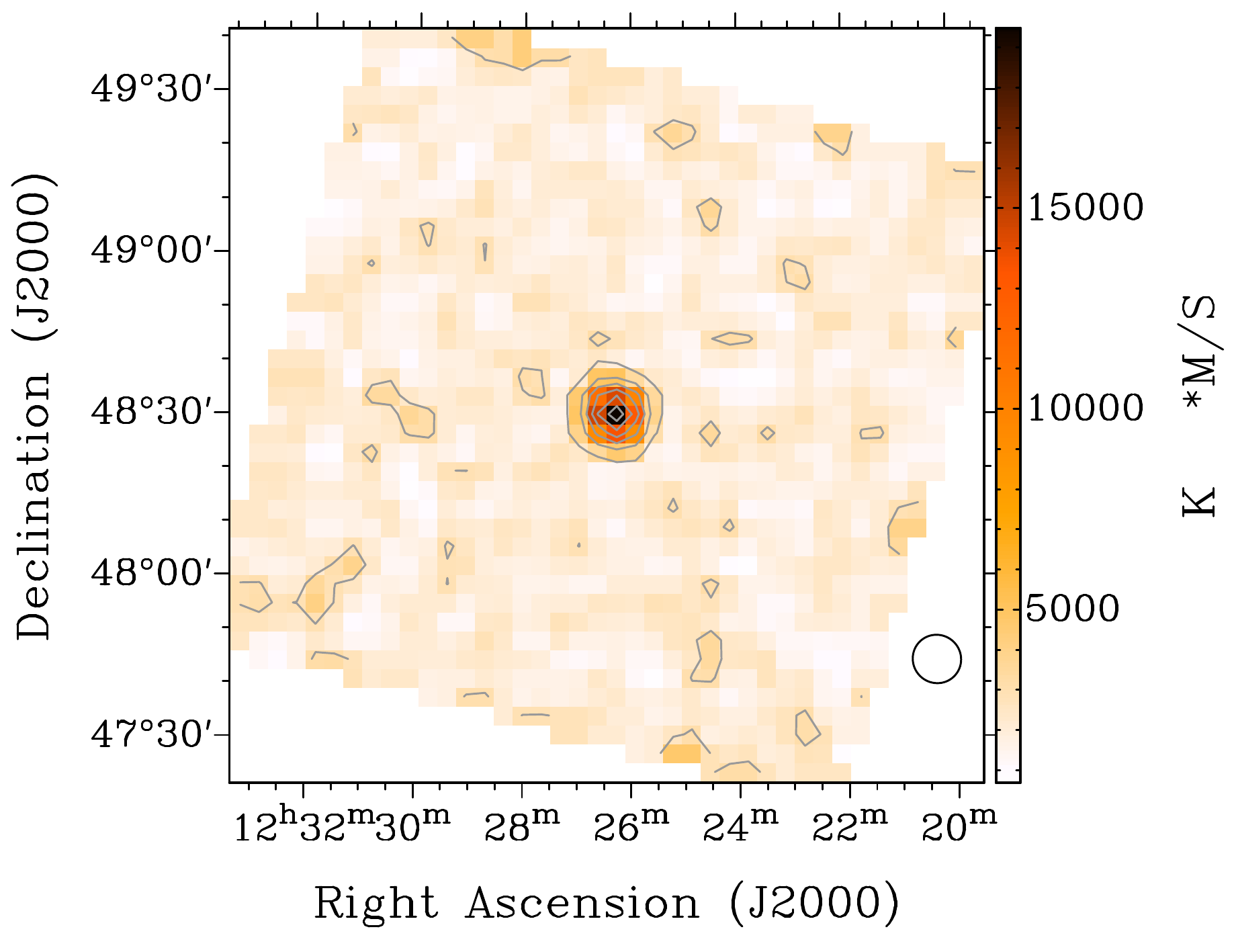}{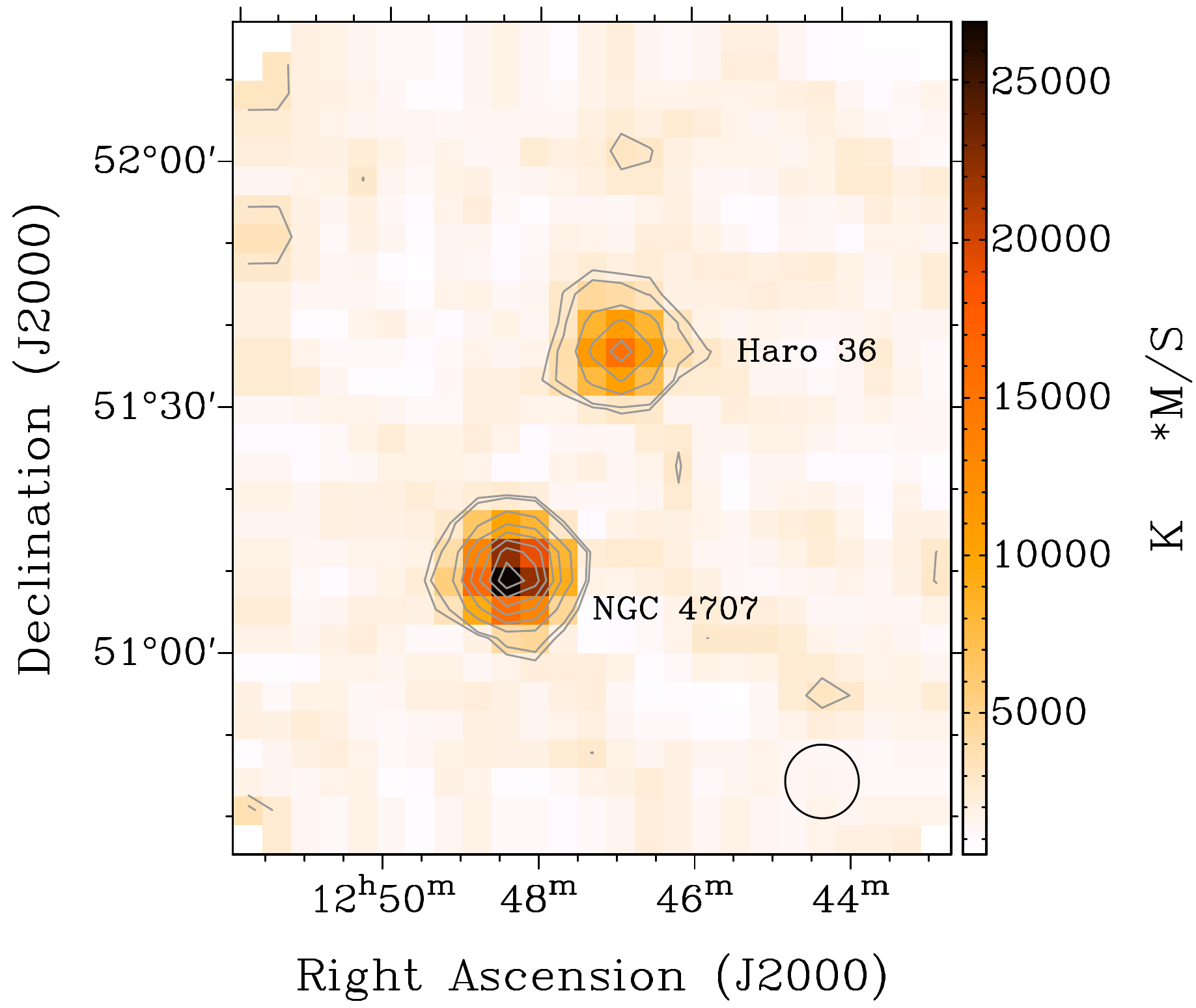}
\caption{Haro 29's and Haro 36's GBT moment maps. Left: Haro 29's integrated \HI\ intensity map; contour levels are 1$\sigma\times$(5, 10, 15, 20, 25, 30) where 1$\sigma= 4.48\times10^{17}\ \rm{atoms}\ \rm{cm}^{-2}$.  The black ellipse represents the synthesized beam.  Right: Haro 36's integrated \HI\ intensity map; contour levels are 1$\sigma\times$(4, 5, 10, 15, 20, 25, 30, 35) where 1$\sigma=5.34\times10^{17}\ \rm{atoms}\ \rm{cm}^{-2}$. The black ellipse represents the synthesized beam. \label{h29_36_gbt}}
\end{figure}

All three of these BCDs appear isolated with respect to other galaxies, but their VLA \HI\ data indicate that they have not likely been evolving in total isolation, either from other galaxies in the past or nearby gas clouds (perhaps with the exception of NGC 3738 which may have experienced strong stellar feedback).  This is the same conclusion made in previous papers of this series: \citet{ashley13} and \citet{ashley14}.  Haro 29 and Haro 36 are BCDs that appear to be advanced mergers or have interacted with a nearby companion \citep{ashley13}.  Like Mrk 178, VII Zw 403, and NGC 3738, there are no companions in the GBT \HI\ maps (shown in Figure~\ref{h29_36_gbt}) of Haro 29 and Haro 36 that are clearly interacting with the BCDs at the sensitivity of the map (these data were taken after \citealp{ashley13} was published).  Haro 36's GBT map appears very suggestive of a potential interaction between NGC 4707 and Haro 36, however, at the sensitivity of the map, this interaction cannot be confirmed because there is no continuous \HI\ bridge connecting them together.  Also, Haro 36 and NGC 4707 are 530 kpc apart with a relative velocity of 34 \kms\ \citep{hunter04}.  Therefore, these galaxies are not likely to have interacted since they would require \s15 Gyr (longer than the age of the universe) to travel that far away from each other (assuming that their line of sight velocity difference is comparable to their transverse velocity).  This could indicate that Haro 29 and Haro 36 are advanced mergers, that their companion is gas poor, or they have formed in some other manner. IC 10 is another BCD that \citet{ashley14} conclude is an advanced merger or accreting IGM, as indicated by an extension of \HI\ to the north \citep{nidever13}.  \citet{vanzee98} also found evidence of tidal features and no potential companion in the BCD II Zw 40.  Several surveys have shown that external gas clouds exist around several other BCDs \citep{taylor94, hoffman03, thuan04, ramya09},  however, it is not clear if these clouds are being expelled from the galaxies or if they are being accreted as is thought to be occurring in the dwarf irregulars NGC 1569 \citep{stil02b, johnson12} and NGC 5253 \citep{lopez12}.  

Other dwarf galaxies in LITTLE THINGS can also provide insight into characteristics of BCDs.  A general feature that is prominent in all of the BCDs (perhaps with the exception of NGC 3738 unless the external gas clouds are counted) is significant morphological asymmetries in the gaseous disk \citep{ashley13, ashley14}.  This is not an uncommon feature in the LITTLE THINGS sample overall, several galaxies also have morphological asymmetries: DDO 69, DDO 155, DDO 167, DDO 210, DDO 216, NGC 4163, LGS 3, and DDO 46 \citep{hunter12}.  Five of these galaxies have low star formation rates: DDO 46, DDO 69, DDO 210, DDO 216, and LGS 3, ranging from $\rm{log\ SFR}_{\rm{D}}$ of \n4.10 to \n2.71 \citep[from H$\alpha$ measurements when possible:][]{hunter12}.  The remaining three dwarf irregulars with significant morphological asymmetries, DDO 155, DDO 167, and NGC 4163, do have higher star formation rates of \n2.28 to \n1.41 \citep{hunter12}.  DDO 155 and NGC 4163 have even been labeled as BCDs in the literature \citep{mcquinn09, e12}.  However, morphological asymmetries in a dwarf galaxy do not appear to imply high star formation rates, a defining feature of BCDs.  This is not unexpected as dwarf irregular galaxies are named for their irregular shape.  DDO 216, for example, has a cometary appearance in \HI\ \citep{mcconnachie07, hunter12}, but it has one of the lowest star formation rates, $\rm{log\ SFR}_{\rm{D}}$=\n4.10.  Therefore morphological asymmetries alone, including cometary appearances, do not imply increased star formation rates.  

Other indications of the disturbed gas in these BCDs are the second moments \HI\ maps.  Haro 36, IC 10, and NGC 3738 all have gas velocity dispersions throughout large regions of their disks reaching at least 15-20 \kms.  Two other dwarf galaxies in the LITTLE THINGS sample also have \HI\ velocity dispersions in excess of 15 \kms: NGC 1569 and NGC 2366  \citep{hunter12}.  Both of these dwarf irregular galaxies have heightened star formation rates.  NGC 1569 is believed to have a gas cloud impacting the disk \citep{stil02b, johnson12}. NGC 2366 is a dwarf irregular with an \HI\ kinematic major axis that is offset from its stellar and \HI\ morphological axes, a supergiant \ion{H}{2} region, and is occasionally referred to as a BCD \citep{hunter01, e12}.    The increased velocity dispersions in NGC 2366's disk do not appear to be associated with the large star forming regions in the disk and therefore their cause is unknown \citep{hunter01}.  Haro 36, IC 10, NGC 3738, and NGC 1569 all have higher velocity dispersions throughout their disk including regions of star formation.  All four of these dwarf galaxies are likely having interactions with their environment through consumption of external gas, mergers, and ram pressure stripping \citep[as shown in this work and in:][]{johnson12, ashley13, ashley14}, therefore, it is possible that their interaction with the environment is the cause of their high velocity dispersions and that their star formation is also contributing to these high dispersions.  

Two distinct kinematic major axes are seen in the disks of Haro 36, Mrk 178, and VII Zw 403 (arguably also in Haro 29) and not in other dwarf galaxies in the LITTLE THINGS sample \citep{ashley13, hunter12}.  This feature is usually indicative of a recent disturbance to the gas.  For Haro 36, Mrk 178, and VII Zw 403, two kinematic major axes indicated that they: were the result of a merger, were experiencing ram pressure stripping and/or had an impacting external gas cloud \citep[as shown in this work and in][]{ashley13}.  Strong interactions with the environment may therefore be playing an important role in fueling the bursts of star formation in BCDs.

\section{Conclusions}

Mrk 178, VII Zw 403, and NGC 3738 all have disturbed \HI\ kinematics and morphology.   Both VII Zw 403 and Mrk 178 have a significant offset of their \HI\ kinematic axis from their stellar morphological major axis, while both Mrk 178 and NGC 3738 have stellar disks that extend beyond their natural-weighted \HI\ maps.  This indicates that all three galaxies have been significantly perturbed in the past.  

Mrk 178 has strange stellar and \HI\ morphologies.  At the sensitivity of the VLA maps, Mrk 178 does not have \HI\ extending as far as the stellar component as indicated in the V-band.  It is possible that Mrk 178's VLA \HI\ morphology is dominated by a large \HI\ hole and an \HI\ extension to the northwest.  However, the hole shape in the VLA \HI\ data may also be a red herring.   A hole cannot easily explain why the northwest region of the \HI\ disk appears to be rotating on a kinematic major axis that is nearly perpendicular to the kinematic major axis for the rest of Mrk 178's \HI\ body. Another scenario that could explain the morphology and kinematics of Mrk 178's VLA maps is a gas cloud running into the galaxy.  This gas cloud would be running into the disk from the southeast and pushing gas in the disk to the west.  Since Mrk 178 appears to be a low mass galaxy, it is very possible for the galaxy to be easily disturbed.  Another possible explanation for Mrk 178 is that it is experiencing ram pressure stripping from ionized intergalactic medium.  Ram pressure stripping would explain the lack of  gas in the southeast region of the galaxy and the overall cometary morphology in Mrk 178.  

VII Zw 403 appears to have a gas cloud in the line of sight that is rotating differently than the main \HI\ body. The low velocity dispersions in the cloud point to it being a cold gas cloud.  The gas cloud is likely an external gas cloud impacting the main \HI\ disk from behind and pushing it to the east.  The underlying disk of VII Zw 403, when most of the emission associated with the cloud in the line of sight has been removed, has a kinematic major axis that is misaligned with the morphological stellar and \HI\ major axes.  It is possible that VII Zw 403 is elongated or bar-like along the line of sight, resulting in the appearance of an offset kinematic major axis. 

NGC 3738 is a BCD with a stellar V-band disk that extends further than the VLA natural-weighted \HI\ disk. The GBT maps also pick up almost two times the \HI\ mass of the VLA natural-weighted maps.  It is therefore possible that NGC 3738 has a tenuous extended \HI\ halo.   It does not have any nearby \HI\ companions at the sensitivity of the GBT \HI\ map, therefore, it did not recently interact with a currently-gas-rich galaxy.  There are also multiple gas clouds around the disk that are moving at approximately the systemic velocity of NGC 3738.  These gas clouds may have been pushed outside of NGC 3738's main \HI\ disk due to stellar feedback.   It is also possible that the gas clouds external to the disk may have been ejected from the main \HI\ disk during a past merger than NGC 3738 has undergone. Another possibility is that NGC 3738 is experiencing face-on ram pressure stripping from ionized IGM.

Whether they have had their starburst triggered through mergers, interactions, or consumption of IGM, it is apparent that each BCD is different and requires individual assessment to understand what has happened to it.  This may be true for modeling what each BCDs' past and future may look like.  If BCDs have been triggered through different means, then there is no guarantee that they will evolve into/from the same type of object.  Different triggers for BCDs may also explain why it is so difficult to derive strict parameters to define this classification of galaxies \citep[for examples of definitions of BCDs see][]{thuan81, gil03}.   If they have been triggered differently, then their parameters likely span a much larger range than if they were all triggered in the same manner.

\acknowledgments
We would like to acknowledge Jay Lockman for his valuable conversations and help with Mrk 178's data.  We would also like to acknowledge Nick Pingel, Spencer Wolfe, and D.J. Pisano for their code and help with the Mrk 178 data.  We would also like to thank the anonymous referee for their many helpful comments that improved this paper.  Trisha Ashley was supported in part by the Dissertation Year Fellowship at Florida International University.  This project was funded in part by the National Science Foundation under grant numbers AST-0707563 AST-0707426, AST-0707468, and AST 0707835 to Deidre A. Hunter, Bruce G. Elmegreen, Caroline E. Simpson, and Lisa M. Young.  This research has made use of the NASA/IPAC Extragalactic Database (NED) which is operated by the Jet Propulsion Laboratory, California Institute of Technology, under contract with the National Aeronautics and Space Administration (NASA).  The National Radio Astronomy Observatory is operated by Associated Universities, Inc., under cooperative agreement with the National Science Foundation.

\appendix
\section{NGC 3738's baseline fitting} \label{appendix:n3738}

  During calibration the off-frequency spectrum is shifted to the on-frequency spectrum's central frequency and subtracted from the on-spectrum.  If a source is within \s3.5 MHz of the target's usable frequency range (also \s3.5 MHz in width, therefore the other source can be up to 5.25 MHz away from the central frequency of the target source), then the unwanted source that was originally 1.75-5.25 MHz away from the spectrum will appear closer in frequency space to the target source as a negative dip in intensity.  The unwanted source reduces the amount of frequency space that can be used to estimate the zero-emission baseline for subtraction.  Two sources were located outside of the target frequency range in NGC 3738's data and with calibration, have been reflected into NGC 3738's target frequency range: NGC 3733 (north of NGC 3738) and NGC 3756 (south of NGC 3738).  The real emission from NGC 3733 and NGC 3756 is very far away from NGC 3738 in velocity space (a difference of \s960 \kms and 1100 \kms, respectively).  It is therefore very unlikely that these sources are interacting with NGC 3738.  
 
\begin{figure}[!ht]
\epsscale{1}
\plotone{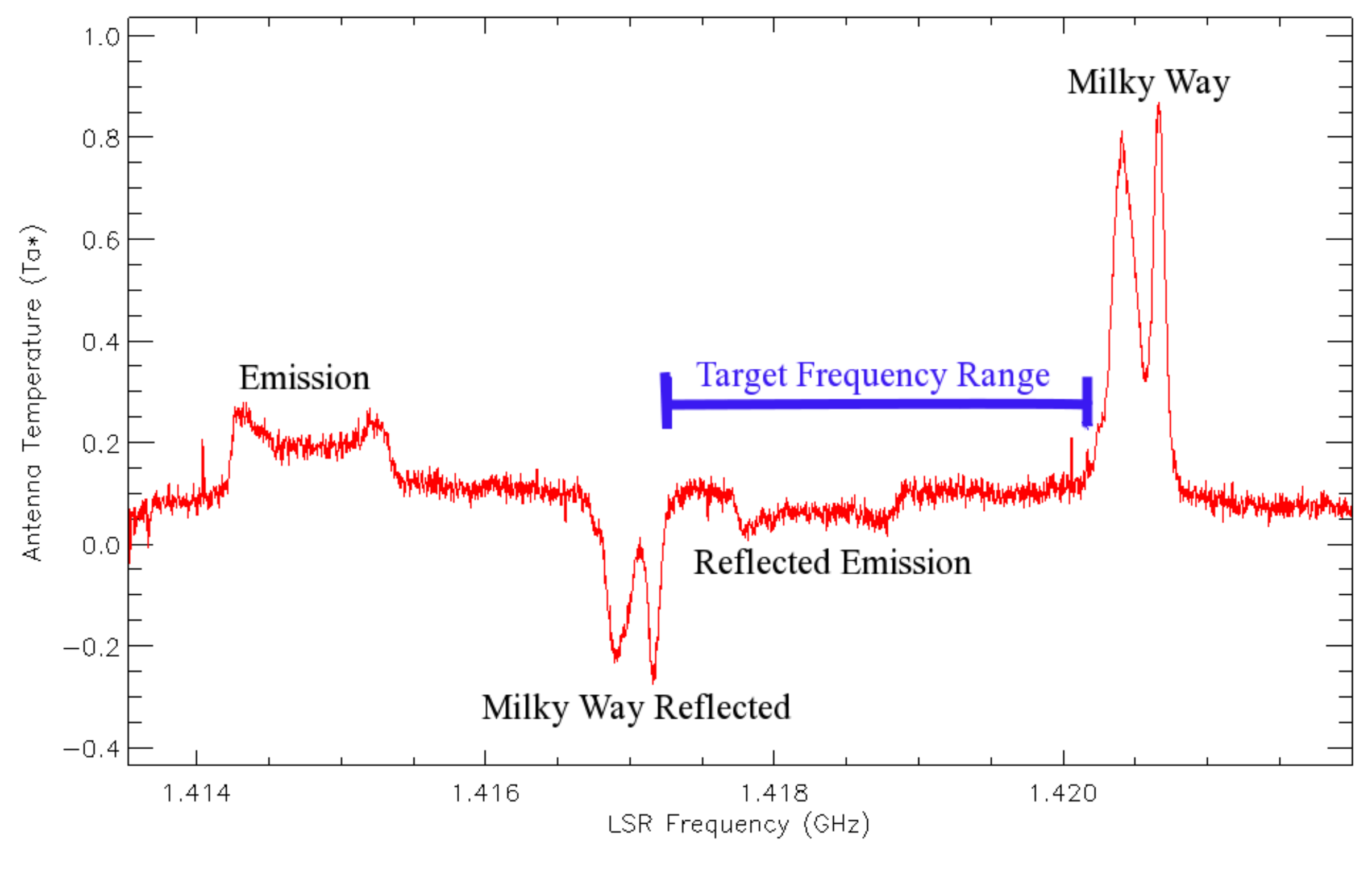}
\caption{The emission from NGC 3733 that was reflected into NGC 3738's GBT target frequency range. \label{n3738_pitag}}
\end{figure}

\begin{figure}[!ht]
\epsscale{0.8}
\plotone{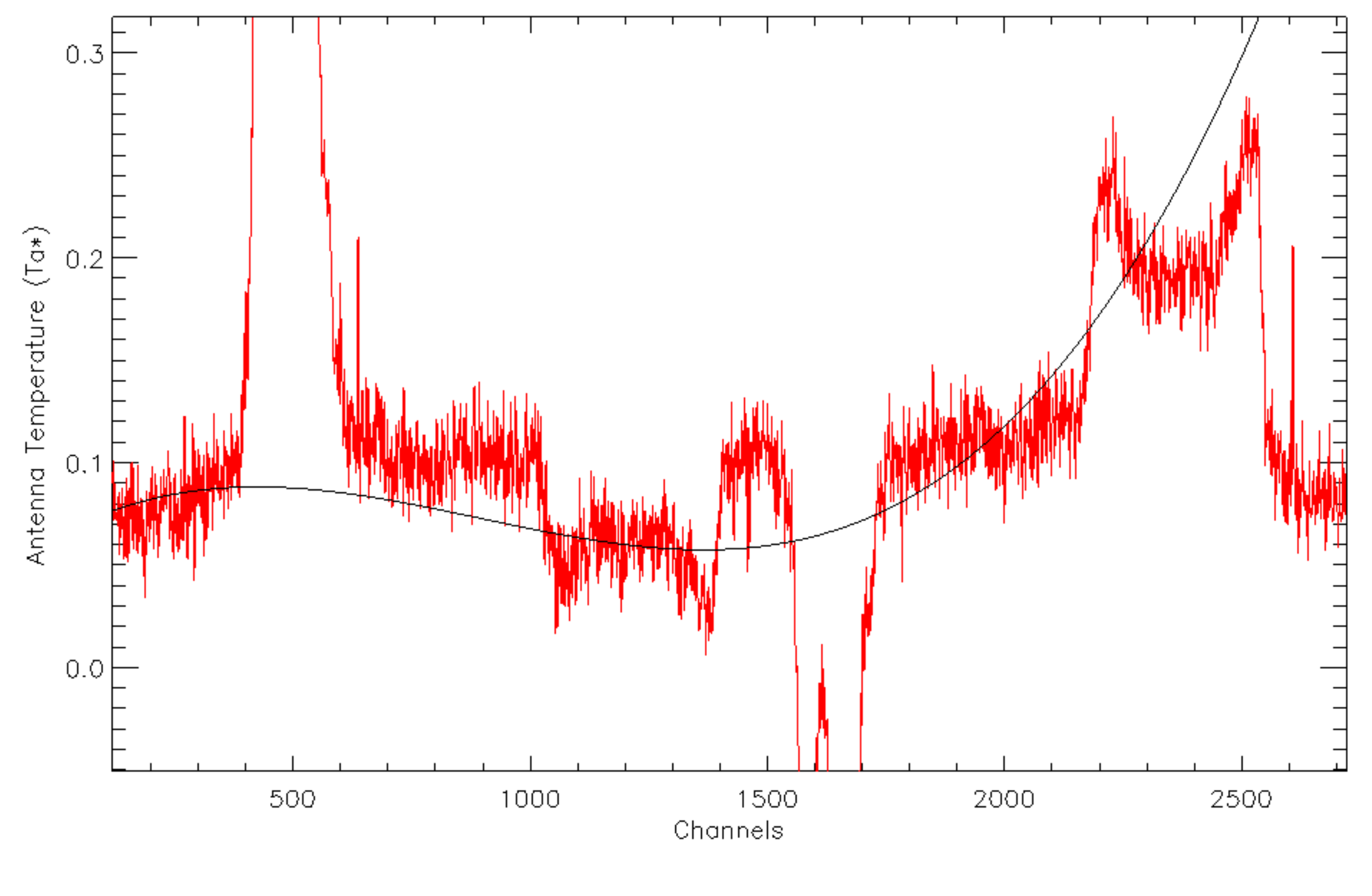}
\plotone{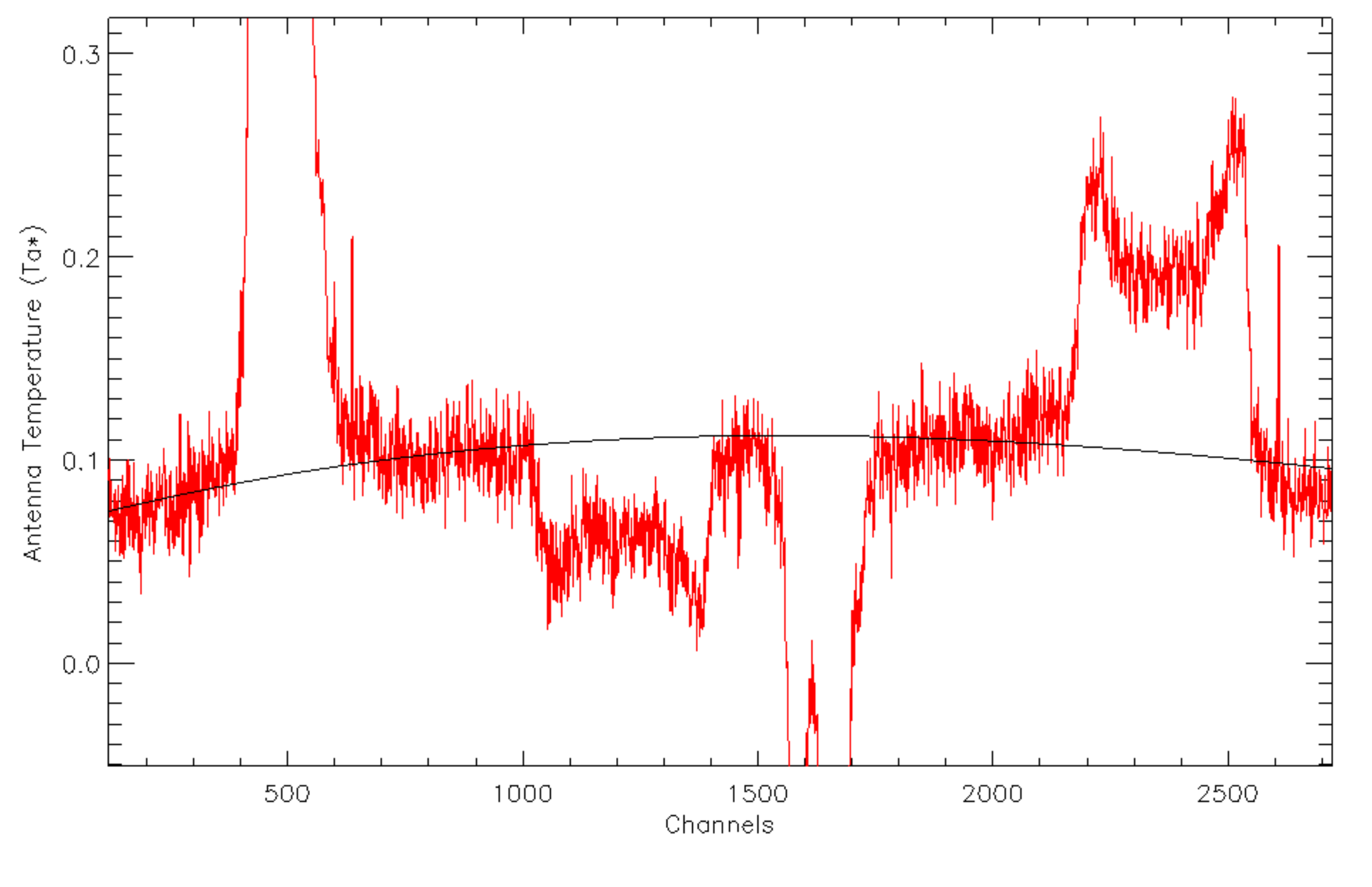}
\caption{\textit{Top:} An example of a bad GBT baseline fit using channels (0-250; 674-720; 1070-1380; 1820-2062).  \textit{Bottom:} An example of a good GBT baseline fit using channels (0-250; 674-900; 1820-2062). \label{n3738_pitag_fit}}
\end{figure}

   The spectra of the approximate space that NGC 3733 covered were summed and averaged to get a clear picture of where it is in frequency space; the resulting labeled spectrum can be seen in Figure~\ref{n3738_pitag}.  This spectrum has not had a baseline fit to it or the RFI removed from it, which is why the spectrum appears to have no zero line of emission and there are also spikes at \s1.414 MHz and \s1.420 MHz. The Milky Way and its reflection have been labeled as such.  The source with emission located at 1414.3-1415.3 MHz is NGC 3733, which has been reflected into the target frequency range at 1417.8-1418.8 MHz.  The frequency range that NGC 3733 occupies in the target frequency range could therefore no longer be used as zero emission space in a baseline fit.  If it was used or partially used, then this could result in a fake source at its location and generally poor fits since a low order polynomial is being fit to the baseline.  An example of a fit using the frequency range that includes the reflections of these sources can be seen in the top of Figure~\ref{n3738_pitag_fit}.  A second example of a fit using frequency ranges that do not overlap with the reflected source can be seen in the bottom of Figure~\ref{n3738_pitag_fit}.  
 
\begin{figure}[!ht]
\epsscale{1}
\plotone{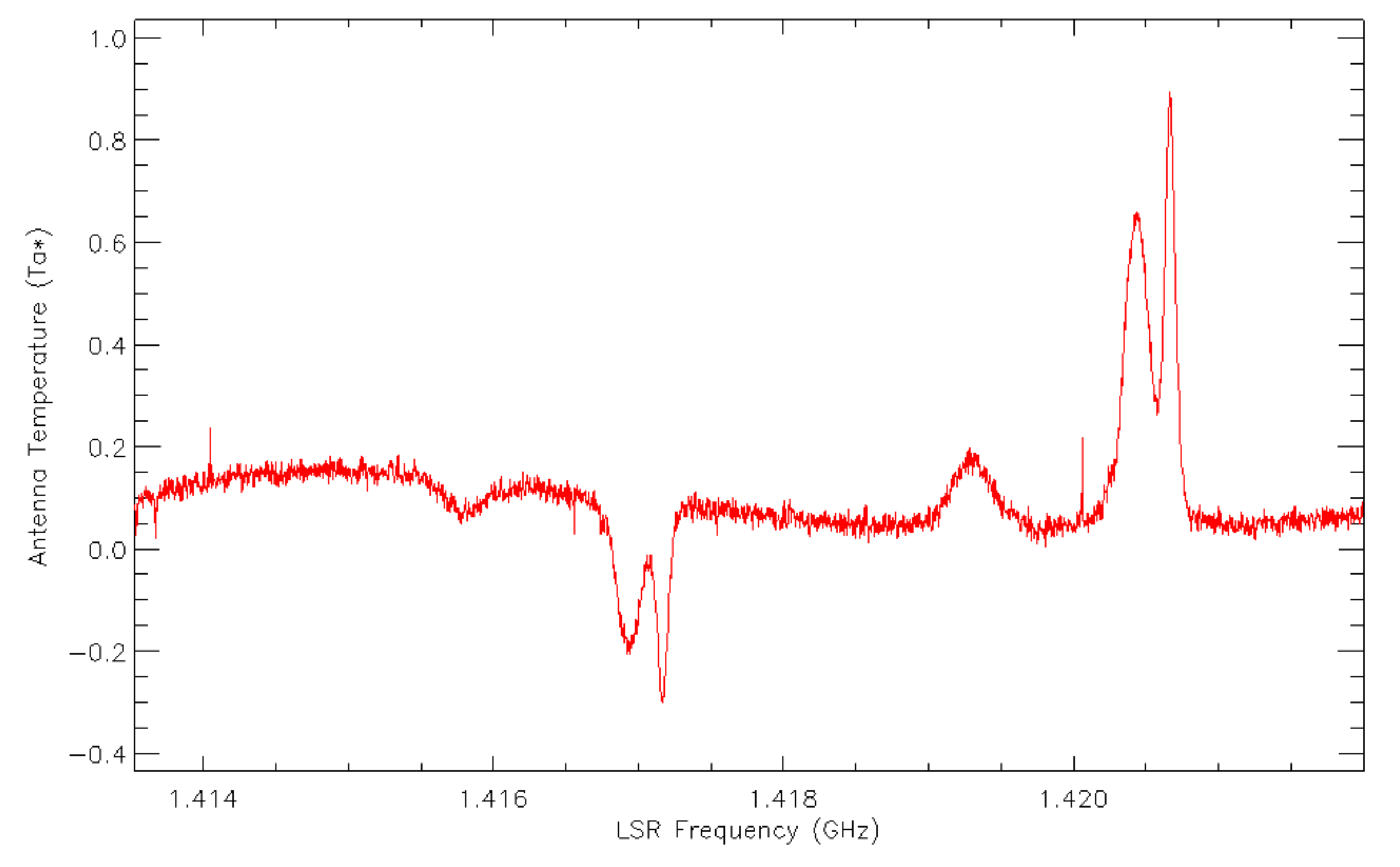}
\caption{The averaged spectra of NGC 3738 from the GBT data. \label{n3738_spectra}}
\end{figure}

\begin{figure}[!ht]
\epsscale{1}
\plotone{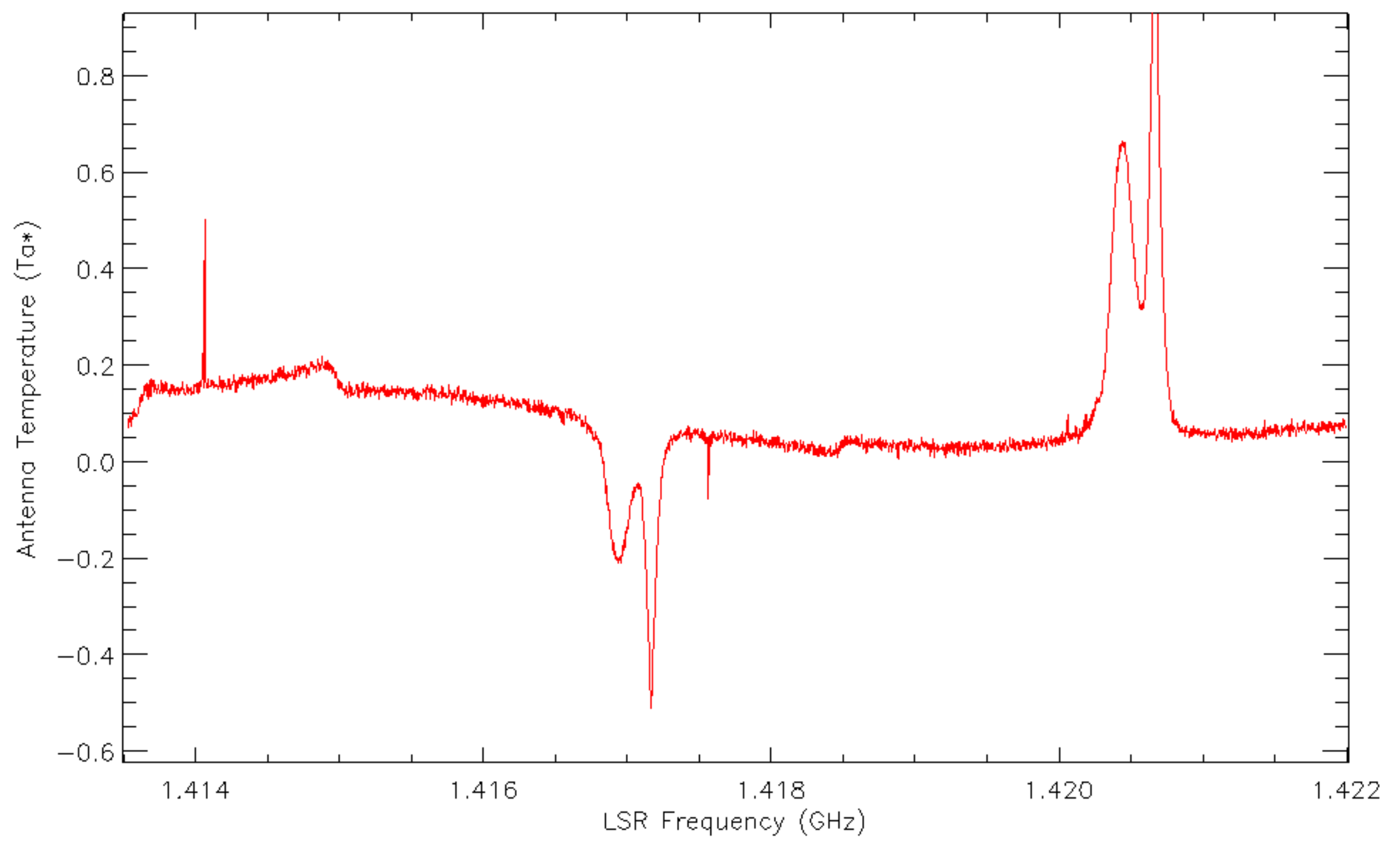}
\caption{The averaged GBT spectra of NGC 3756 that is reflected into the target frequency range of NGC 3738.  \label{n3738_topsource}}
\end{figure}

An averaged spectrum of NGC 3738's emission is shown in Figure~\ref{n3738_spectra}.  In this spectrum, NGC 3738 peaks at a frequency of 1419.3 MHz, and its reflection appears at a frequency of 1415.8 MHz.  The reflection now occurs outside of the target frequency range.  The emission of NGC 3738 and the reflection of NGC 3733 consume most of the target frequency range, leaving very little room to fit to a baseline.  Some of the emission and reflection-free space outside of the target frequency range could be used for baseline fitting when the baseline appears continuous, but using a significant portion of the target frequency range is necessary for a good fit.  There is also the added problem of another source, NGC 3756, being reflected into the target frequency range. NGC 3756's averaged spectrum is shown in Figure~\ref{n3738_topsource}.  In this figure, NGC 3756 has emission at 1415 MHz and has a reflection at 1418.5 MHz.

With NGC 3733, NGC 3756, and NGC 3738 on the GBT maps stretching throughout the target frequency range, it was impossible to fit one single baseline to all three sources.  Luckily, all three sources are separated in Galactic Latitude.  Therefore, the solution to the baseline problem was to split the GBT map into four spatial regions, with different baseline fit parameters for each region.  The first region included the raster scan rows (in Galactic Latitude) that contained only NGC 3733.  The second region included only the raster scan rows that contained NGC 3738.  The third region included NGC 3756.  The fourth region included everything else.  The baseline fit used for NGC 3738 was also used for this fourth region so that any sources close to NGC 3738 in velocity could be fit properly.  The rows with NGC 3733 and NGC 3756 also had their emission-free pixels searched for any possible emission that could be related to NGC 3738 or a companion.  This was done by averaging 10 spectra at a time to look for possible emission in the resulting averaged spectrum.  The exceptions to this 10-spectra-averaging were the spectra close to the reflected sources and the last four spectra of each row, where fewer spectra were available for averaging, but still averaged for inspection. No sources related to NGC 3738 were found in the target frequency range.  This method gives the four different regions slightly different noise levels, but the effect is relatively small with the rms values of the regions over \s10 \kms\ ranging from 3.8 to 4.5 $\times$ 10$^{16}$ cm$^{-2}$.    These rms values are likely higher than the overall rms over 10 \kms\ for the entirety of NGC 3738's GBT map because they were measured over smaller regions.

{}

\end{document}